\documentclass[11pt,english]{article}
\usepackage[T1]{fontenc}
\usepackage[latin9]{inputenc}
\usepackage{bm}
\usepackage{relsize}
\usepackage{amsmath}
\usepackage{graphicx}
\usepackage{amssymb}

\makeatletter

\DeclareRobustCommand{\greektext}{%
  \fontencoding{LGR}\selectfont\def\encodingdefault{LGR}}
\DeclareRobustCommand{\textgreek}[1]{\leavevmode{\greektext #1}}
\DeclareFontEncoding{LGR}{}{}

\DeclareRobustCommand{\lyxmathsym}[1]{\ifmmode\begingroup\def\b@ld{bold}
  \def\rmorbf##1{\ifx\math@version\b@ld\textbf{##1}\else\textrm{##1}\fi}
  \mathchoice{\hbox{\rmorbf{#1}}}{\hbox{\rmorbf{#1}}}
  {\hbox{\smaller[2]\rmorbf{#1}}}{\hbox{\smaller[3]\rmorbf{#1}}}
  \endgroup\else#1\fi}

\providecommand{\tabularnewline}{\\}

\newcommand{\lyxaddress}[1]{
\par {\raggedright #1
\vspace{1.4em}
\noindent\par}
}

\usepackage{psfrag}
\usepackage{bm}

\makeatother

\usepackage{babel}

\begin{document}

\title{Some aspects of color superconductivity: an introduction}

\author{Qun Wang}

\maketitle

\lyxaddress{Interdisciplinary Center for Theoretical Study and Department of
Modern Physics, University of Science and Technology of China, Anhui
230026, People's Republic of China}
\begin{verse}
\textbf{Abstract}: A pedagogical introduction to color superconductivity
in the weak coupling limit is given. The focus is on the basic tools
of thermal field theory necessary to compute observables of color
superconductivity. The rich symmetry structure and symmetry breaking
patterns are analyzed on the basis of the Anderson-Higgs mechanism.
Some techniques can also be applied for computing neutrino processes
in compact stars. As an example, we show how to obtain the neutrino
emissivity for Urca processes in neutron stars by computing the polarization
tensor of the W-boson. We also illustrate how a spin-1 color superconducting
phase generates an anisotropic neutrino emissions in compact stars. 

\end{verse}

\section{What is color superconductivity?}

Nowadays, superconductivity in a wide range of materials and superfluidity
of Helium-3 can be generated with practical equipments in laboratories.
It is well known that the underlying mechanism of all these 'super'
phenomena is the so-called Cooper paring of fermions (electrons in
superconductors, fermionic $^{3}\mathrm{He}$ atoms in Helium superfluids)
near the Fermi surface of the corresponding fermions. These phenomena
are all of electromagnetic nature and at atomic level. In the regime
of strong interaction where quarks and gluons are elementary particles
described by quantum chromodynamics (QCD), a similar phenomena, the
so called \emph{color superconductivity} was first proposed in cold
dense quark matter by Barrois \cite{Barrois:1977xd}, Bailin and Love
\cite{Bailin:1983bm}, and further developed by many others (for reviews,
see \cite{Rajagopal:2000wf,Alford:2001dt,Rischke:2003mt,Casalbuoni:2003wh,Schafer:2003vz,Buballa:2003qv,Ren:2004nn,Shovkovy:2004me,Huang:2004ik,Gerhold:2005nt,Rebhan:2005hw,Alford:2007xm}).
However color superconducting quark matter cannot be produced in any
laboratory in the foreseeable future. The reason is that this exotic
phase of matter requires extremely high baryonic densities and relatively
low temperatures. In nature, such conditions are only realized in
the cores of cold compact stars, i.e. in relatively old remnants of
supernova explosions. 

In a color superconductor, quarks form Cooper pairs similarly to electrons
in a normal superconductor or to fermionic $^{3}\mathrm{He}$ atoms
in a Helium superfluid. Quarks carry color charges and interact via
gluon exchange (the gluon is the QCD gauge boson). At very high baryonic
densities the quark chemical potential $\mu$ is much larger than
the QCD scale $\Lambda_{QCD}\sim200\;\mathrm{MeV}$, the running coupling
constant $\alpha_{S}(\mu)$ is thus small, because QCD is an asymptotically
free theory for quarks and gluons. This was first shown by Gross,
Politzer and Wilczek in 1973 \cite{Gross:1973id,Politzer:1973fx}
and awarded the Nobel Prize in 2004 \cite{gross-nobellecture,wilczek-nobellecture}.
In this case, the interaction between quarks is dominated by one-gluon
exchange. As will be demonstrated explicitly at the end of this section
there is an attractive channel in one-gluon exchange providing the
binding agent for the quark Cooper pair. This is in contrast to normal
superconductors where an electron Cooper pair is formed indirectly
via electron-lattice interaction (direct electron-electron interaction
is repulsive). In this sense color superconductivity is simpler than
the normal one. In hadronic matter quarks are confined in color-neutral
hadrons. Only at sufficiently high baryonic densities when these hadrons
overlap and eventually free the quarks out of their hadronic domains,
the quarks can move in much larger space-time volumes than the size
of hadrons. This phenomenon is called deconfinement. It is expected
to occur in the core of some compact stellar objects as neutron stars,
where the baryonic densities are a few times larger than the nuclear
matter saturation density $\rho_{0}\sim0.14\;\mathrm{fm}^{-3}$, which
corresponds to a quark chemical potential $\mu\sim400-500\;\mathrm{MeV}$. 

Since quarks are fermions, the Pauli exclusion principle does not
allow two quarks to occupy the same quantum state. At zero temperature,
they fill up states up to the so-called Fermi momentum $k_{F}$ (or
the Fermi surface) above which all states are empty. Due to Pauli
blocking, only quarks near the Fermi surface are able to interact
and exchange momenta. The exchanged momenta in quark-quark scattering
are of order $k_{F}$. In the ultra-relativistic limit, we have $E_{F}=\sqrt{k_{F}^{2}+m^{2}}\approx k_{F}\approx\mu$.
We know that the quark density $\rho\sim k_{F}^{3}\approx\mu^{3}$,
the high baryonic density means large Fermi momentum $k_{F}$ or chemical
potential $\mu$. We will consider asymptotically large quark chemical
potentials, $\mu\gg\Lambda_{QCD}$, the so-called weak coupling approach,
which means high density and large mementum transfer. The weak coupling
approach enables one to study the phenomena in a rigorous or well-controlled
way. For reviews of the weak coupling appoach, see, for example, \cite{Rischke:2003mt,Rebhan:2005hw}.
The physical predictions of such an approach certainly have to be
interpreted very carefully, since for realistic $\mu$ one has $\alpha_{s}(\mu)\sim1$.
But aided by the resummation methods, the perturbative analysis can
be extrapolated even to realistic densities such as those in the cores
of neutron stars \cite{Rebhan:2005hw}. Support also comes from the
NJL model \cite{Alford:1997zt,Rapp:1997zu}, a simple effective theory
of QCD without gluons but with a local four-quark interaction. It
predicts color superconductivity also at moderate baryonic densities
with gaps of order of $\phi\sim100\;\mathrm{MeV}$, which matches
the gap value extropolated from the weak coupling approach. Recently
there are many developments in the NJL model \cite{Ruster:2004eg,Fukushima:2004zq,Ruster:2005jc,Blaschke:2005uj}
and in the random matrix model \cite{Vanderheyden:2000ti,Vanderheyden:2001gx},
a more rigorous effective model, to describe the color superconducting
phase diagram.

\begin{figure}
\caption{\label{fig:One-gluon-exchange}One gluon exchange}

\vspace{0.5cm}

\includegraphics[scale=1.2]{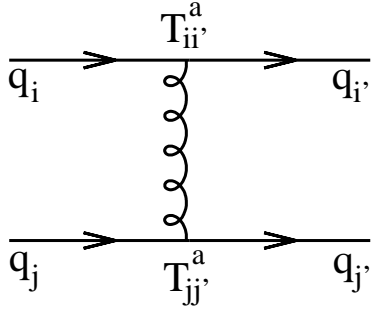}
\end{figure}

So at high baryonic densities where the strong coupling constant is
small, the single gluon exchange dominates the quark-quark scattering,
as shown in Fig. (\ref{fig:One-gluon-exchange}), whose amplitude
is proportional to \begin{eqnarray}
\sum\limits _{a=1}^{N_{c}^{2}-1}T_{a}^{ii^{\prime}}T_{a}^{jj^{\prime}} & = & -\frac{N_{c}+1}{4N_{c}}\left(\delta_{ii^{\prime}}\delta_{jj^{\prime}}-\delta_{ij^{\prime}}\delta_{i^{\prime}j}\right)\nonumber \\
 &  & +\frac{N_{c}-1}{4N_{c}}\left(\delta_{ii^{\prime}}\delta_{jj^{\prime}}+\delta_{ij^{\prime}}\delta_{i^{\prime}j}\right)\;.\label{eq:antitrip}\end{eqnarray}
 Here, $N_{c}=3$ is the number of colors and $T_{a}$ are generators
of the $SU(3)_{c}$ gauge group. Here $T_{a}\equiv\lambda_{a}/2$
where $\lambda_{a}$ are the Gell-Mann matrices. The indices $i,j$
are the fundamental colors of two quarks in the incoming channel,
while $i^{\prime},j^{\prime}$ their resprective colors in the outgoing
channel, see Fig. (\ref{fig:One-gluon-exchange}). Interchanging two
color indices in the incoming or outgoing channel changes the sign
of the first term while the second term remains intact. The minus
sign in front of the asymmetric term indicates that this channel is
attractive, as it is similar to the Coulomb interaction between a
negative electric charge and a positive one (the product of two charges
is negative). This finding is crucial, as due to Cooper's theorem
any arbitrarily weak attractive interaction will destabilize the Fermi
surface in favor of the formation of Cooper pairs. Since the pairs
are of bosonic nature they will condense at sufficiently low temperatures
into a Bose-Einstein condensate (in a more rigorous way, Cooper pairs
are Bose-Einstein condensate in strong coupling limit). 

In group theory, Eq. (\ref{eq:antitrip}) corresponds to deducing
the direct product of two triplets into one color antitriplet (anti-symmetric)
and one color sextet (symmetric) \begin{eqnarray}
[3]_{c}\otimes[3]_{c} & = & [\bar{3}]_{c}^{a}\oplus[6]_{c}^{s}\;.\label{antitriprep}\end{eqnarray}
Now we illustrate how two triplets makes an \emph{anti}-triplet. Denote
$\psi_{i}$ the quark field following the $SU(3)$ transformation:
\begin{eqnarray}
\psi'_{i} & = & g_{ij}\psi_{j},\end{eqnarray}
where $g$ is a $SU(3)$ element. An anti-triplet can be written by
$\widetilde{\psi}_{k}=\epsilon_{ijk}\psi_{i}\psi_{j}$, which transforms
as \begin{eqnarray}
\widetilde{\psi}'_{k} & = & \epsilon_{ijk}\psi'_{i}\psi'_{j}\nonumber \\
 & = & \epsilon_{ijk}g_{ii'}g_{jj'}\psi_{i'}\psi_{j'}\nonumber \\
 & = & \delta_{k'k}\epsilon_{ijk'}g_{ii'}g_{jj'}\psi_{i'}\psi_{j'}\nonumber \\
 & = & g_{k'l}g_{lk}^{\dagger}\epsilon_{ijk'}g_{ii'}g_{jj'}\psi_{i'}\psi_{j'}\nonumber \\
 & = & g_{lk}^{\dagger}\mathrm{det}(g)\epsilon_{i'j'l}\psi_{i'}\psi_{j'}\nonumber \\
 & = & \widetilde{\psi}_{l}g_{lk}^{\dagger},\end{eqnarray}
where we have used $\epsilon_{ijk'}g_{ii'}g_{jj'}g_{k'l}=\mathrm{det}(g)\epsilon_{i'j'l}$
and $\mathrm{det}(g)=1$. We see that the anti-symmetric diquark field
$\widetilde{\psi}_{k}$ transforms like an anti-quark.  

From the anti-symmetry of the attractive channel it follows that the
two quarks in a Cooper pair must carry different colors. In addition
to colors, quarks also come along with spin and flavor quantum numbers.
Since the total wave function of a Cooper pair has to be anti-symmetric
under the exchange of two quarks, the combined spin-flavor part has
to be symmetric. This requirement will be used in the following section
to classify various possible color superconducting phases.

\section{Structure of diquark condensate}

In this section we discuss the structure of diquark condensate \cite{Buballa:2003qv,Pisarski:1999bf}.
A diquark condensate is defined as an expectation value \begin{equation}
\mathcal{M}=\left\langle \overline{\psi}_{c}\mathcal{O}\psi\right\rangle ,\end{equation}
where $\psi$ is the quark field with all Dirac, color and flavor
indices, $\overline{\psi}_{c}$ is its charge conjugate field defined
by $\overline{\psi}_{c}=\psi^{T}C$ with the charge conjugate operator
$C=i\gamma^{2}\gamma_{0}$. The operator $\mathcal{O}$ has Dirac,
color and flavor parts: \begin{eqnarray}
\mathcal{O} & = & \mathcal{O}_{Dirac}\otimes\mathcal{O}_{color}\otimes\mathcal{O}_{flavor}.\end{eqnarray}
The charge conjugate operator can be determined by transforming the
Dirac equation of a fermion to that of an anti-fermion, \begin{eqnarray}
 &  & (p_{\mu}\gamma^{\mu}-m)\psi=0\nonumber \\
 & \rightarrow & \psi^{T}(p_{\mu}\gamma^{\mu T}-m)=0\nonumber \\
 & \rightarrow & \psi^{T}C(p_{\mu}C^{-1}\gamma^{\mu T}C-m)=0\nonumber \\
 & \rightarrow & \overline{\psi}_{c}(p_{\mu}C^{-1}\gamma^{\mu T}C-m)=0\nonumber \\
 & \rightarrow & \overline{\psi}_{c}(p_{\mu}\gamma^{\mu}+m)=0.\end{eqnarray}
One can verify that $C=i\gamma^{2}\gamma_{0}$ satisfies $C^{-1}\gamma^{\mu T}C=-\gamma^{\mu}$
or $C\gamma^{\mu T}C=\gamma^{\mu}$ using $\gamma^{\mu T}=\gamma^{0},-\gamma^{1},\gamma^{2},-\gamma^{3}$
for $\mu=0,1,2,3$. Exploiting the anti-symmetric property of the
fermionic quark fields, one finds \begin{eqnarray}
\mathcal{M} & = & \left\langle \overline{\psi}_{c}\mathcal{O}\psi\right\rangle =\left\langle \psi^{T}C\mathcal{O}\psi\right\rangle \nonumber \\
 & = & \left\langle \psi_{i}(C\mathcal{O})_{ij}\psi_{j}\right\rangle \nonumber \\
 & = & -\left\langle \psi_{j}(C\mathcal{O})_{ji}^{T}\psi_{i}\right\rangle =-\left\langle \psi^{T}(C\mathcal{O})^{T}\psi\right\rangle .\end{eqnarray}
So the operator $C\mathcal{O}$ obeys \begin{equation}
(C\mathcal{O})^{T}=-C\mathcal{O}.\end{equation}
This means that the operator $C\mathcal{O}$ is anti-symmetric under
transposition. We can choose all three parts of $C\mathcal{O}$, i.e.
Dirac, color and flavor parts, anti-symmetric or two of them symmetric
and the third one anti-symmetric. In Tab. (\ref{tab:sym-transposition})
a variety of choices to build up the operator $\mathcal{O}$ is given
together with the corresponding symmetric properties under transposition. 

\begin{table}
\caption{\label{tab:sym-transposition}Symmetric properties of $\mathcal{O}$
under transposition. In the row of Dirac part, 'S', 'P', 'V', 'A',
'T' means that the operator $C\mathcal{O}$ is a scalar, pseudo-scalar,
vector, axial-vector and tensor, respectively. In the U(2) row, $\sigma_{i}$
are Pauli matrices, while in the U(3) row, $\lambda_{i}$ are Gell-Mann
matrices. }

\vspace{0.5cm}

\begin{tabular}{|c|c|c|}
\hline 
 & Anti-symmetric & Symmetric\tabularnewline
\hline
\begin{tabular}{c}
Dirac\tabularnewline
(combined with $C$)\tabularnewline
\end{tabular} & \begin{tabular}{ccc}
$\gamma_{5}$ & 1 & $\gamma^{\mu}\gamma_{5}$\tabularnewline
S & P & V\tabularnewline
\end{tabular} & \begin{tabular}{cc}
$\gamma^{\mu}$ & $\sigma^{\mu\nu}$\tabularnewline
A & T\tabularnewline
\end{tabular}\tabularnewline
\hline 
U(2) & \begin{tabular}{c}
$\sigma_{2}$\tabularnewline
singlet {[}1{]}\tabularnewline
\end{tabular} & \begin{tabular}{c}
1, $\sigma_{1}$, $\sigma_{3}$\tabularnewline
triplet {[}3{]}\tabularnewline
\end{tabular}\tabularnewline
\hline 
U(3) & \begin{tabular}{c}
$\lambda_{2}$, $\lambda_{5}$, $\lambda_{7}$\tabularnewline
anti-triplet {[}$\overline{3}${]}\tabularnewline
\end{tabular} & \begin{tabular}{c}
1, $\lambda_{1}$, $\lambda_{3}$, $\lambda_{4}$, $\lambda_{6}$,
$\lambda_{8}$ \tabularnewline
sextet {[}6{]}\tabularnewline
\end{tabular}\tabularnewline
\hline
\end{tabular}
\end{table}

Now we show how to determine the parity of the condensate. Suppose
under parity transformation $\mathcal{P}$, the field $\psi(t,\mathbf{x})$
changes as follows\begin{eqnarray}
\mathcal{P}\psi(t,\mathbf{x})\mathcal{P} & = & \eta\gamma_{0}\psi(t,-\mathbf{x})\nonumber \\
\mathcal{P}\overline{\psi}(t,\mathbf{x})\mathcal{P} & = & \eta^{*}\overline{\psi}(t,-\mathbf{x})\gamma_{0},\end{eqnarray}
where $\eta$ is a complex number. Then the conjugate fields transform
as\begin{eqnarray}
\mathcal{P}\psi_{c}(t,\mathbf{x})\mathcal{P} & = & \eta_{c}\gamma_{0}\psi_{c}(t,-\mathbf{x})\nonumber \\
\mathcal{P}\overline{\psi}_{c}(t,\mathbf{x})\mathcal{P} & = & \eta_{c}^{*}\overline{\psi}_{c}(t,-\mathbf{x})\gamma_{0}.\end{eqnarray}
So the condensate transforms as\begin{eqnarray}
\mathcal{P}\overline{\psi}_{c}(t,\mathbf{x})\mathcal{O}\psi(t,\mathbf{x})\mathcal{P} & = & \eta_{c}^{*}\eta\overline{\psi}_{c}(t,-\mathbf{x})\gamma_{0}\mathcal{O}\gamma_{0}\psi(t,-\mathbf{x})\nonumber \\
 & = & -\overline{\psi}_{c}(t,-\mathbf{x})\gamma_{0}\mathcal{O}\gamma_{0}\psi(t,-\mathbf{x}),\end{eqnarray}
where use was made of the fact that the parity of the conjugate particle
is opposite to that of the particle, i.e. $\eta_{c}^{*}=-\eta$. Then
$\mathcal{O}=\gamma_{5},1,\gamma^{\mu}\gamma_{5},\gamma^{\mu},\sigma^{\mu\nu}$
correspond to a scalar, pseudo-scalar, vector, pseudo-vector and tensor,
respectively, as shown in the second row of Tab. (\ref{tab:sym-transposition}).
For example, $\mathcal{O}=\gamma^{\mu}\gamma_{5}$ gives a vector,
since \begin{eqnarray}
\mathcal{P}\overline{\psi}_{c}(t,\mathbf{x})\gamma^{0}\gamma_{5}\psi(t,\mathbf{x})\mathcal{P} & = & \overline{\psi}_{c}(t,-\mathbf{x})\gamma^{0}\gamma_{5}\psi(t,-\mathbf{x}),\nonumber \\
\mathcal{P}\overline{\psi}_{c}(t,\mathbf{x})\gamma^{i}\gamma_{5}\psi(t,\mathbf{x})\mathcal{P} & = & -\overline{\psi}_{c}(t,-\mathbf{x})\gamma^{i}\gamma_{5}\psi(t,-\mathbf{x}),\end{eqnarray}
which transforms as a vector. 

From spin-0 pairing (see, e.g. \cite{Alford:1997zt,Rapp:1997zu,Pisarski:1999bf,Hong:1999fh}),
we require that the color part is anti-symmetric with respect to exchanging
two quarks, i.e. the color part is in the anti-triplet channel, and
that the Dirac part must be a Lorentz scalar or pseudo-scalar, i.e.
the Dirac part of $\mathcal{O}$ must be $\gamma_{5}$ or $1$ which
is anti-symmetric combined with $C$. Therefore the flavor part should
be anti-symmetric too, i.e. it must be a flavor singlet in the two-flavor
case or a flavor anti-triplet in the three-flavor case. The condensate
is then a $J^{P}=0^{+}$(spin-parity) or $J^{P}=0^{-}$ bound state
of two quarks depending on whether the Dirac part is $\gamma_{5}$
or $1$. For the spin-1 pairing (see, e.g. \cite{Schafer:2000tw,Schmitt:2002sc,Alford:2002rz,Schmitt:2003xq,Schmitt:2004hg,Schmitt:2004et,Aguilera:2005tg,Brauner:2008ma,Brauner:2009df}),
the color part must be anti-triplet (anti-symmetric) and the Dirac
part can be a vector (symmetric) or an axial-vector (anti-symmetric),
so the flavor part can be symmetric or anti-symmetric respectively.
There is another example for the spin-1 pairing in the single-color
case where the color part is symmetric and flavor one anti-symmetric
\cite{Buballa:2002wy}. If the single flavor pairing occurs (the flavor
part is symmetric), the Dirac part must be $\gamma^{\mu}$. Therefore
a spin-1 Cooper pair is a $J^{P}=1^{-}$ state. Tab. (\ref{tab:condensates-structure})
summarizes the structure of diquark condensates in spin-0 and spin-1
pairings. 

\begin{table}
\caption{\label{tab:condensates-structure}Structure of diquark condensates
$\mathcal{M}$, the Dirac part is refered to that of the operator
$\mathcal{O}$ (not that of $C\mathcal{O}$). $J^{P}$ is the spin-parity.
Note that the color part is always in anti-triplet for all condensates.}

\vspace{0.5cm}

\begin{tabular}{|c|c|c|}
\hline 
pairing & Dirac part ($J^{P}$) & flavor multiplet (number of flavor)\tabularnewline
\hline
spin-0 & $\gamma_{5}$($0^{+}$),1 ($0^{-}$) & 1 ($N_{f}=2$), $\overline{3}$ ($N_{f}=3$)\tabularnewline
\hline
spin-1 & $\gamma^{\mu}$ ($1^{-}$)  & 1 ($N_{f}=1$), 3 ($N_{f}=2$), 6 ($N_{f}=3$)\tabularnewline
\hline
spin-1 & $\gamma^{\mu}\gamma_{5}$ ($1^{+}$) & 1 ($N_{f}=2$), $\overline{3}$ ($N_{f}=3$)\tabularnewline
\hline
\end{tabular}
\end{table}

Generally the gap of a spin-0 pairing is much larger than that of
a spin-1 pairing. In the ideal case there is more energy benefit for
the spin-0 pairing compared to the spin-1 pairing, so the spin-0 pairing
is favorable. But in the real world, there are many factors to disfavor
the spin-0 pairing. For example, when electric neutrality is taken
into account, there is a difference between the chemical potential
of the strange quark and that of light quarks $u$ and $d$ due to
the large strange quark mass. Additionally, the chemical potentials
of $u$ and $d$ quarks differ due to $\beta$-equilibrium. If the
differences of the chemical potentials are large enough, the spin-0
pairing, e.g. CFL phase, is not favorable \cite{Huang:2004bg,Huang:2004am,Casalbuoni:2004tb,Alford:2005qw},
and the spin-1 pairing or LOFF pairing \cite{Alford:2000ze} might
be the true ground state \cite{Giannakis:2004pf,Giannakis:2005vw}.
Another possibility to kill the spin-0 pairing is the presence of
a strong magnetic field which favors the symmetric spin wave function
\cite{Schmitt:2003xq}.

\section{Spontaneous symmetry breaking}

\label{sec:Sym1}A symmetry of the Lagrangian is said to be spontaneously
broken if the ground state or the vacuum of the system is not invariant
under the operations of that symmetry. We call this phenomenon spontaneous
symmetry breaking. 

\begin{figure}
\caption{\label{fig:ferromagnetism}The ferromagnetism as an example of spontaneous
symmetry breaking.}

\vspace{0.5cm}

\includegraphics[scale=0.8]{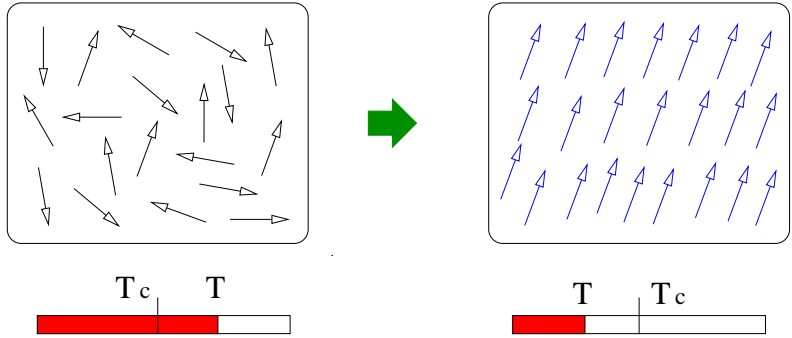}
\end{figure}

A well-known example is the ferromagnetism arising from the spin-spin
coupling, see Fig. (\ref{fig:ferromagnetism}). The Lagrangian describing
the ferromagnetism is invariant with respect to a $SO(3)$ rotation.
Above the transition temperature $T_{c}$, the vacuum state of the
spin system is invariant with respect to $SO(3)$ rotation since the
spin orientations are totally random and any rotation does not change
the spin state macroscopically. Below $T_{c}$ the vacuum spontaneously
chooses one magnetization direction, so the groud state is no longer
invariant under $SO(3)$ rotation. The $SO(3)$ symmetry of the vacuum
is spontaneously broken to $SO(2)$ which describes the rotational
symmetry around the total spin direction. We thus write this symmetry
breaking pattern as $G=SO(3)\rightarrow H=SO(2)$, where $G$ is the
total symmetry and $H$ the residual symmetry. 

\begin{figure}
\caption{\label{fig:locking-state}The locking state as an example of spontaneous
symmetry breaking. }

\vspace{0.5cm}

\includegraphics[scale=0.5]{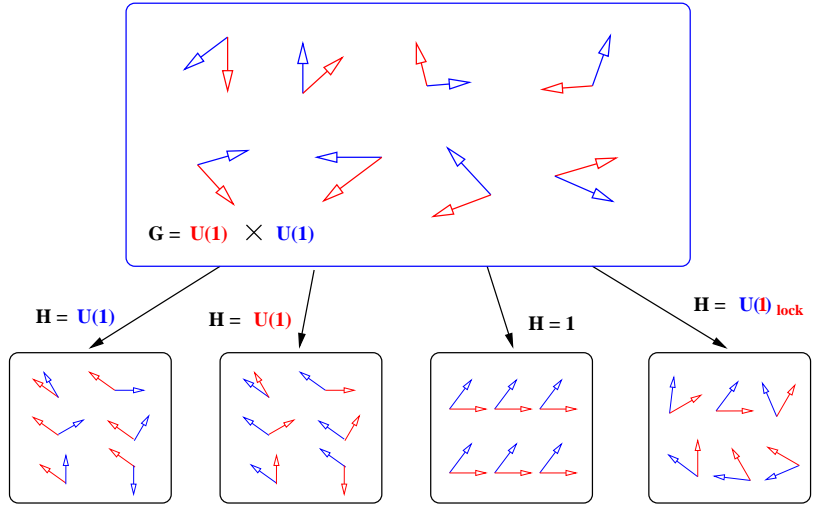}
\end{figure}

In the second example of spontaneous symmetry breaking the concept
of the locking state is explained. Suppose we have a system composed
of red and blue spins of $U(1)$ symmetry, see Fig. (\ref{fig:locking-state}).
Above a transition temperature $T_{c}$, the red and blue spin orientations
are totally random. Below $T_{c}$, there are four possibilities of
breaking the total symmetry spontaneously:\begin{eqnarray}
G=U(1)_{red}\otimes U(1)_{blue} & \rightarrow & H=U(1)_{blue}\nonumber \\
G=U(1)_{red}\otimes U(1)_{blue} & \rightarrow & H=U(1)_{red}\nonumber \\
G=U(1)_{red}\otimes U(1)_{blue} & \rightarrow & H=1\nonumber \\
G=U(1)_{red}\otimes U(1)_{blue} & \rightarrow & H=U(1)_{red+blue}\end{eqnarray}
 In the first one the red spin symmetry is broken while the blue one
remains. In the second one the blue spin symmetry is broken but the
red one remains. In the third breaking pattern, both the red and blue
spin symmetries are spontaneously broken. The fourth pattern is special
in that both the red and blue spins point in any directions but their
relative angle is fixed. This symmetry breaking pattern is called
locking in the sense that the state is invariant under the joint rotation
of the red and blue spins with the same rotational angle. The locking,
for example, is the underlying mechanism for various phases in the
superfluid Helium-3 \cite{Vollhardt1990}. 

\begin{figure}
\caption{\label{fig:goldstone}Potential for complex scalar field}

\psfrag{potential}{$V(|\phi |^2)$}
\psfrag{Rephi}{Re($\phi$)}
\psfrag{Imphi}{Im($\phi$)}
\psfrag{TgreaterTc}{$T>T_c$}
\psfrag{TlessTc}{$T<T_c$}
\psfrag{massive}{\small massive}
\psfrag{massless}{\small massless}\vspace{0.5cm}

\includegraphics[scale=0.8]{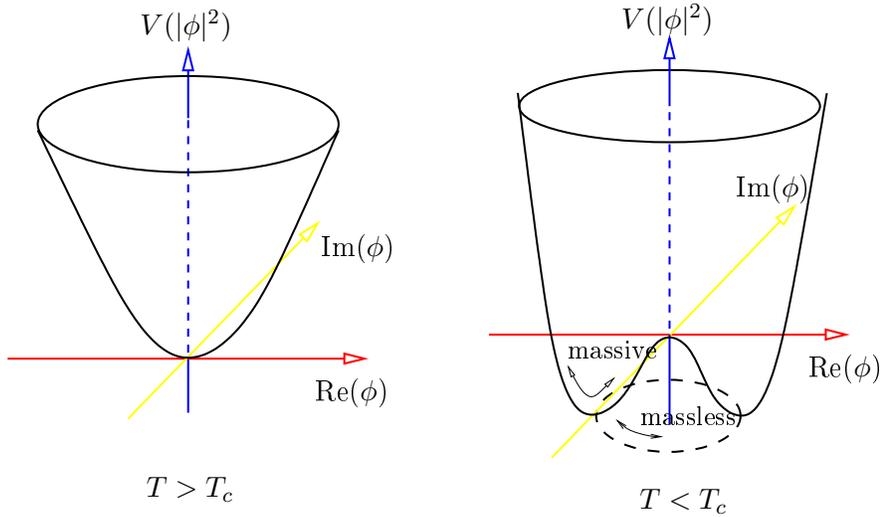}
\end{figure}

The well-known Nambu-Goldstone theorem \cite{Nambu:1961tp,Goldstone:1961eq}
states that the spontaneous breaking of a continuous global symmetry
implies a massless spin zero or scalar particle, and that each broken
generator of the symmetry group gives rise to one such particle. Consider
the complex scalar field as an example. The potential can be written
as \begin{eqnarray}
V(|\phi|^{2}) & = & a|\phi|^{2}+b|\phi|^{4},\end{eqnarray}
where $a\sim a'(T-T_{c})$ and $a',b>0$, see Fig. (\ref{fig:goldstone}).
The above potential is invariant under $U(1)$ transformation, $\phi'=e^{i\theta}\phi$.
Above $T_{c}$ the potential exhibits a minimum at $\phi=0$ which
is intact under $U(1)$ transformation. Below $T_{c}$ the potential
is minimal along the circle with radius $|\phi|=\phi_{0}$, because
the coefficient $a$ has different signs below and above $T_{c}$.
The symmetry of $U(1)$ is spontaneously broken because the vacuum
is changed after any $U(1)$ transformation $\phi'=e^{i\theta}\phi=e^{i\theta'}\phi_{0}$.
One sees that below $T_{c}$ there is a massless mode (the so-called
Goldstone mode) along the circle of minima. 

A general proof of the Goldstone theorem can be given as follows.
Suppose a given potential is invariant under the transformation of
a continuous group: \begin{equation}
\delta\Phi=i\theta^{a}C^{a}\Phi,\end{equation}
where $\Phi=[\phi_{1},\cdots,\phi_{n_{G}}]^{T}$ with $\phi_{i}$
being the real scalar field component and $n_{G}$ the number of real
scalar fields. $C^{a}$ are operators of the corresponding Lie algebra
of the group. The infinitesimal change of the potential due to $\delta\Phi$
should vanish \begin{eqnarray}
\delta V(\Phi) & = & \frac{\delta V}{\delta\phi_{i}}\delta\phi_{i}=i\frac{\delta V}{\delta\phi_{i}}\theta^{a}(C^{a})_{ij}\phi_{j}=0.\end{eqnarray}
Since $\theta^{a}$ are arbitrary we have \begin{eqnarray}
\frac{\delta V}{\delta\phi_{i}}(C^{a})_{ij}\phi_{j} & = & 0\end{eqnarray}
for $a=1,\cdots,n_{G}$. Taking an additional derivative with respect
to $\phi_{k}$, we obtain \begin{eqnarray}
\frac{\delta^{2}V}{\delta\phi_{k}\delta\phi_{i}}(C^{a})_{ij}\phi_{j}+\frac{\delta V}{\delta\phi_{i}}(C^{a})_{ik} & = & 0.\end{eqnarray}
Evaluating the above equation at the vacuum $\Phi=\Phi_{0}$, the
second term vanishes, and we have \begin{eqnarray}
M_{ki}^{2}(C^{a})_{ij}\phi_{0j} & = & 0,\label{eq:mass-matrix}\end{eqnarray}
where the mass matrix is defined by $M_{ki}^{2}=\left.\frac{\delta^{2}V}{\delta\phi_{k}\delta\phi_{i}}\right|_{\Phi=\Phi_{0}}$.
If the ground state is invariant under a sub-group $H$ of $G$, for
each generator belonging to this sub-group $H$, we have \begin{eqnarray}
(C^{a})_{ij}\phi_{0j} & = & 0\end{eqnarray}
for $a=1,\cdots,n_{H}$. For the remaining $n_{G}-n_{H}$ generators,
we have \begin{eqnarray}
(C^{a})_{ij}\phi_{0j} & \neq & 0.\end{eqnarray}
 So Eq. (\ref{eq:mass-matrix}) shows that there are $n_{G}-n_{H}$
zero eigenvalues for the mass matrix which means $n_{G}-n_{H}$ massless
Goldstone bosons. 

Note that the above argument holds for a Lorentz invariant system.
In the context of dense quark matter where the Lorentz invariance
is lost, the Nambu-Goldstone bosons have many fine structures, see,
e.g. Ref. \cite{Miransky:2001sw,Rischke:2002rz,Miransky:2001tw}. 

\section{Scalar QED as an example of superconductor}

In the above section we discussed the spontaneous breaking of a global
symmetry. In this section we will investigate the spontaneous breaking
of a local gauge symmetry. We will show that it is also a good example
for superconductivity. To this end, we consider a theory of a complex
scalar and a U(1) gauge field where they couple to each other in a
minimal way, the so-called scalar quantum electromagnetism or scalar
QED. The Lagrangian is \begin{eqnarray}
\mathcal{L} & = & (D_{\mu}\phi)^{\dagger}D^{\mu}\phi-V(|\phi|^{2})-\frac{1}{4}F_{\mu\nu}F^{\mu\nu},\end{eqnarray}
where the covariant derivative is defined by $D_{\mu}\equiv\partial_{\mu}+ieA_{\mu}$
and the guage field stength tensor by $F_{\mu\nu}\equiv\partial_{\mu}A_{\nu}-\partial_{\nu}A_{\mu}$.
The potential is given by $V(|\phi|^{2})=\lambda(|\phi|^{2}-\phi_{0}^{2})$.
The Lagrangian $\mathcal{L}$ is invariant under a U(1) local gauge
transformation, \begin{eqnarray}
\phi' & = & e^{ie\theta(x)}\phi,\nonumber \\
A'_{\mu} & = & A_{\mu}-\partial_{\mu}\theta.\end{eqnarray}
One can verify that under the above transformation $D'_{\mu}\phi'=e^{ie\theta(x)}D_{\mu}\phi$.
Therefore the term $(D_{\mu}\phi)^{\dagger}D^{\mu}\phi$ is invariant. 

The Lagrangian equation for the gauge field reads \begin{eqnarray}
\partial^{\mu}F_{\mu\nu} & = & j_{\nu},\nonumber \\
j_{\nu} & = & ie(\phi^{*}\partial_{\nu}\phi-\phi\partial_{\nu}\phi^{*})-2e^{2}|\phi|^{2}A_{\nu}.\end{eqnarray}
The current $j_{\mu}$ satisfies the conservation law $\partial^{\mu}j_{\mu}=0$
due to $\partial^{\mu}\partial^{\nu}F_{\mu\nu}=0$. The Lagrangian
equation for the scalar field reads \begin{eqnarray}
(\partial_{\mu}+ieA_{\mu})(\partial^{\mu}+ieA^{\mu})\phi & = & 2\lambda\phi(\phi_{0}^{2}-|\phi|^{2}).\end{eqnarray}
In the covariant gauge $\partial_{\mu}A^{\mu}=0$ and choosing the
real scalar field (unitary gauge) $\phi^{*}=\phi=\rho$, the equation
for the gauge field becomes \begin{eqnarray}
(\partial^{\mu}\partial_{\mu}+2e^{2}\rho^{2})A_{\nu} & = & 0.\end{eqnarray}
Suppose $\rho$ is very close to the vacuum value $\phi_{0}$, we
can expand $\rho=\phi_{0}+\eta$, the linearized equation of motion
for the scalar field become \begin{eqnarray}
(\partial^{\mu}\partial_{\mu}+4\lambda\phi_{0}^{2})\eta & = & 0,\end{eqnarray}
where we have dropped higher order terms like $A^{\mu}\partial_{\mu}\eta$
and $A^{\mu}A_{\mu}$ etc..

\section{Invariant subgroup for order parameter}

\label{sec:Sym2}Consider a symmetry group in a theory\begin{equation}
G=G_{1}\otimes G_{2},\end{equation}
where $G_{1}$ and $G_{2}$ are Lie groups (a generalization to more
than two groups is straightforward). The group elements can be expressed
in terms of the generators of the respective Lie algebras \begin{eqnarray}
g & = & g_{1}\otimes g_{2}\nonumber \\
g_{1} & = & e^{iC_{1}^{a}\theta_{1}^{a}}\in G_{1}\nonumber \\
g_{2} & = & e^{iC_{2}^{a}\theta_{2}^{a}}\in G_{2},\end{eqnarray}
where $C_{1}^{a}\;(a=1\cdots n_{1})$ and $C_{2}^{a}\;(a=1\cdots n_{2})$
are the generators of the Lie algebras $G_{1}$ and $G_{2}$, respectively.
$\theta_{1}^{a}$ and $\theta_{2}^{a}$ are rotational angles associated
with the operators $C_{1}^{a}$ and $C_{2}^{a}$ respectively. 

Now we investigate how the order parameter changes under the transformation
of the group $G$. After choosing one representation for the group
$G$, the order parameter $\mathcal{M}$ can be written on its basis
\begin{eqnarray}
\mathcal{M} & \sim & \Delta_{ij}\mathbf{e}_{1i}\otimes\mathbf{e}_{2j},\label{eq:order1}\end{eqnarray}
where the coefficient $\Delta_{ij}$ is the gap matrix, and $\mathbf{e}_{1i}$
and $\mathbf{e}_{2j}$ are basis vectors for $G_{1}$ and $G_{2}$
representations, respectively. After transformation, the order parameter
becomes\begin{eqnarray}
\mathcal{M}' & = & g_{1}\otimes g_{2}\mathcal{M}g_{1}^{T}\otimes g_{2}^{T}\nonumber \\
 & \sim & \Delta_{ij}g_{1}\mathbf{e}_{1i}g_{1}^{T}\otimes g_{2}\mathbf{e}_{2j}g_{2}^{T}.\label{eq:order2}\end{eqnarray}
For infinitesimal transformation\begin{eqnarray}
g_{1} & \approx & 1+iC_{1}^{a}\theta_{1}^{a}\nonumber \\
g_{2} & \approx & 1+iC_{2}^{a}\theta_{2}^{a},\end{eqnarray}
Eq. (\ref{eq:order2}) becomes\begin{eqnarray}
\mathcal{M}' & = & \mathcal{M}+i\theta_{1}^{a}\Delta_{ij}(C_{1}^{a}\mathbf{e}_{1i}+\mathbf{e}_{1i}C_{1}^{aT})\otimes\mathbf{e}_{2j}\nonumber \\
 &  & +i\theta_{2}^{a}\Delta_{ij}\mathbf{e}_{1i}\otimes(C_{2}^{a}\mathbf{e}_{2j}+\mathbf{e}_{2j}C_{2}^{aT})\end{eqnarray}
To find the invariant subgroup for the order parameter, we require
$\mathcal{M}'=\mathcal{M}$ and then \begin{eqnarray}
 &  & i\theta_{1}^{a}\Delta_{ij}(C_{1}^{a}\mathbf{e}_{1i}+\mathbf{e}_{1i}C_{1}^{aT})\otimes\mathbf{e}_{2j}\nonumber \\
 &  & +i\theta_{2}^{a}\Delta_{ij}\mathbf{e}_{1i}\otimes(C_{2}^{a}\mathbf{e}_{2j}+\mathbf{e}_{2j}C_{2}^{aT})=0\label{eq:eqn-subgroup}\end{eqnarray}
From this equation we can find the subset of operators $C_{1,2}^{a}$
which make up the sub-algebra of the residual symmetry. 

For the $SU(3)$ group, we have a simpler version of Eq. (\ref{eq:eqn-subgroup}).
We know that the color part of the order parameter is the anti-triplet.
So the basis of the color part now should be $J_{i}$, the color anti-symmetric
matrices defined by $(J_{i})_{jk}=i\epsilon_{jik}=-i\epsilon_{ijk}$.
Now we look at $g_{1}J_{i}g_{1}^{T}$, where $g_{1}$ is a $SU(3)$
group element in fundamental representation (a 3 by 3 unitary matrix),
which satisfies $g_{1}g_{1}^{\dagger}=1$ and $\mathrm{det}(g_{1})=1$.
Using these properties we can simplify $g_{1}J_{i}g_{1}^{T}$ as \begin{eqnarray}
g_{1ji}[g_{1}J_{i}g_{1}^{T}]_{mn} & = & g_{1ji}g_{1ml}(J_{i})_{lp}g_{1np}\nonumber \\
 & = & -i\epsilon_{ilp}g_{1ji}g_{1ml}g_{1np}\nonumber \\
 & = & -i\epsilon_{jmn}\mathrm{det}(g_{1}).\end{eqnarray}
Multiplying $(g_{1}^{\dagger})_{kj}$ on both sides, we get \begin{eqnarray}
[g_{1}J_{k}g_{1}^{T}]_{mn} & = & -i\epsilon_{jmn}(g_{1}^{\dagger})_{kj}\nonumber \\
 & = & (g_{1}^{\dagger})_{kj}(J_{j})_{mn},\end{eqnarray}
or in operator form \begin{eqnarray}
g_{1}J_{k}g_{1}^{T} & = & (g_{1}^{\dagger})_{kj}J_{j}.\label{eq:trans-basis1}\end{eqnarray}

In the case of the $SU(2)$ group, for example, the number of flavors
is 2, we will encounter $g_{1}\sigma_{2}g_{1}^{T}$, where $g_{1}$
is a $SU(2)$ group element in fundamental representation (a 2 by
2 unitary matrix). Using $g_{1}g_{1}^{\dagger}=1$ and $\mathrm{det}(g_{1})=1$,
we have \begin{eqnarray}
[g_{1}\sigma_{2}g_{1}^{T}]_{mn} & = & g_{1ml}(\sigma_{2})_{lp}g_{1np}\nonumber \\
 & = & -i\epsilon_{lp}g_{1ml}g_{1np}\nonumber \\
 & = & -i\epsilon_{mn}\mathrm{det}(g_{1})\nonumber \\
 & = & (\sigma_{2})_{mn}\label{eq:transform-su2}\end{eqnarray}
Therefore we obtain $g_{1}\sigma_{2}g_{1}^{T}=\sigma_{2}$. 

We can also simplify $g_{2}\kappa_{j}g_{2}^{T}$ for $SO(3)_{J}$.
Here $\kappa_{j}$ with $j=1,2,3$ are the basis for the spin triplet.
Because $g_{2}$ is a group element of $SO(3)$, we have $g_{2}^{T}=g_{2}^{-1}$.
Then we obtain\begin{eqnarray}
g_{2}\kappa_{j}g_{2}^{T} & = & g_{2}\kappa_{j}g_{2}^{-1}\nonumber \\
 & \approx & (1+i\theta_{2}^{m}\kappa_{m})\kappa_{j}(1-i\theta_{2}^{n}\kappa_{n})\nonumber \\
 & \approx & \kappa_{j}+i\theta_{2}^{m}\kappa_{m}\kappa_{j}-i\theta_{2}^{n}\kappa_{j}\kappa_{n}\nonumber \\
 & = & \kappa_{j}+i\theta_{2}^{m}i\epsilon_{mjl}\kappa_{l}\nonumber \\
 & = & \kappa_{j}-i\theta_{2}^{m}(J_{m})_{jl}\kappa_{l}\nonumber \\
 & \rightarrow & [e^{-i\theta_{2}^{m}J_{m}}]_{jl}\kappa_{l}=(g_{2}^{\dagger})_{jl}\kappa_{l}.\label{eq:trans-basis2}\end{eqnarray}

\subsection{Symmetry patterns: 2SC phase}

Our first example for the spin-0 pairing is the 2SC phase with $u$
and $d$ quarks. For simplicity we consider the positive parity channel,
for the complete case, see, e.g., Eq. (9) in Ref. \cite{Alford:2007xm}.
The order parameter for this color-flavor coupling phase can be written
in the following form \begin{eqnarray}
\mathcal{M} & = & J_{i}\Delta_{ij}I_{j}\gamma_{5},\label{eq:order-para1}\end{eqnarray}
where $J_{i}$ is the color anti-symmetric matrices given by $(J_{i})_{jk}=i\epsilon_{jik}=-i\epsilon_{ijk}$.
We can also express $J_{i}$ in terms of Gell-Mann matrices $J_{1}=\lambda_{7}$,
$J_{2}=-\lambda_{5}$ and $J_{3}=\lambda_{2}$. $I_{i}$ are the flavor
anti-symmetric matrices defined in the same way as $J_{i}$. The coefficient
of the order parameter $\Delta_{ij}$ for 2SC is $\Delta_{ij}=\Delta\delta_{j3}\delta_{i3}$.
Incorporating this expression for $\Delta_{ij}$ into Eq. (\ref{eq:order-para1}),
we obtain \begin{eqnarray}
\mathcal{M}_{2SC} & = & \Delta J_{3}\sigma_{2}\gamma_{5},\end{eqnarray}
where $(\sigma_{2})_{ij}=(I_{3})_{ij}=-i\epsilon_{ij}$ with $i,j=1,2$. 

The symmetry group for the 2SC phase is \begin{eqnarray}
G & = & G_{1}\otimes G_{2}\otimes G_{3},\end{eqnarray}
where $G_{1}=SU(3)_{c}$ is the color group, $G_{2}=SU(2)_{f}$ the
flavor group, and $G_{3}=U(1)_{B}$ the baryon number group. Note
that the electromagnetism group $U(1)_{em}$ is a subgroup of $SU(2)_{f}$
and $U(1)_{B}$, since the electric charge satisfies \begin{equation}
Q=\mathrm{diag}(Q_{u},Q_{d})=\mathrm{diag}(2/3,-1/3)=T_{3}+B/2,\end{equation}
where $B=1/3$ is the baryon number. 

First we want to find the invariant subgroup of $SU(3)_{c}$ and $SU(2)_{f}$,
so we can safely drop $G_{3}$ for simplification of the notation.
The symmetry group now is \begin{equation}
G=G_{1}\otimes G_{2}=SU(3)_{c}\otimes SU(2)_{f}.\end{equation}
The elements of $SU(3)_{c}$ and $SU(2)_{f}$ are \begin{eqnarray}
g & = & g_{1}\otimes g_{2}\nonumber \\
g_{1} & = & e^{iT_{a}\theta_{1}^{a}}\in SU(3)_{c}\nonumber \\
g_{2} & = & e^{i\sigma_{a}\theta_{2}^{a}}\in SU(2)_{f},\end{eqnarray}
where $\sigma_{a}$ with $a=1,2,3$ are Pauli matrices. According
to Eq. (\ref{eq:order2}), we obtain\begin{eqnarray}
\mathcal{M}' & = & g_{1}\otimes g_{2}\mathcal{M}g_{1}^{T}\otimes g_{2}^{T}\nonumber \\
 & = & \Delta\gamma_{5}g_{1}J_{3}g_{1}^{T}\otimes g_{2}\sigma_{2}g_{2}^{T}\nonumber \\
 & = & \Delta\gamma_{5}(g_{1}^{\dagger})_{3i}J_{i}\otimes\sigma_{2}\end{eqnarray}
where we have used Eq. (\ref{eq:trans-basis1}) and (\ref{eq:transform-su2}).
By expanding $g_{1}$ and $g_{2}$ to the linear term of $\theta_{1}^{a}$
and $\theta_{2}^{a}$ respectively, and requiring $\mathcal{M}'=\mathcal{M}$,
we have \begin{eqnarray}
\theta_{1}^{a}(T_{a})_{3i}J_{i}\otimes\sigma_{2} & = & 0,\label{eq:subgroup-2sc}\end{eqnarray}
For $a=1,2,3$ Eq. (\ref{eq:subgroup-2sc}) holds automatically, because
elements on the third row of $T_{1},T_{2},T_{3}$ are zero. This means
that there is freedom to choose any values of $\theta_{1}^{a}$ with
$a=1,2,3$. For $a=4,5,6,7,8$, we have \begin{eqnarray}
\theta_{1}^{a}(T_{a})_{3i}J_{i} & = & (\theta_{1}^{4}+i\theta_{1}^{5})T_{7}-(\theta_{1}^{6}+i\theta_{1}^{7})T_{5}-\frac{2}{\sqrt{3}}\theta_{1}^{8}T_{2}\label{eq:op1}\end{eqnarray}
where we see that $\theta_{1}^{a}(T_{a})_{3i}J_{i}$ is not zero.
Therefore to make Eq. (\ref{eq:subgroup-2sc}) hold, $\theta_{1}^{a}$
must vanish. We see that the orginal $SU(3)_{c}$ symmetry is spontaneously
broken to $SU(2)_{c}$ spanned by operators $T_{1}$, $T_{2}$ and
$T_{3}$ as follows \begin{eqnarray}
SU(3)_{c}\otimes SU(2)_{f} & \rightarrow & SU(2)_{c}\otimes SU(2)_{f}.\end{eqnarray}

In order to get the invariant $U(1)$ group for the order parameter,
we consider the transfromation of $U(1)_{em}$ for electromagnetism
in the flavor sector. The elements of $SU(3)_{c}$ and $U(1)_{em}$
are \begin{eqnarray}
g & = & g_{1}\otimes g_{2}\nonumber \\
g_{1} & = & e^{iT_{a}\theta_{1}^{a}}\in SU(3)_{c}\nonumber \\
g_{2} & = & e^{iQ\theta_{2}}\in U(1)_{em},\end{eqnarray}
Eq. (\ref{eq:eqn-subgroup}) becomes \begin{eqnarray}
 &  & -\theta_{1}^{a}(T_{a})_{3i}J_{i}\otimes\sigma_{2}+J_{3}\otimes\sigma_{2}\frac{\theta_{2}}{3}=0.\label{eq:mixing-2sc}\end{eqnarray}
where we have used \begin{eqnarray}
[e^{iQ\theta}\sigma_{2}e^{iQ\theta}]_{kl} & = & -i\epsilon_{ij}[e^{iQ\theta}]_{ki}[e^{iQ\theta}]_{lj}\nonumber \\
 & = & -i\epsilon_{kl}\det[e^{iQ\theta}]\nonumber \\
 & = & \sigma_{2}e^{i(Q_{u}+Q_{d})\theta}=\sigma_{2}e^{i\theta/3}\end{eqnarray}
The above equation becomes \begin{eqnarray}
 &  & \left[(-\theta_{1}^{4}-i\theta_{1}^{5})T_{7}+(\theta_{1}^{6}+i\theta_{1}^{7})T_{5}\right.\nonumber \\
 &  & \left.+\frac{2}{\sqrt{3}}\theta_{1}^{8}T_{2}\right]\otimes\sigma_{2}+\frac{2}{3}\theta_{2}T_{2}\otimes\sigma_{2}=0.\end{eqnarray}
We obtain \begin{eqnarray}
\theta_{1}^{a} & = & 0,\;\; a=4,5,6,7\nonumber \\
\theta_{1}^{8} & = & -\frac{1}{\sqrt{3}}\theta_{2}.\end{eqnarray}
With this relation between $\theta_{1}^{8}$ and $\theta_{2}$ we
can combine the group elements as \begin{equation}
e^{iT_{8}\theta_{1}^{8}}\otimes e^{iQ\theta_{2}}=e^{i\widetilde{Q}\theta_{2}},\end{equation}
with the new $\widetilde{U}(1)$ charge given by \begin{equation}
\widetilde{Q}=1\otimes Q-\frac{1}{\sqrt{3}}T_{8}\otimes1,\end{equation}
where $T_{8}=\frac{1}{2\sqrt{3}}\mathrm{diag}(1,1,-2)$. So the original
symmetry is broken as \begin{eqnarray}
SU(3)_{c}\otimes U(1)_{em} & \rightarrow & SU(2)_{c}\otimes\widetilde{U}(1).\end{eqnarray}
The $\widetilde{Q}$ charge in the 2SC phase for $u$ and $d$ quarks
with different colors are listed in Tab. (\ref{tab:charges}). We
see that the condensate made of $(u_{1}d_{2},u_{2}d_{1})$ is $\widetilde{Q}$
neutral because $\widetilde{Q}_{cond}=\widetilde{Q}(u_{1})+\widetilde{Q}(d_{2})=\widetilde{Q}(u_{2})+\widetilde{Q}(d_{1})=0$. 

\begin{table}
\caption{\label{tab:charges}$\widetilde{Q}$ charge in 2SC and CFL phases.
The colors are labelled by 1,2 and 3.}

\vspace{0.5cm}

\begin{tabular}{|c|c|c|c|}
\hline 
2SC & 1 & 2 & 3\tabularnewline
\hline
\hline 
u & 1/2 & 1/2 & 1\tabularnewline
\hline 
d & -1/2 & -1/2 & 0\tabularnewline
\hline
\end{tabular} \hspace{1cm}\begin{tabular}{|c|c|c|c|}
\hline 
CFL & 1 & 2 & 3\tabularnewline
\hline
\hline 
u & 0 & 1 & 1\tabularnewline
\hline 
d & -1 & 0 & 0\tabularnewline
\hline 
s & -1 & 0 & 0\tabularnewline
\hline
\end{tabular}
\end{table}

\subsection{Symmetry pattern: CFL phase}

Our second example for the spin-0 pairing CFL phase. For simplicity
we consider the positive parity channel, for the full case, see, e.g.,
Eq. (6) in Ref. \cite{Alford:2007xm}. The order parameter can be
written as \begin{eqnarray}
\mathcal{M} & = & J_{i}\Delta_{ij}I_{j}\gamma_{5}=\Delta\mathbf{J}\cdot\mathbf{I}\gamma_{5},\end{eqnarray}
with $\Delta_{ij}=\Delta\delta_{ij}$. The symmetry group is \begin{eqnarray}
G & = & G_{1}\times G_{2}\times G_{3}=SU(3)_{c}\times SU(3)_{f}\times U(1)_{B}.\end{eqnarray}
Now we try to find the invariant subgroup for the order parameter.
We will see $U(1)_{B}$ plays no role, so we only $G_{1}$ and $G_{2}$
in the following discussion. Then the group elements are \begin{eqnarray}
g & = & g_{1}\otimes g_{2}\nonumber \\
g_{1} & = & e^{iT_{a}\theta_{1}^{a}}\in SU(3)_{c}\nonumber \\
g_{2} & = & e^{iT_{a}\theta_{2}^{a}}\in SU(3)_{f}.\end{eqnarray}
Eq. (\ref{eq:eqn-subgroup}) for the CFL phase becomes \begin{eqnarray}
 &  & \theta_{1}^{a}(T_{a})_{ij}J_{j}\otimes I_{i}\nonumber \\
 &  & +\theta_{2}^{a}J_{i}\otimes(T_{a})_{ij}I_{j}=0.\label{eq:subgroup-cfl1}\end{eqnarray}
We use following formula to simplify the above equation \begin{eqnarray}
\theta^{a}(T_{a})_{1j}J_{j} & = & (-\theta^{1}+i\theta^{2})T_{5}+(\theta^{4}-i\theta^{5})T_{2}\nonumber \\
 &  & +(\theta^{3}+\frac{1}{\sqrt{3}}\theta^{8})T_{7}\nonumber \\
\theta^{a}(T_{a})_{2j}J_{j} & = & (\theta^{1}+i\theta^{2})T_{7}+(\theta^{6}-i\theta^{7})T_{2}\nonumber \\
 &  & +(\theta^{3}-\frac{1}{\sqrt{3}}\theta^{8})T_{5}\nonumber \\
\theta^{a}(T_{a})_{3j}J_{j} & = & (\theta^{4}+i\theta^{5})T_{7}-(\theta^{6}+i\theta^{7})T_{5}\nonumber \\
 &  & -\frac{2}{\sqrt{3}}\theta^{8}T_{2}\end{eqnarray}
Then we obtain following conditions from Eq. (\ref{eq:subgroup-cfl1})
\begin{eqnarray}
\theta_{1}^{a} & = & \eta_{a}\theta_{2}^{a}\end{eqnarray}
where $\eta_{a}=1$ for $a=2,5,7$ and $\eta_{a}=-1$ for $a=1,3,4,6,8$.
We see that in order to make the order parameter invariant, we have
to lock the rotaional angles for $SU(3)_{c}$ and $SU(3)_{f}$. The
orginal symmetry $SU(3)_{c}\times SU(3)_{f}$ is broken to $SU(3)_{c+f}$.

To find the new $U(1)$ charge, we consider the transfromation of
$U(1)_{em}$ in the flavor sector with the generator of $U(1)_{em}$
given by \begin{eqnarray}
Q & = & \left(\begin{array}{ccc}
Q_{u} & 0 & 0\\
0 & Q_{d} & 0\\
0 & 0 & Q_{s}\end{array}\right)=\left(\begin{array}{ccc}
2/3 & 0 & 0\\
0 & -1/3 & 0\\
0 & 0 & -1/3\end{array}\right).\end{eqnarray}
Note that $U(1)_{em}$ is a subgroup of $SU(3)_{f}$ since $Q=T_{3}+T_{8}/\sqrt{3}$.
We use \begin{eqnarray}
[e^{iQ\theta}]_{ii'}[e^{iQ\theta}J_{i'}e^{iQ\theta}]_{kl} & = & -i\epsilon_{i'mn}[e^{iQ\theta}]_{ii'}[e^{iQ\theta}]_{km}[e^{iQ\theta}]_{ln}\nonumber \\
 & = & -i\epsilon_{ikl}\det[e^{iQ\theta}]\nonumber \\
 & = & J_{i}e^{i(Q_{u}+Q_{d}+Q_{s})\theta}=J_{i}\end{eqnarray}
to obtain \begin{eqnarray}
e^{iQ\theta}J_{i}e^{iQ\theta} & = & e^{-iQ\theta}J_{i}.\end{eqnarray}
Then we derive the equation for determing the residue symmetry \begin{eqnarray}
 &  & \theta_{1}^{a}(T_{a})_{ij}J_{j}\otimes I_{i}+\theta_{2}J_{i}\otimes Q_{ii}I_{i}=0.\label{eq:cfl-new-photon}\end{eqnarray}
Finally we obtain following conditions \begin{eqnarray}
\theta_{1}^{a} & = & 0,\;\; a\neq3,8\nonumber \\
\theta_{1}^{3} & = & -\theta_{2}\nonumber \\
\theta_{1}^{8} & = & -\frac{1}{\sqrt{3}}\theta_{2}.\end{eqnarray}
We can rewrite the group element as\begin{equation}
e^{iT_{3}\theta_{1}^{3}+iT_{8}\theta_{1}^{8}}\otimes e^{iQ\theta_{2}}=e^{i\widetilde{Q}\theta_{2}},\end{equation}
where the new $U(1)$ charge is \begin{eqnarray}
\widetilde{Q} & = & 1\otimes Q-\left(T_{3}+\frac{1}{\sqrt{3}}T_{8}\right)\otimes1\end{eqnarray}
The $\widetilde{Q}$ charge in the CFL phase for $u$, $d$ and $s$
quarks with different colors are listed in Tab. (\ref{tab:charges}).
One can check that the condensate made of $(u_{1}d_{2},u_{2}d_{1})$,
$(d_{2}s_{3},d_{3}s_{2})$ and $(s_{3}u_{1},s_{1}u_{3})$ is neutral
because \begin{eqnarray}
\widetilde{Q}_{cond} & = & \widetilde{Q}(u_{1})+\widetilde{Q}(d_{2})=\widetilde{Q}(u_{2})+\widetilde{Q}(d_{1})\nonumber \\
 & = & \widetilde{Q}(d_{2})+\widetilde{Q}(s_{3})=\widetilde{Q}(d_{3})+\widetilde{Q}(s_{2})\nonumber \\
 & = & \widetilde{Q}(s_{3})+\widetilde{Q}(u_{1})=\widetilde{Q}(s_{1})+\widetilde{Q}(u_{3})=0.\end{eqnarray}

\subsection{Symmetry pattern: spin-1 phases}

The condensate of spin-1 CSC transforms as an anti-triplet under $SU(3)_{c}$
transformation, and as a vector under spatial $SO(3)$ rotation. It
carries a $U_{b}(1)$ charge of the baryon number. The symmetry group
of the theory is then $G=SU(3)_{c}\times U_{b}(1)\times SO(3)$. The
condensate can be written as $\mathcal{M}=\Delta_{ai}J_{a}\otimes\gamma^{i}$,
where $\gamma^{i}$ with $i=1,2,3$ are Dirac matrices. The order
parameter $\Delta_{ai}$ is a complex matrix of $3\times3$ dimension
and transforms as, $\Delta\rightarrow g_{1}\Delta g_{2}$, where $g_{1}\in U(3)\equiv SU(3)_{c}\times U_{b}(1)$
and $g_{2}\in SO(3)$. This can be seen by the transformation of the
condensate, \begin{eqnarray}
\Delta_{ai}J_{a}\otimes\gamma^{i} & \rightarrow & \Delta_{ai}g_{1}J_{a}g_{1}^{T}\otimes g_{2}\gamma^{i}g_{2}^{T}=\Delta_{ai}J_{b}(g_{1})_{ba}\otimes(g_{2})_{ij}\gamma^{j}\nonumber \\
 &  & =[(g_{1})_{ba}\Delta_{ai}(g_{2})_{ij}]J_{b}\otimes\gamma^{j}.\end{eqnarray}
Here $g_{2}=\exp(i\frac{1}{2}\epsilon_{ijk}\theta_{k}\gamma^{i}\gamma^{j})$
in $g_{2}\gamma^{i}g_{2}^{T}$ is the $SO(3)$ representation in Dirac
space and $g_{2}=\exp(i\theta_{k}J_{k})$ in $(g_{2})_{ij}\gamma^{j}$
is the $SO(3)$ representation in vector space. By a suitable transformation,
$\Delta$ can be cast into the following form, $\Delta=S+iA$, where
$S$ and $A$ are real symmetric and anti-symmetric matrices respectively.
Since orthogonal rotation does not change the symmety of $S$ and
$A$, they can be after transformation written in the form $S=\mathrm{diag}(\Delta_{1},\Delta_{2},\Delta_{3})$
and $A_{ij}=\epsilon_{ijk}\omega_{k}$, where $\Delta_{i},\omega_{i}$
are real numbers. The condensate then has the form, \begin{equation}
\Delta=\left(\begin{array}{ccc}
\Delta_{1} & i\omega_{3} & -i\omega_{2}\\
-i\omega_{3} & \Delta_{2} & i\omega_{1}\\
i\omega_{2} & -i\omega_{1} & \Delta_{3}\end{array}\right).\end{equation}
Generally a complex matrix has 18 real parameters, 12 of which can
be fixed by a transformation of symmetry group $G$. We can classify
the order parameter $\Delta$ by its invariant subgroups of $G$,
which can be found by looking for $g_{1}$ and $g_{2}$ satisfying
$g_{1}g_{2}^{T}\Delta g_{2}=\Delta$, which is equivalent to $g_{1}\Delta=\Delta$
and $g_{2}^{T}\Delta g_{2}=\Delta$. Since the matrix $A$ defines
a vector $\overrightarrow{\omega}=\omega_{i}\mathbf{e}_{i}$, $A$
is invariant under a rotation $g_{2}=e^{iJ_{i}\theta_{2}^{i}}\in SO(3)$
with the axis in the direction of $\overrightarrow{\omega}$, i.e.
$\overrightarrow{\theta_{2}}\parallel\overrightarrow{\omega}$, i.e.
$g_{2}^{T}Ag_{2}=A$. One can verify that the same rotation $g_{2}$
also makes $S$ invariant if $S$ is in the form, \begin{equation}
S_{ij}=c_{1}\delta_{ij}+c_{2}\left(\delta_{ij}-\frac{\omega_{i}\omega_{j}}{|\overrightarrow{\omega}|^{2}}\right),\end{equation}
To see that we note that $M_{i}\omega_{i}\omega_{j}\omega_{k}=0$.
The $U(3)$ invariant subgroup whose element $g_{1}$ satisfies $g_{1}\Delta=\Delta$
is determined by the number of zero modes of $\Delta$. The subgroup
is then $U(n)$ for the case of $n$ zero modes of $\Delta$. The
classification of the order parameters is listed in Tab. \ref{tab:classification},
see \cite{Schmitt:2004et,Brauner:2008ma}. 

\begin{table}
\begin{tabular}{|c|c|c|}
\hline 
order parameter  & unbroken symmetry  & name \tabularnewline
\hline
\hline 
$\begin{pmatrix}\Delta_{1} & +ia & 0\\
-ia & \Delta_{1} & 0\\
0 & 0 & \Delta_{2}\end{pmatrix}$  & $SO(2)_{v}$  & oblate \tabularnewline
\hline 
$\begin{pmatrix}\Delta & +ia & 0\\
-ia & \Delta & 0\\
0 & 0 & 0\end{pmatrix}$  & $SO(2)_{v}\times U(1)_{L}$  & cylindrical \tabularnewline
\hline 
$\begin{pmatrix}\Delta_{1} & +i\Delta_{1} & 0\\
-i\Delta_{1} & \Delta_{1} & 0\\
0 & 0 & \Delta_{2}\end{pmatrix}$  & $SO(2)_{v}\times U(1)_{L}$  & $\varepsilon$ \tabularnewline
\hline 
$\begin{pmatrix}1 & +i & 0\\
-i & 1 & 0\\
0 & 0 & 0\end{pmatrix}$  & $SU(2)_{L}\times SO(2)_{v}\times U(1)_{L}$  & A \tabularnewline
\hline 
$\begin{pmatrix}1 & 0 & 0\\
0 & 1 & 0\\
0 & 0 & 1\end{pmatrix}$  & $SO(3)_{v}$  & CSL \tabularnewline
\hline 
$\begin{pmatrix}0 & 0 & 0\\
0 & 0 & 0\\
0 & 0 & 1\end{pmatrix}$  & $SU(2)_{L}\times SO(2)_{R}\times U(1)_{L}$  & polar \tabularnewline
\hline 
$\begin{pmatrix}0 & 0 & 0\\
z_{1} & z_{2} & z_{3}\\
z_{4} & z_{5} & z_{6}\end{pmatrix}$  & $U(1)_{L}$  & $N_{1}$ \tabularnewline
\hline 
$\begin{pmatrix}0 & 0 & 0\\
0 & 0 & 0\\
z_{1} & z_{2} & z_{3}\end{pmatrix}$  & $SU(2)_{L}\times U(1)_{L}$  & $N_{2}$ \tabularnewline
\hline 
$\begin{pmatrix}\Delta_{1} & 0 & 0\\
0 & \Delta_{1} & 0\\
0 & 0 & \Delta_{2}\end{pmatrix}$  & $SO(2)_{v}$  & axial \tabularnewline
\hline 
$\begin{pmatrix}1 & 0 & 0\\
0 & 1 & 0\\
0 & 0 & 0\end{pmatrix}$  & $SO(2)_{v}\times U(1)_{L}$  & planar \tabularnewline
\hline
\end{tabular}

\caption{Classification and the symmetry breaking patterns of the order parameters,
see Fig.1 of \cite{Schmitt:2004et,Brauner:2008ma}. L/R denote the
transformation matrix acting on the left/right hand side of the condensate,
while $SO(2)_{v}$ denotes the transformation that acts on both right
and left hand side of the condensate. }

\label{tab:classification} 
\end{table}

We take a few examples. For the oblate phase, one can verify that
the order parameter has $SO(2)$ residue symmetry, $g_{2}^{T}\Delta g_{2}=\Delta$,
for $g_{2}(oblate)=e^{iJ_{3}\theta_{2}^{3}}$. Note that $\det\Delta\neq0$,
so there is no residue symmetry inherited from $U(3)$. For the cylindrical
phase, the residue symmetry is $SO(2)\times U(1)$, whose transformation
is \begin{equation}
g_{1}=\mathrm{diag}(1,1,e^{i\theta}),\; g_{2}=e^{iJ_{3}\theta_{2}^{3}}.\end{equation}
Note that there is one zero eigenvalue for $\Delta$ so the invariant
subgroup is $U(1)$. For the CSL phase, the residue symmetry is $SO(3)$,
whose transformation matrix obeying $g_{2}^{T}\Delta g_{2}=\Delta$
is any rotational matrix since $\Delta$ is a unit matrix, i.e. $g_{2}(CSL)=e^{iJ_{i}\theta_{2}^{i}}$
with $i=1,2,3$. 

A more convenient way of looking for invariant subgroup is by using
infinitesimal transformation, \begin{eqnarray}
\Delta & = & g_{1}\Delta g_{2}=(1+i\theta+iT^{a}\theta_{1}^{a})\Delta(1+iJ_{i}\theta_{2}^{i})\nonumber \\
 & \approx & \Delta+i\theta\Delta+i\theta_{1}^{a}T^{a}\Delta+i\theta_{2}^{i}\Delta J_{i},\end{eqnarray}
which requires \begin{equation}
\theta\Delta+\theta_{1}^{a}T^{a}\Delta+\theta_{2}^{i}\Delta J_{i}=0\end{equation}
for any subset of non-vanishing angles $\theta$, $\theta_{1}^{a}$
and $\theta_{2}^{i}$. 

We analyze a few phases. The forms of order parameters $\Delta$ are
listed in Tab \ref{tab:classification}. (1) The oblate phase. Since
$J_{3}\Delta=\Delta J_{3}$, we obtain the only non-vansihing angles
are $\theta_{1}^{2}$ and $\theta_{1}^{3}$ and obey $\theta_{1}^{2}=-2\theta_{2}^{3}$.
So we have verified that the invariant subgroup is $SO(2)$, whose
generator is $2T^{2}\otimes1-1\otimes J_{3}^{T}$. (2) The cylindrical
phase. We have $\theta_{1}^{8}=-2\sqrt{3}\theta$ and $\theta_{1}^{2}=-2\theta_{2}^{3}$,
which corresponds to residue symmetry $SO(2)\times U(1)$. The unbroken
generators are $2T^{2}\otimes1-1\otimes J_{3}^{T}$ for $SO(2)$ and
$-\frac{1}{2\sqrt{3}}+T^{8}$ for $U(1)$. (3) The $\varepsilon$
phase. We obtain $\theta_{1}^{2}=-2\theta_{2}^{3}$ for $SO(2)$ and
$(\theta=\frac{1}{3}\theta_{1}^{2},\theta_{1}^{8}=\frac{1}{\sqrt{3}}\theta_{1}^{2})$
for $U(1)$. The invariant generators are $2T^{2}\otimes1-1\otimes J_{3}^{T}$
for $SO(2)$ and $\frac{1}{3}+T^{2}+\frac{1}{\sqrt{3}}T^{8}$ for
$U(1)$ (to see this one can verify that the generator behaves like
a projector). (4) The A phase. It is a special case of the cylindrical
phase. We obtain $\theta_{1}^{2}=-2\theta_{2}^{3}$ for $SO(2)$ and
$\theta_{1}^{8}=-2\sqrt{3}\theta$ for $U(1)$, whose generators are
$2T^{2}\otimes1-1\otimes J_{3}^{T}$ and $-\frac{1}{2\sqrt{3}}+T^{8}$
respectively. Additionally we have $\theta_{1}^{1}=-i\theta_{1}^{3}$,
$\theta_{1}^{2}=\frac{1}{\sqrt{3}}\theta_{1}^{8}$ and $\theta_{1}^{4}=i(\theta_{1}^{6}-\theta_{1}^{5})-\theta_{1}^{7}$
for $SU(2)$, whose generators are $\frac{1}{2\sqrt{2}}(\frac{1}{\sqrt{3}}T^{2}+T^{8})$,
$\frac{\sqrt{3}}{4}(T^{5}+T^{6})$, and $\frac{1}{2\sqrt{2}}(T^{7}-T^{4})$.
It is more convenient to transform the order parameter $\Delta$ which
is Hermitian in Tab \ref{tab:classification} by a unitary matrix
$\mathcal{U}$ into, \begin{equation}
\Delta=\left(\begin{array}{ccc}
0 & 0 & 0\\
0 & 0 & 0\\
1 & i & 0\end{array}\right)=\frac{1}{\sqrt{2}}\mathcal{U}\left(\begin{array}{ccc}
1 & i & 0\\
-i & 1 & 0\\
0 & 0 & 0\end{array}\right),\;\mathcal{U}=\frac{1}{\sqrt{2}}\left(\begin{array}{ccc}
0 & 0 & -\sqrt{2}\\
i & 1 & 0\\
1 & i & 0\end{array}\right).\end{equation}
It is easy to check that the unbroken generators are, $T^{i}$ with
$i=1,2,3$ for $SU(2)$, $\frac{1}{\sqrt{3}}+T^{8}=\frac{\sqrt{3}}{2}P_{12}$
for $U(1)$, and $1+J_{3}$ for $SO(2)$. Here we define $P_{12}=\mathrm{diag}(1,1,0)$
and $P_{3}=\mathrm{diag}(0,0,1)$. (5) The polar phase. The unbroken
generators are $T^{i}$ with $i=1,2,3$ for $SU(2)$, $\frac{1}{\sqrt{3}}+T^{8}=\frac{\sqrt{3}}{2}P_{12}$
for $U(1)$, and $J_{3}$ for $SO(2)$. For more comprehensive analysis
of symmetry breaking features of the spin-1 pairings, see \cite{Schmitt:2004et,Brauner:2008ma,Brauner:2009df}.

\subsection{More physics related to symmetry patterns}

There are a lot of interesting phenomena arising from symmetry patterns
in color superconductivity. For example, the new $U(1)$ charge corresponds
to the new photon in some superconducting phases, which resembles
$W^{\pm}$ and $Z^{0}$ bosons in electroweak theory, see Tab. (\ref{tab:comparison})
for the comparison between the electroweak theory and color superconductivity.
One can do electroweak physics with the new photon \cite{Casalbuoni:2000cn,Casalbuoni:2000jn}.
In presence of the new photon, magnetic fields show a special behavior
in the neutron star core \cite{Alford:1999pb}. One can study the
transimission and reflection of the new photon at the interface between
different phases to reveal its connection to confinement \cite{Alford:2004ak}.
Especially the new photon in the CFL phase can propagate in the color
superconductor. Therefore the external magnetic field can penetrate
the color superconducting quark core modifying the gap structure \cite{Ferrer:2005vd,Ferrer:2006vw},
and producing different phases with different symmetries and low-energy
physics \cite{Ferrer:2006ie,Ferrer:2007iw}. Another interesting thing
related to the residual $SU(2)_{c}$ symmetry in the 2SC phase is
that one can construct $SU(2)_{c}$ effective theory, where the elementary
excitation is the glueball \cite{Ouyed:2001fv}. Through decaying
to photons these glueballs can be possible source of Gamma Ray Burst
\cite{Ouyed:2001cg}. 

\begin{table}
\caption{\label{tab:comparison}Comparison to the standard model.}

\vspace{0.5cm}

\begin{tabular}{|c|c|c|}
\hline 
 & Weinberg-Salam & CSC with $N_{f}=2,3$\tabularnewline
\hline 
group & \begin{tabular}{c}
$SU(2)\times U(1)$\tabularnewline
isospin, hypercharge\tabularnewline
\end{tabular} & \begin{tabular}{c}
$SU(3)_{c}\times U(1)_{em}$\tabularnewline
color, electromagnetism\tabularnewline
\end{tabular}\tabularnewline
\hline 
field & $W_{1},W_{2},W_{3},W_{0}$ & \begin{tabular}{c}
$A_{1},\cdots,A_{8},A$\tabularnewline
gluons, photon\tabularnewline
\end{tabular}\tabularnewline
\hline 
\begin{tabular}{c}
coupling\tabularnewline
constant\tabularnewline
\end{tabular} & $g,g'$ & $g,e$\tabularnewline
\hline 
\begin{tabular}{c}
symmetry\tabularnewline
breaking\tabularnewline
\end{tabular} & \begin{tabular}{c}
$SU(2)\times U(1)$\tabularnewline
$\rightarrow U(1)_{em}$\tabularnewline
\end{tabular} & \begin{tabular}{c}
$SU(3)_{c}\times U(1)_{em}$\tabularnewline
$\rightarrow\widetilde{U}(1)$\tabularnewline
\end{tabular}\tabularnewline
\hline 
new field & $W^{+},W^{-},Z^{0},A$ & $A_{1},\cdots,A_{7},\widetilde{A}_{8},\widetilde{A}$\tabularnewline
\hline 
\begin{tabular}{c}
new coupling\tabularnewline
constant\tabularnewline
\end{tabular} & $e=g'\cos\theta_{W}$ & $\widetilde{e}=e\cos\theta$\tabularnewline
\hline 
\begin{tabular}{c}
massless\tabularnewline
field\tabularnewline
\end{tabular} & $A$ & $\widetilde{A}$\tabularnewline
\hline
\end{tabular}
\end{table}

\section{Deriving the gap equation}

In the above symmetry analysis it was demonstrated that a multitude
of phases are possible when quarks form Cooper pairs. To decide which
of those will be favorable at given $T$ and $\mu$ one has to determine
the phase with the largest pressure. Generally, the pressure will
grow with the gain of condensation energy of the Cooper pairs. This
in turn will depend on the magnitude of the respective color superconducting
order or gap parameters. In this section we will derive the QCD gap
equation and solve it in the weak coupling limit. 

We will use units in which $\hbar=c=k_{B}=1$. We denote a 4-vector
by capital letters, $X^{\mu}=(x_{0},\mathbf{x})$, with $\mathbf{x}$
being a 3-vector of modulus $|\mathbf{x}|\equiv x$ and direction
$\widehat{\mathbf{x}}\equiv\mathbf{x}/x$. For the summation over
Lorentz indices, we use a notation familiar from Minkowski space,
with metric $g^{\mu\nu}=\mathrm{diag}(+,-,-,-)$, although we exclusively
work in compact Euclidean space-time with volume $V/T$, where $V$
is the 3-volume and $T$ the temperature of the system. Space-time
integrals are denoted as $\int_{0}^{1/T}d\tau\int_{V}d^{3}\mathbf{x}\equiv\int_{X}$. 

The partition function for QCD in absence of external sources reads
\begin{eqnarray}
\mathcal{Z} & = & \int\mathcal{D}A\,\exp\left\{ S_{A}[A]\right\} \,\mathcal{Z}_{q}[A]\;.\label{ZQCD}\end{eqnarray}
 Here the (gauge-fixed) gluon action is \begin{eqnarray}
S_{A}[A] & = & \int_{X}\left[-\frac{1}{4}F_{a}^{\mu\nu}(X)\, F_{\mu\nu}^{a}(X)\right]+S_{gf}[A]+S_{ghost}[A]\;,\label{SA}\end{eqnarray}
 where $F_{\mu\nu}^{a}=\partial_{\mu}A_{\nu}^{a}-\partial_{\nu}A_{\mu}^{a}+gf^{abc}A_{\mu}^{b}A_{\nu}^{c}$
is the gluon field strength tensor, $S_{gf}$ is the gauge-fixing
part, and $S_{ghost}$ the ghost part of the action. 

The partition function for quarks in presence of gluon fields is \begin{eqnarray}
\mathcal{Z}_{q}[A] & = & \int\mathcal{D}\overline{\psi}\,\mathcal{D}\psi\,\exp\left\{ S_{q}[A,\overline{\psi},\psi]\right\} \,\,,\label{Zq}\end{eqnarray}
 where the quark action is \begin{eqnarray*}
S_{q}[A,\overline{\psi},\psi] & = & \int_{X}\overline{\psi}(X)\,\left(i\gamma_{\mu}D_{X}^{\mu}+\mu\gamma_{0}-m\right)\,\psi(X)\,\,,\end{eqnarray*}
 with the covariant derivative $D_{X}^{\mu}=\partial_{X}^{\mu}-igA_{a}^{\mu}(X)T_{a}$.
Note that we focus on a single chemical poteintial and a single mass
for simplicity. In fermionic systems at nonzero density, it is advantageous
to additionally introduce charge-conjugate fermionic degrees of freedom,
\begin{eqnarray}
\psi_{C}(X) & = & C\overline{\psi}^{T}(X)\nonumber \\
\overline{\psi}_{C}(X) & = & \psi^{T}(X)C\nonumber \\
\psi(X) & = & C\overline{\psi}_{C}^{T}(X)\nonumber \\
\overline{\psi}(X) & = & \psi_{C}^{T}(X)C\label{eq:cc}\end{eqnarray}
 where the charge-conjugation matrix satisfies $C^{-1}=C^{\dagger}=C^{T}=-C$,
$C^{-1}\gamma_{\mu}^{T}C=-\gamma_{\mu}$. We may then rewrite the
quark action in the form \begin{eqnarray}
S_{q}[A,\overline{\Psi},\Psi] & = & \frac{1}{2}\int_{X,Y}\overline{\Psi}(X)\,\mathcal{G}_{0}^{-1}(X,Y)\,\Psi(Y)\nonumber \\
 &  & +\frac{g}{2}\int_{X}\overline{\Psi}(X)\,\hat{\Gamma}_{a}^{\mu}A_{\mu}^{a}(X)\,\Psi(X)\,\,,\label{eq:quarkaction}\end{eqnarray}
 where we defined the Nambu-Gor'kov quark spinors \begin{eqnarray*}
 &  & \Psi\equiv\left(\begin{array}{c}
\psi\\
\psi_{C}\end{array}\right),\;\;\;\overline{\Psi}\equiv(\overline{\psi},\overline{\psi}_{C})\,\,,\end{eqnarray*}
 and the free inverse quark propagator in the Nambu-Gor'kov basis
\begin{eqnarray*}
\mathcal{G}_{0}^{-1}(X,Y) & = & \left(\begin{array}{cc}
[G_{0}^{+}]^{-1}(X,Y) & 0\\
0 & [G_{0}^{-}]^{-1}(X,Y)\end{array}\right)\,\,,\end{eqnarray*}
 with the free inverse propagator for quarks and charge-conjugate
quarks \begin{eqnarray*}
[G_{0}^{\pm}]^{-1}(X,Y) & \equiv & (i\gamma_{\mu}\partial_{X}^{\mu}\pm\mu\gamma_{0}-m)\,\delta^{(4)}(X-Y)\,\,,\end{eqnarray*}
 The quark-gluon vertex in the Nambu-Gor'kov basis is defined as \begin{eqnarray}
\hat{\Gamma}_{a}^{\mu} & \equiv & \left(\begin{array}{cc}
\gamma^{\mu}T_{a} & 0\\
0 & -\gamma^{\mu}T_{a}^{T}\end{array}\right)\,\,.\label{NGvertex}\end{eqnarray}
 The factors $1/2$ in Eq. (\ref{eq:quarkaction}) compensate the
doubling of quark degrees of freedom in the Nambu-Gor'kov basis. 

As we shall work in momentum space, we Fourier-transform all fields,
as well as the free inverse quark propagator. Since space-time is
compact, energy-momentum space is discretized, with sums $(T/V)\sum_{K}\equiv T\sum_{n}(1/V)\sum_{\mathbf{k}}$.
For a large volume $V$, the sum over 3-momenta can be approximated
by an integral, $(1/V)\sum_{\mathbf{k}}\simeq\int d^{3}\mathbf{k}/(2\pi)^{3}$.
For bosons, the sum over $n$ runs over bosonic Matsubara frequencies
$\omega_{n}^{\mathrm{b}}=2n\pi T$, while for fermions, it runs over
fermionic Matsubara frequencies $\omega_{n}^{\mathrm{f}}=(2n+1)\pi T$.
In our Minkowski-like notation for four-vectors, $x_{0}\equiv t\equiv-i\tau$,
$k_{0}\equiv-i\omega_{n}^{\mathrm{b}/\mathrm{f}}$. The 4-dimensional
delta-function is conveniently defined as $\delta^{(4)}(X)\equiv\delta(\tau)\delta^{(3)}(\mathbf{x})=-i\delta(x^{0})\delta^{(3)}(\mathbf{x})$.
The Fourier transformations for the fields and free inverse quark
propagator are given by \begin{eqnarray}
\Psi(X) & = & \frac{1}{\sqrt{V}}\sum_{K}e^{-iK\cdot X}\,\Psi(K)\,\,,\nonumber \\
\overline{\Psi}(X) & = & \frac{1}{\sqrt{V}}\sum_{K}e^{iK\cdot X}\,\overline{\Psi}(K)\,\,,\nonumber \\
\mathcal{G}_{0}^{-1}(X,Y) & = & \frac{T^{2}}{V}\sum_{K,Q}e^{-iK\cdot X}\, e^{iQ\cdot Y}\,\mathcal{G}_{0}^{-1}(K,Q)\,\,,\nonumber \\
A_{a}^{\mu}(X) & = & \frac{1}{\sqrt{TV}}\sum_{P}e^{-iP\cdot X}\, A_{a}^{\mu}(P)\,\,.\label{FTA}\end{eqnarray}
 The normalization factors are chosen such that fields in momentum
space are dimensionless. The free inverse quark propagator is diagonal
in momentum space, too, \begin{eqnarray}
\mathcal{G}_{0}^{-1}(K,Q) & = & \frac{1}{T}\left(\begin{array}{cc}
[G_{0}^{+}]^{-1}(K) & 0\\
0 & [G_{0}^{-}]^{-1}(K)\end{array}\right)\delta_{K,Q}^{(4)}\,\,,\label{G0FT}\end{eqnarray}
 where $[G_{0}^{\pm}]^{-1}(K)\equiv\gamma_{\mu}K^{\mu}\pm\mu\gamma_{0}-m$. 

Due to the relations (\ref{eq:cc}), the charge-conjugate quark field
in momentum space is related to the original field via $\psi_{C}(K)=C\overline{\psi}^{T}(-K)$
and $\overline{\psi}_{C}(K)=\psi^{T}(-K)C$. The measure of the functional
integral over quark fields can then be rewritten in the form \begin{eqnarray}
\mathcal{D}\overline{\psi}\,\mathcal{D}\psi & \equiv & \prod_{K}d\overline{\psi}(K)\, d\psi(K)\nonumber \\
 & = & \mathcal{N}\prod_{(K,-K)}d\overline{\psi}(K)\, d\psi(K)\, d\overline{\psi}(-K)\, d\psi(-K)\nonumber \\
 & = & \mathcal{N}'\prod_{(K,-K)}d\overline{\psi}(K)\, d\psi(K)\, d\overline{\psi}_{C}(K)\, d\psi_{C}(K)\nonumber \\
 & = & \mathcal{N}''\prod_{(K,-K)}d\overline{\Psi}(K)\, d\Psi(K)\equiv\mathcal{D}\overline{\Psi}\,\mathcal{D}\Psi\,\,,\label{measure}\end{eqnarray}
 with the constant normalization factors $\mathcal{N}$, $,\mathcal{N}'$
and $\mathcal{N}''$. The last identity has to be considered as a
definition for the expression on the right-hand side. 

\begin{figure}
\caption{\label{Gamma2general}$\Gamma_{2}$ is the sum of all QCD two-particle
irreducible vacuum diagrams. Here, only the two-loop contributions
are shown explicitly. Wavy lines correspond to gluon propagation and
straight lines to quark propagation.}

\vspace{0.5cm}

\includegraphics[scale=0.6]{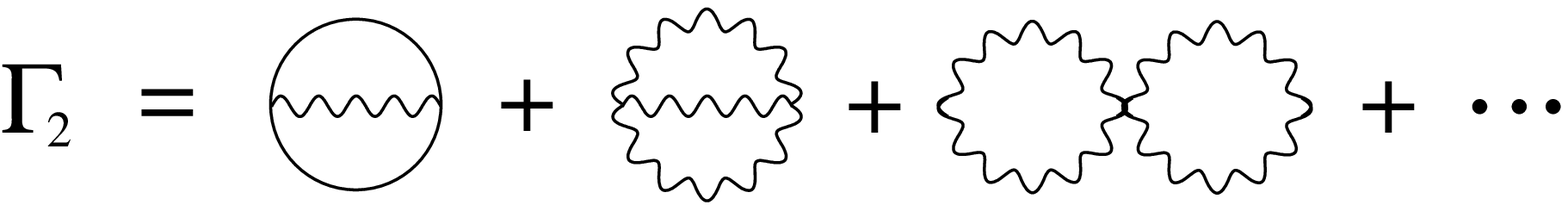}
\end{figure}

Inserting Eqs. (\ref{FTA}) - (\ref{measure}) into Eq. (\ref{Zq}),
the partition function for quarks becomes \begin{eqnarray*}
\mathcal{Z}_{q}[A] & = & \int\mathcal{D}\overline{\Psi}\,\mathcal{D}\Psi\exp\left[\frac{1}{2}\,\overline{\Psi}\left(\mathcal{G}_{0}^{-1}+g\mathcal{A}\right)\Psi\right]\,\,.\end{eqnarray*}
 Here, we employ a compact matrix notation, \begin{eqnarray}
S_{q}[A,\overline{\Psi},\Psi] & \equiv & \frac{1}{2}\overline{\Psi}\,\left(\mathcal{G}_{0}^{-1}+g\mathcal{A}\right)\,\Psi\nonumber \\
 & \equiv & \frac{1}{2}\sum_{K,Q}\overline{\Psi}(K)\left[\mathcal{G}_{0}^{-1}(K,Q)+g\mathcal{A}(K,Q)\right]\Psi(Q)\;\;,\label{eq:compact}\end{eqnarray}
 with the definition \begin{eqnarray}
\mathcal{A}(K,Q) & \equiv & \frac{1}{\sqrt{VT^{3}}}\,\hat{\Gamma}_{a}^{\mu}A_{\mu}^{a}(K-Q)\,\,.\label{calA}\end{eqnarray}

Since the binding of Cooper pairs is a non-perturbative effect we
have to resort to a non-perturbative, self-consistent, many-body resummation
techniques to calculate the gap parameter. For this purpose, it is
convenient to employ the CJT formalism \cite{CJT}. The first step
is to add source terms to the QCD action,\begin{eqnarray}
S[A,\overline{\Psi},\Psi] & \equiv & S_{A}[A]+S_{q}[A,\overline{\Psi},\Psi]\nonumber \\
 & \rightarrow & S[A,\overline{\Psi},\Psi]+JA+\frac{1}{2}\, AKA\nonumber \\
 &  & +\frac{1}{2}\left(\overline{\Psi}H+\overline{H}\Psi+\overline{\Psi}\mathcal{K}\Psi\right)\;,\label{eq:Ssources}\end{eqnarray}
 where we employed the compact matrix notation defined in Eq. (\ref{eq:compact}).
$J$, $\overline{H}$, and $H$ are local source terms for the soft
gluon and relevant quark fields, respectively, while $K$ and $\mathcal{K}$
are bilocal source terms. The bilocal source $\mathcal{K}$ for quarks
is also a matrix in Nambu-Gor'kov space. Its diagonal components are
source terms which couple quarks to antiquarks, while its off-diagonal
components couple quarks to quarks. The latter have to be introduced
for systems which can become superconducting, i.e. where the ground
state has a non-vanishing diquark expectation value, $\langle\psi^{T}C\mathcal{O}\psi\rangle\neq0$.
One then performs a Legendre transformation with respect to all sources
and arrives at the CJT effective action \cite{CJT,kleinert}\begin{eqnarray}
\Gamma\left[A,\overline{\Psi},\Psi,\Delta,\mathcal{G}\right] & = & S\left[A,\overline{\Psi},\Psi\right]-\frac{1}{2}\,\mathrm{Tr}_{g}\ln\Delta^{-1}-\frac{1}{2}\,\mathrm{Tr}_{g}\left(\Delta_{0}^{-1}\Delta-1\right)\nonumber \\
 &  & +\frac{1}{2}\,\mathrm{Tr}_{q}\ln\mathcal{G}^{-1}+\frac{1}{2}\,\mathrm{Tr}_{q}\left(\mathcal{G}_{0}^{-1}\mathcal{G}-1\right)\nonumber \\
 &  & +\Gamma_{2}\left[A,\overline{\Psi},\Psi,\Delta,\mathcal{G}\right]\;.\label{Gamma}\end{eqnarray}
 The tree-level action $S[A,\overline{\Psi},\Psi]$ now depends on
the \emph{expectation values} $A\equiv\langle A\rangle$, $\overline{\Psi}\equiv\langle\overline{\Psi}\rangle$,
and $\Psi\equiv\langle\Psi\rangle$ for the one-point functions of
gluon and quark fields. The traces denoted by $\mathrm{Tr}_{g}$ run
over gluonic degrees of freedom (i.e. over adjoint color) and momenta,
while the traces $\mathrm{Tr}_{q}$ run over quark degrees of freedom(i.e.
fundamental color), flavor, Dirac and Nambu-Gor'kov indices as well
as over momenta. The quantity \begin{eqnarray*}
\Delta_{0}^{-1} & \equiv & -\frac{\delta^{2}S_{A}[A]}{\delta A\delta A}\end{eqnarray*}
 in Eq. (\ref{Gamma}) denotes the free inverse gluon propagator.
$\Delta$ and $\mathcal{G}$ are the \textit{expectation values} for
the two-point functions, i.e., the \emph{full} propagators, of gluons
and quarks, respectively. The functional $\Gamma_{2}$ is the sum
of all two-particle irreducible (2PI) diagrams. These diagrams are
vacuum diagrams, so they have no external legs. They are constructed
from the vertices defined by $S[A,\overline{\Psi},\Psi]$, linked
by full propagators $\Delta$, $\mathcal{G}$, cf. Fig. (\ref{Gamma2general}).

The expectation values for the one- and two-point functions of the
theory are determined from the stationarity conditions \begin{eqnarray}
0 & = & \frac{\delta\Gamma}{\delta A}=\frac{\delta\Gamma}{\delta\overline{\Psi}}\nonumber \\
 & = & \frac{\delta\Gamma}{\delta\overline{\Psi}}=\frac{\delta\Gamma}{\delta\Delta}=\frac{\delta\Gamma}{\delta\mathcal{G}}\;.\label{eq:statcond}\end{eqnarray}
 The first condition yields the Yang-Mills equation for the expectation
value $A$ of the gluon field. The second and third condition correspond
to the Dirac equation for $\Psi$ and $\overline{\Psi}$, respectively.
Since $\overline{\Psi}$ and $\Psi$ are Grassmann-valued fields,
their expectation values must vanish identically, $\overline{\Psi}=\langle\overline{\Psi}\rangle=\Psi=\langle\Psi\rangle\equiv0$.
On the other hand, for the Yang-Mills equation, the solution $A$
is in general non-zero but, at least for the two-flavor color superconductor
considered here, it was shown \cite{Gerhold:2003js,Dietrich:2003nu}
to be parametrically small, $A\sim\phi^{2}/(g^{2}\mu)$, where $\phi$
is the color-superconducting gap parameter. Therefore, to subleading
order in the gap equation it can be neglected. 

The fourth and fifth condition (\ref{eq:statcond}) are Dyson-Schwinger
equations for the gluon and quark propagator, respectively, \begin{eqnarray}
\Delta^{-1} & = & \Delta_{0}^{-1}+\Pi\;,\label{DSEgluon}\\
\mathcal{G}^{-1} & = & \mathcal{G}_{0}^{-1}+\Sigma\;,\label{DSEquark}\end{eqnarray}
 where \begin{eqnarray}
\Pi & \equiv & -2\,\frac{\delta\Gamma_{2}}{\delta\Delta^{T}}\;,\label{selfenergygluon}\\
\Sigma & \equiv & 2\,\frac{\delta\Gamma_{2}}{\delta\mathcal{G}^{T}}\label{selfenergyquark}\end{eqnarray}
are the gluon and quark self-energies, respectively. The Dyson-Schwinger
equation for the quark propagator (\ref{DSEquark}) is a $2\times2$
matrix equation in Nambu-Gor'kov space, since \begin{eqnarray}
\mathcal{G}_{0}^{-1} & \equiv & \left(\begin{array}{cc}
[G_{0}^{+}]^{-1} & 0\\
0 & [G_{0}^{-}]^{-1}\end{array}\right)\;,\nonumber \\
\Sigma & \equiv & \left(\begin{array}{cc}
\Sigma^{+} & \Phi^{-}\\
\Phi^{+} & \Sigma^{-}\end{array}\right)\;,\label{eq:NGsigma}\end{eqnarray}
 where $\Sigma^{+}$ is the regular self-energy for quarks and $\Sigma^{-}$
the corresponding one for charge-conjugate quarks. The off-diagonal
self-energies $\Phi^{\pm}$, the so-called \emph{gap matrices}, connect
regular with charge-conjugate quark degrees of freedom. A non-zero
$\Phi^{\pm}$ corresponds to the diquark condensate. Only two of the
four components of this matrix equation are independent, say $[G^{+}]^{-1}+\Sigma^{+}$
and $\Phi^{+}$. Charge conjugation invariance of the action requires
$[G^{-}]^{-1}+\Sigma^{-}=C\{[G^{+}]^{-1}+\Sigma^{+}\}^{T}C^{-1}$.
Furthermore, the action has to be real valued, yielding the relation
$\Phi^{-}\equiv\gamma_{0}[\Phi^{+}]^{\dagger}\gamma_{0}$. The quark
propagator $\mathcal{G}$ can be formally inverted, \begin{eqnarray}
\mathcal{G} & \equiv & \left(\begin{array}{cc}
\mathcal{G}^{+} & \Xi^{-}\\
\Xi^{+} & \mathcal{G}^{-}\end{array}\right)\;,\label{NGquarkprop}\end{eqnarray}
 where \begin{eqnarray}
\mathcal{G}^{\pm} & \equiv & \left\{ [G_{0}^{\pm}]^{-1}+\Sigma^{\pm}\right.\nonumber \\
 &  & \left.-\Phi^{\mp}\left([G_{0}^{\mp}]^{-1}+\Sigma^{\mp}\right)^{-1}\Phi^{\pm}\right\} ^{-1}\label{eq:G}\end{eqnarray}
 is the propagator describing normal propagation of quasiparticles
and their charge-conjugate counterparts in presence of diquark condensates,
while \begin{eqnarray}
\Xi^{\pm} & \equiv & -\left([G_{0}^{\mp}]^{-1}+\Sigma^{\mp}\right)^{-1}\Phi^{\pm}\mathcal{G}^{\pm}\label{eq:Xi}\end{eqnarray}
 describes the anomalous propagation of quasiparticles. It can be
interpreted as the absorption (+) and the emission ($-$) of a quasiparticle-pair
by the condensate, for details, see Ref. \cite{Rischke:2003mt}. 

\begin{figure}
\caption{\label{quarkselfenergy}The quark self-energy $\Sigma$ written in
the Nambu-Gor'kov basis, cf. Eq. (\ref{eq:NGsigma}). Single straight
lines directing to the left/right correspond to  $\left([G_{0}^{\mp}]^{-1}+\Sigma^{\mp}\right)^{-1}$.
Double straight lines to the left/right correspond to $\mathcal{G}^{\pm}$.
Full/empty dots symbolizes  the Cooper pair condensate $\Phi^{\pm}$
absorbing/emitting a quasiparticle pair.}

\vspace{0.5cm}

\includegraphics[scale=0.7]{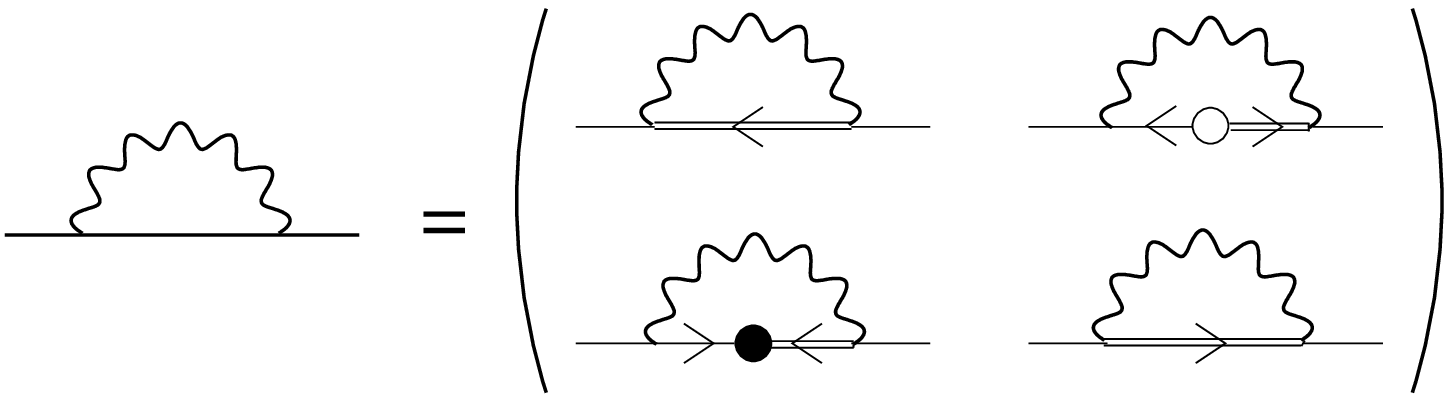}
\end{figure}

In order to proceed one has to make an approximation for $\Gamma_{2}$.
We will restrict to two-loop 2PIs. Furthermore, pure gluonic loops
are proportional to $T^{2}$ \cite{LeBellac} and can be safely neglected
in the limit of small temperatures. Consequently, we will consider
only the first diagramm in Fig. \ref{Gamma2general}. It is an advantage
of the CJT formalism, that truncating $\Gamma_{2}$ will not destroy
the self-consistency of the solution. The gluon self-energy is computed
as $\Pi=-2\,\delta\Gamma_{2}/\delta\Delta^{T}$, i.e., by cutting
a gluon line in the first diagram of Fig. (\ref{Gamma2general}).
Thus, in our approximation $\Pi$ is simply a quark loop with Nambu-Gor'kov
propagators $\mathcal{G}$. Accordingly, the quark self-energy $\Sigma\equiv2\,\delta\Gamma_{2}/\delta\mathcal{G}^{T}$,
is obtained by cutting a quark-line in the same diagram. Its Nambu-Gor'kov
structure is shown in Fig. \ref{quarkselfenergy}. The diagonal components
$\Sigma^{\pm}$ correspond to the ordinary self-energies for particles
and charge-conjugate particles. The self-energies $\Phi^{\pm}$ symbolize
the Cooper pair condensate connecting particles and charge-conjugate
particles. In the following, also the term \emph{gap matrix} will
be used for $\Phi^{+}$. Explicitly, the Nambu-Gor'kov components
of $\Sigma$ read \begin{eqnarray}
\Sigma^{+}(K) & = & -g^{2}\,\frac{T}{V}\sum\limits _{Q}\gamma^{\mu}T_{a}\,\mathcal{G}^{+}(Q)\,\gamma^{\nu}T_{b}\,\Delta_{\mu\nu}^{ab}(K-Q)\,\,,\label{eq:Sigma+}\\
\Sigma^{-}(K) & = & -g^{2}\,\frac{T}{V}\sum\limits _{Q}\gamma^{\mu}T_{a}^{T}\,\mathcal{G}^{-}(Q)\,\gamma^{\nu}T_{b}^{T}\,\Delta_{\mu\nu}^{ab}(K-Q)\,\,,\label{eq:Sigma-}\\
\Phi^{+}(K) & = & g^{2}\,\frac{T}{V}\sum\limits _{Q}\gamma^{\mu}T_{a}^{T}\,\Xi^{+}(Q)\,\gamma^{\nu}T_{b}\,\Delta_{\mu\nu}^{ab}(K-Q)\,\,,\label{Phi+}\\
\Phi^{-}(K) & = & g^{2}\,\frac{T}{V}\sum\limits _{Q}\gamma^{\mu}T_{a}\,\Xi^{-}(Q)\,\gamma^{\nu}T_{b}^{T}\,\Delta_{\mu\nu}^{ab}(K-Q)\,\,.\label{Phi-}\end{eqnarray}
 Inserting these self-energies (and the corresponding one for the
gluons) into the Dyson-Schwinger equations (\ref{DSEgluon}) and (\ref{DSEquark})
one obtains a coupled set of integral equations which has to be solved
self-consistently. In particular, the Dyson-Schwinger equations for
the off-diagonal components $\Phi^{\pm}$ of the inverse propagator
$\mathcal{G}^{-1}$, i.e., Eqs. (\ref{Phi+}) and (\ref{Phi-}), are
the \emph{gap equations} for the color-superconducting condensate. 

Even in the mean-field approximation applied here a completely self-consistent
solution of the coupled set of Dyson-Schwinger equations is not feasible.
Therefore, we will refrain from determining the gluon propagator self-consistently
but resort to its hard-dense loop approximation \cite{LeBellac}.
Thus we will obtain a well-controlled, approximate solution of the
gap-equation as it will be demonstrated in next section.

\section{Quasi-particle excitations in color superconductor}

\label{sec:excite}In this section we analyse how the presence of
a Cooper pair condensate affects the excitation spectrum of quasiparticles.
To this end, the poles of the propagator $\mathcal{G}$ have to be
determined. The excitation spectrum depends on the specific symmetry
broken pattern of the superconducing phase considered here, cf. Sec.
\ref{sec:Sym1} and \ref{sec:Sym2}. To this end, we first analyse
the Dirac, color and flavor structure of the varios terms contained
in $\mathcal{G}$, cf. Eq. (\ref{eq:G}). As we work in the ultra-relativistic
limit, $m=0$, we may expand the free (charge conjugate) quark propagator
$G_{0}^{\pm}$ in terms of positive and negative energy states using
the projection operator $\Lambda_{\mathbf{k}}^{e}\equiv(1+e\,\gamma_{0}\,\bm{\mbox{$\gamma$}}\cdot\widehat{\mathbf{k}})/2$,
where $e=+$ projects on (charge conjugate) quarks and $e=-$ projects
on (charge conjugate) antiquarks. One obtains for the respective free
propagators \begin{equation}
[G_{0}^{\pm}]^{-1}(K)=\gamma_{0}\sum_{e=\pm}\left[k_{0}-(ek\mp\mu)\right]\,\Lambda_{\mathbf{k}}^{e}\,\,.\end{equation}
Furthermore, $[G_{0}^{\pm}]^{-1}$ is diagonal in color and flavor
space. To simplify the further analysis we anticipate that in weak
coupling limit adopted here and to subleading order, the regular self-energy
$\Sigma^{\pm}$ may be approximated by its one-loop approximation
\cite{Brown:2000eh,Wang:2001aq,Schafer:2004zf} in the leading logarithmic
order (cf. Ref. \cite{Gerhold:2005uu} for result beyond the leading
logarithmic order). In this approximation, the self-consistent propagators
in Eqs. (\ref{eq:Sigma+},\ref{eq:Sigma-}) are replaced by free propagators.
In particular, this amounts to neglecting the Cooper pair condensate
in $\Sigma^{\pm}$, so that $\Sigma^{\pm}$ is diagonal in color and
flavor space. In Dirac space one finds \cite{Brown:2000eh,Wang:2001aq,Schafer:2004zf}\begin{equation}
\Sigma^{+}(K)=\Sigma^{-}(K)\approx\gamma_{0}\;\bar{g}^{2}\left(k_{0}\ln\frac{M^{2}}{k_{0}^{2}}+i\pi|k_{0}|\right)\;,\label{Sigmanormal}\end{equation}
 where $\bar{g}\equiv g/(3\sqrt{2}\pi)$ and $M^{2}\equiv(3\pi/4)m_{g}^{2}$
(with $m_{g}^{2}=N_{f}\frac{g^{2}\mu^{2}}{6\pi^{2}}$). To subleading
order it is sufficient to include only the real part of $\Sigma^{\pm}(K)$
\cite{Wang:2001aq}. Its effect is simply a shift of the quasiparticle
poles, $k_{0}\rightarrow k_{0}/Z(k_{0})$, where we introduced the
quark wave function renormalization function \begin{equation}
Z(k_{0})\equiv\left(1+\bar{g}^{2}\ln\frac{M^{2}}{k_{0}^{2}}\right)^{-1}\;.\label{quarkrenormfunc}\end{equation}
 The imaginary part of $\Sigma^{\pm}(K)$ can be shown to contribute
beyond subleading order \cite{Manuel:2000nh} and therefore will be
neglected in the following. Physically, the imaginary part of $\Sigma^{\pm}(K)$
gives rise to a finite life-time of the quasi-particles off the Fermi
surface, which reduces the magnitude of the gap at sub-subleading
order \cite{Manuel:2000nh}. 

Turning to the structure of $\Phi^{\pm}$ we introduce a compact notation
following \cite{Schmitt:2002sc}, \begin{equation}
\Phi^{+}(K)=\sum_{e=\pm}\phi^{e}(K)\,\mathcal{M}_{\mathbf{k}}\,\Lambda_{\mathbf{k}}^{e}\,\,,\label{eq:Phi+decomp}\end{equation}
 where $\phi^{e}$, the so-called \emph{gap function}, is a scalar
function of 4-momentum $K$. In general, the quantity $\mathcal{M}_{\mathbf{k}}$
is a matrix in color, flavor, and Dirac space, which is determined
by the symmetries of the color-superconducting order parameter. It
can be chosen such that \begin{equation}
[\mathcal{M}_{\mathbf{k}},\Lambda_{\mathbf{k}}^{e}]=0\;.\end{equation}
Note that the above is only valid in massless case. For the spin-0
phases 2SC and CFL the strucure of $\mathcal{M}_{\mathbf{k}}$ has
been determined in Sec. \ref{sec:Sym2}. In Tab. \ref{tableM} the
corresponding structures for three spin-1 phases, the CSL, the polar
and the A-phase, are shown. Note that these spin-1 phases are actually
the mixtures of longitudinal and transverse phases, and that the transverse
phases have the largest gaps. 

\begin{table}
\caption{\label{tableM}The structure of the matrices $\mathcal{M}_{\mathbf{k}}$
in various color-superconducting phases. The last two columns show
the two eigenvalues $\lambda_{r}$ of $L_{\mathbf{k}}$ and their
degeneracy $d_{r}$ (counting color-flavor degrees of freedom in the
2SC and CFL phases, and color-Dirac degrees of freedom in the three
spin-one phases).}

\vspace{0.5cm}

\begin{tabular}{|l|c|c|c|}
\hline
~phase~ & ${\cal M}_{\bf k}$ & $\lambda_1\; (d_1)$
& $\lambda_2\; (d_2)$ \\
\hline
~2SC & $\gamma_5 \, \sigma_2 \, J_3$ 
& 1 (4) & 0 (2) \\
\hline
~CFL & $\gamma_5 \, {\bf I} \cdot {\bf J} $ 
& 4 (8) & 1 (1) \\
\hline
~CSL & ${\bf J} \cdot \left[ \hat{\bf k} + \mbox{\boldmath
$\gamma$}_\perp({\bf k}) \right]$ 
& 4 (4) & 1 (8) \\
\hline
~polar & $J^3 \left[ \hat{k}^z + 
\gamma ^z_\perp ({\bf k})\right]$ & 1 (8) & 0 (4) \\
\hline
~A & $J^3 \left[ \hat{k}^x + i\hat{k}^y +
\gamma ^x_\perp({\bf k})+i\gamma ^y_\perp ({\bf k})\right]$ 
& 2 (4) & 0 (8) \\
\hline
\hline
\end{tabular}
\end{table}

With representation (\ref{eq:Phi+decomp}) for $\Phi^{+}$ one may
rewrite the second term on the r.h.s. of Eq. (\ref{eq:G}) \[
\Phi^{-}G^{-}\Phi^{+}[G^{-}]^{-1}=\sum_{e=\pm}|\phi^{e}(K)|^{2}L_{\mathbf{k}}\Lambda_{\mathbf{k}}^{-e},\]
 where we introduced \begin{equation}
L_{\mathbf{k}}\equiv\gamma_{0}\mathcal{M}_{\mathbf{k}}^{\dagger}\mathcal{M}_{\mathbf{k}}\gamma_{0}\;.\label{L}\end{equation}
 Since $L_{\mathbf{k}}$ is hermitian, it has real eigenvalues, $\lambda_{r}$,
and can be expanded in terms of a complete set of orthogonal projectors,
$\mathcal{P}_{\mathbf{k}}^{r}$, \begin{equation}
L_{\mathbf{k}}=\sum_{r}\lambda_{r}\mathcal{P}_{\mathbf{k}}^{r}\,\,.\end{equation}
In the five phases considered here, there are only two distinct eigenvalues
and therefore two distinct projectors. The eigenvalues are also listed
in Tab. \ref{tableM}, and the projectors can be expressed in terms
of $L_{\mathbf{k}}$ via \begin{equation}
\mathcal{P}_{\mathbf{k}}^{1,2}=\frac{L_{\mathbf{k}}-\lambda_{2,1}}{\lambda_{1,2}-\lambda_{2,1}}\,\,.\label{P}\end{equation}
 Note, that $[L_{\mathbf{k}},\Lambda_{\mathbf{k}}^{e}]=[\mathcal{P}_{\mathbf{k}}^{1,2},\Lambda_{\mathbf{k}}^{e}]=0$.
Since the projectors $\mathcal{P}^{r}\,\Lambda_{\mathbf{k}}^{e}$
form a basis in color, flavor, and Dirac space, inverting the term
in the curly brackets in Eq. (\ref{eq:G}) is straightforward, yielding
\begin{equation}
\mathcal{G}^{+}(K)=[G^{-}]^{-1}(K)\sum_{e,r}\mathcal{P}_{\mathbf{k}}^{r}\,\Lambda_{\mathbf{k}}^{-e}\,\frac{1}{[k_{0}/Z(k_{0})]^{2}-\left[\epsilon_{\mathbf{k},r}^{e}(\phi^{e})\right]^{2}}\,\,,\label{eq:G+}\end{equation}
 where \begin{equation}
\epsilon_{\mathbf{k},r}^{e}(\phi^{e})=\left[(ek-\mu)^{2}+\lambda_{r}\,|\phi^{e}|^{2}\right]^{1/2}\,\,.\label{disp}\end{equation}
 Inserting Eqs. (\ref{eq:Phi+decomp},\ref{eq:G+}) into Eq. (\ref{eq:Xi})
yields \begin{equation}
\Xi^{+}(K)=-\sum_{e,r}\gamma_{0}\,\mathcal{M}_{\mathbf{k}}\,\gamma_{0}\,\mathcal{P}_{\mathbf{k}}^{r}\,\Lambda_{\mathbf{k}}^{-e}\,\frac{\phi^{e}(K)}{[k_{0}/Z(k_{0})]^{2}-\left[\epsilon_{\mathbf{k},r}^{e}(\phi^{e})\right]^{2}}\,\,.\label{eq:Xi+}\end{equation}
 Obviously, the poles of the propagators $\mathcal{G}^{+}$ and $\Xi^{+}$
are located at $k_{0}\equiv\pm Z(k_{0})\,\epsilon_{\mathbf{k},r}^{e}(\phi^{e})$
and the same result can be obtained for the propagators $\mathcal{G}^{-}$
and $\Xi^{-}$. Eq. (\ref{disp}) is the relativistic analogue of
the standard BCS dispersion relation. The presence of the Cooper pair
condenstate has generated a \textit{gap} in the excitation spectrum
of quasiquarks, $e=+$. Hence, even on the Fermi surface, $k=\mu$,
they carry a non-zero amount of energy, which is of the order of the
gap function, $\epsilon_{\mu,r}^{+}=\sqrt{\lambda_{r}}\,\phi^{+}$.
To excite (generate) a pair of a quasiparticle and a quasiparticle
hole therefore costs at least the energy $2\sqrt{\lambda_{r}}\,\phi^{+}$.
Such an excitation process can be interpreted as a break-up of a Cooper
pair and the required energy as the Cooper pair binding energy. To
investigate which quasiparticles exhibit a gap in their excitation
spectrum one has to determine the eigenvalues $\lambda_{r}$ of the
operator $L_{\mathbf{k}}$. For the phases considered here, $L_{\mathbf{k}}$
has one or two non-zero eigenvalues. For $\lambda_{r}=0$ the respective
quasi-particles remain ungapped. The degeneracy $d_{r}$ gives the
number of excitation branches with the same gap, $\sqrt{\lambda_{r}}\,\phi^{+}$. 

Finally, for quasi anti-quarks, $e=-$, the dispersion relation hardly
differs from the non-interacting case in the limit of large $\mu$,
$\epsilon_{\mathbf{k}}^{-}\approx k+\mu$. 

In the next section, the gap equation (\ref{Phi+}) for the gap function
$\phi^{e}(K)$ will be solved for quasi-particle excitations, $e=+$,
at zero temperature in order to determine the magnitude of the gap
in their excitation spectrum.

\section{Solving the gap equation}

\label{sec:gapequation}To obtain the Dyson-Schwinger equation for
the scalar gap function $\phi^{+}(K)$ one inserts Eq. (\ref{eq:Xi+})
into Eq. (\ref{Phi+}), multiplies from the right with $\mathcal{M}_{\mathbf{k}}^{\dagger}\Lambda_{\mathbf{k}}^{+}$,
and traces over color, flavor, and Dirac space. The result is the
so-called \textit{gap equation}\begin{eqnarray}
\phi^{+}(K) & = & g^{2}\, T\sum_{n}\int\frac{d^{3}\mathbf{q}}{(2\pi)^{3}}\sum_{e',s}\frac{\phi^{e'}(Q)}{\left[q_{0}/Z(q_{0})\right]^{2}-\left[\epsilon_{\mathbf{q},s}^{e'}(\phi^{e'})\right]^{2}}\nonumber \\
 &  & \times\Delta_{\mu\nu}^{ab}(K-Q)\mathcal{T}_{ab,s}^{\mu\nu,ee'}(\mathbf{k},\mathbf{q})\,\,,\label{eq:GEphie}\end{eqnarray}
 where \begin{equation}
\mathcal{T}_{ab,s}^{\mu\nu,ee'}(\mathbf{k},\mathbf{q})\equiv-\frac{\mathrm{Tr}\left[\gamma^{\mu}T_{a}^{T}\gamma_{0}\mathcal{M}_{\mathbf{q}}\gamma_{0}\mathcal{P}_{\mathbf{q}}^{s}\Lambda_{\mathbf{q}}^{-e'}\gamma^{\nu}T_{b}\mathcal{M}_{\mathbf{k}}^{\dagger}\Lambda_{\mathbf{k}}^{e}\right]}{\mathrm{Tr}\left[\mathcal{M}_{\mathbf{k}}\mathcal{M}_{\mathbf{k}}^{\dagger}\Lambda_{\mathbf{k}}^{e}\right]}\,\,.\label{T}\end{equation}
 The power counting scheme for the various contributions arising on
the r.h.s. of the gap equation (\ref{eq:GEphie}) is the same for
all phases considered here. Due to the factor $g^{2}$ the integral
must give terms proportional to $\phi/g^{2}$ in order to fulflill
the equality. They contribute to the gap function at the largest possible
order, $O(\phi)$, and are therefore referred to as \textit{leading
order} contributions. Terms of order $\phi/g$ from the integral contribute
at the order $O(g\phi)$ to the gap function and are called \textit{subleading
order} corrections. \textit{Sub-subleading order} corrections contributing
at $O(g^{2}\phi)$ to the gap function have not been calculated yet
and will be neglected in the following. 

In principle it is possible to proceed without deciding on one of
the specific phases considered here. However, to keep the calculation
as transparent as possible we restrict to the 2SC phase. The results
for the other phases will be summarized at the end of this section. 

As already mentioned in the previous section, the Dyson-Schwinger
equation for the gluon propagator will not be solved self-consistently.
Instead we will adopt the gluon propagator in its HDL approximation.
In pure Coulomb gauge they can be decomposed into a electric (or longitudinal)
and a magnetic (or transversal) part as \cite{Rischke:2003mt,LeBellac}\begin{eqnarray*}
\Delta_{00}(P) & = & \Delta_{\ell}(P)\\
\Delta_{0i}(P) & = & 0\\
\Delta_{ij}(P) & = & \Delta_{t}(P)\left(\delta_{ij}-\hat{p}_{i}\hat{p}_{j}\right)\,\,.\end{eqnarray*}
 with $P\equiv K-Q$ and the longitudinal and transverse propagators
$\Delta_{\ell,t}$. Since the HDL propagators are diagonal in adjoint
color space, $\Delta_{ab}^{\mu\nu}=\delta_{ab}\Delta^{\mu\nu}$, the
color indices can safely be suppressed. The explicit forms of the
propagators $\Delta_{\ell,t}$ can be found in Ref. \cite{Rischke:2003mt,LeBellac}.
For our purposes the approximative forms of $\Delta_{\ell,t}$ given
below are sufficient. As will be substantiated in the following the
\textit{leading} order contribution to the gap arises from almost
static magnetic gluons with \begin{equation}
\Delta_{t}^{\mathrm{LDM}}(P)\simeq\frac{p^{4}}{p^{6}+M^{4}\,\omega^{2}}\Theta(M-p).\label{LDMprop}\end{equation}
 The term $M^{4}\omega^{2}$ arises from Landau-damping which dynamically
screens magnetic gluons at the scale $p\sim M^{2/3}\omega^{1/3}$.
Consequently almosr static magnetic gluons dominate the interaction
between quarks at large distances. \textit{Subleading} order corrections
come from non-static magnetic gluons with \begin{equation}
\Delta_{t}^{\mathrm{NSM}}(p)\simeq\frac{1}{p^{2}}\Theta(p-M)\,\,,\label{NSMprop}\end{equation}
 and from static electric gluons with \begin{equation}
\Delta_{\ell}^{\mathrm{SE}}(p)\simeq-\frac{1}{p^{2}+3m_{g}^{2}}\,\,,\label{SEprop}\end{equation}
 where the constant mass term $m_{g}^{2}=N_{f}\frac{g^{2}\mu^{2}}{6\pi^{2}}$
provides Debye-screening at the scale $p\sim g\mu$. As the gap equation
(\ref{eq:GEphie}) requires a self-consistent solution, we may proceed
by first giving the known result and confirming it in the following.
At zero temperature the 2SC gap function assumes for on-shell quasiparticles
($e=+$) with momenta $k\approx\mu$ the value \begin{equation}
\phi_{0}^{\mathrm{2SC}}=2\tilde{b}b_{0}'\mu\exp\left(-\frac{\pi}{2\bar{g}}\right)\,\,,\label{phi02SC}\end{equation}
 where \begin{eqnarray}
\tilde{b} & \equiv & 256\pi^{4}\left(\frac{2}{N_{f}g^{2}}\right)^{5/2}\nonumber \\
b_{0}' & \equiv & \exp\left(-\frac{\pi^{2}+4}{8}\right)\,\,.\label{constants}\end{eqnarray}
 The term in the exponent of Eq. (\ref{phi02SC}) was first determined
by Son \cite{Son:1998uk} via a renormalization group analysis taking
into account the long-range nature of almost static, Landau-damped
magnetic gluons, cf. Eq. (\ref{LDMprop}). Before that, a non-zero
magnetic mass (analogous to the Debye-mass for electric gluons) was
assumed leading to the standard BCS exponent, $\exp\left(-\frac{3\pi^{2}\Lambda^{2}}{2\mu^{2}g^{2}}\right)$,
where $\Lambda^{2}$ is the the typical momentum of the exchanged
gluon and $g^{2}/3$ the effective coupling in the anti-triplet channel.
This finding was crucial since in the correct exponential (\ref{phi02SC})
the parametric suppression becomes much weaker for small values of
$g$, i.e. for large values of $\mu$. The factor $\tilde{b}$ in
front of the exponential is generated by the exchange of non-static
magnetic (\ref{NSMprop}) and static electric (\ref{SEprop}) gluons
\cite{Hong:1999fh,Pisarski:1999tv,Schafer:1999jg}. The prefactor
$b_{0}'$ is due to the real part of the quark self-energy as given
by Eq. (\ref{quarkrenormfunc}) \cite{Brown:2000eh,Wang:2001aq}. 

From Tab. (\ref{tableM}) we read off \begin{eqnarray*}
\mathcal{M}_{\mathbf{k}} & = & \gamma_{5}\sigma_{2}J_{3}\end{eqnarray*}
 and hence with Eqs. (\ref{L},\ref{P}) one finds the projectors
$\mathcal{P}_{\mathbf{k}}^{1,2}$ to be \begin{eqnarray*}
\mathcal{P}_{\mathbf{k}}^{1} & = & J_{3}^{2}\;,\\
\mathcal{P}_{\mathbf{k}}^{2} & = & 1-J_{3}^{3}\;,\end{eqnarray*}
 with the eigenvalues $\lambda_{1}=1$ (4-fold) and $\lambda_{2}=0$
(2-fold). This yields \begin{eqnarray}
\mathcal{T}_{1}^{\ell,ee'}(\mathbf{k},\mathbf{q}) & = & \frac{1}{3}\,\left(1+ee'\hat{\mathbf{k}}\cdot\hat{\mathbf{q}}\right)\,\,,\nonumber \\
\mathcal{T}_{1}^{t,ee'}(\mathbf{k},\mathbf{q}) & = & \frac{1}{3}\,\left[3-ee'\hat{\mathbf{k}}\cdot\hat{\mathbf{q}}\right.\,\,.\nonumber \\
 &  & \left.-\frac{(ek-e'q)^{2}}{p^{2}}\left(1+ee'\hat{\mathbf{k}}\cdot\hat{\mathbf{q}}\right)\right]\,\,.\label{T2SC2}\end{eqnarray}
 and $\mathcal{T}_{2}^{\mu\nu,ee'}(\mathbf{k},\mathbf{q})\equiv0$.
Hence, the gapless quasi-particles corresponding to $\mathcal{P}_{\mathbf{k}}^{2}$
do not enter the gap equation. Performing the Matsubara sum in Eq.
(\ref{eq:GEphie}) one then obtains for the gapped branch with the
projector $\mathcal{P}_{\mathbf{k}}^{1}$ to subleading order \begin{eqnarray}
\phi^{+}(\epsilon_{\mathbf{k}}^{e},k) & = & \frac{g^{2}}{16\pi^{2}k}\int_{\mu-\delta}^{\mu+\delta}dq\, q\sum_{e'}Z(\epsilon_{\mathbf{q}}^{e'})\,\frac{\phi^{e'}(\epsilon_{\mathbf{q}}^{e'},q)}{\epsilon_{\mathbf{q}}^{e'}}\nonumber \\
 &  & \times\tanh\left(\frac{\epsilon_{\mathbf{q}}^{e'}}{2T}\right)\sum\limits _{m=0}^{1}\int_{|k-q|}^{k+q}dp\, p\left(\frac{p^{2}}{kq}\right)^{m}\nonumber \\
 &  & \left\{ -2\,\Delta_{00}^{\mathrm{SE}}(p)\,\eta_{2m}^{\ell}+\left[\;2\,\Delta_{t}^{\mathrm{NSM}}(p)\right.\right.\nonumber \\
 &  & +\Delta_{t}^{\mathrm{LDM}}\left(\epsilon_{\mathbf{q}}^{e'}+\epsilon_{\mathbf{k}}^{e},p\right)\nonumber \\
 &  & \left.\left.+\Delta_{t}^{\mathrm{LDM}}\left(\epsilon_{\mathbf{q}}^{e'}-\epsilon_{\mathbf{k}}^{e},p\right)\right]\eta_{2m}^{t}\right\} .\label{phi1}\end{eqnarray}
 To subleading order the coefficients $\eta_{2m}^{\ell,t}$ are given
by $\eta_{0}^{\ell}=\eta_{0}^{t}=2/3$ and $\eta_{2}^{\ell}=-\eta_{2}^{t}=-1/6\,$.
(Other phases than 2SC have different coefficients $\eta_{2m}^{\ell,t}$
and possibly have an additional excitation branch. To subleading order,
however, the coefficients $\eta_{2m}^{\ell,t}$ (with $-1\leq m\leq2$)
become simple numbers of order $O(1)$ for all phases considered here.
It follows that the values of the respective gap functions differ
only at subleading order \cite{Schmitt:2002sc}.) To obtain Eq. (\ref{phi1})
several sub-subleading contributions are dropped. For example, the
integration over $q$ can be confined to a narrow interval of length
$2\delta$ around the Fermi surface, $\delta\sim g\mu$, since the
gap function peaks strongly around the Fermi surface. Furthermore,
use was made of the fact, that the gap function must be an even function
in $K$, $\phi^{e}(K)=\phi^{e}(-K)$. This property follows from the
antisymmetry of the fermionic quark wave-functions. One has with $\psi_{C}(K)=C\bar{\psi}^{T}(-K)$,
$\bar{\psi}_{C}(K)=\psi^{T}(-K)C$ and $C=-C^{-1}=-C^{T}$\begin{eqnarray*}
\sum\limits _{K}\bar{\psi}_{C}(K)\Phi^{+}(K)\psi(K) & \equiv & -\sum\limits _{K}\psi^{T}(K)\left[\Phi^{+}(K)\right]^{T}\bar{\psi}_{C}^{T}(K)\\
 & = & \sum\limits _{K}\bar{\psi}_{C}(K)C^{-1}\left[\Phi^{+}(-K)\right]^{T}C\psi(K)\;.\end{eqnarray*}
 Hence, the gap matrix must fulfill \begin{eqnarray*}
C\Phi^{+}(K)C^{-1} & = & \left[\Phi^{+}(-K)\right]^{T}.\end{eqnarray*}
 Since in the 2SC case we have $C\gamma_{5}\Lambda_{\mathbf{k}}^{e}C^{-1}=[\gamma_{5}\Lambda_{-\mathbf{k}}^{e}]^{T}$
and $[J_{3}\tau_{2}]^{T}=J_{3}\tau_{2}$, it follows that $\phi^{e}(K)=\phi^{e}(-K)$.
Moreover, the gap function was assumed to be isotropic in momentum
space, $\phi^{e}(\mathbf{k})\equiv\phi^{e}(k)$ \cite{Schmitt:2002sc}.
Finally, all imaginary contributions arising from the energy dependence
of the gluon propagators and the gap function itself are neglected.
It follows, that the energy dependence of the gap function occuring
in the final result has not been determined self-consistently. In
Ref. \cite{Pisarski:1999tv} it is argued that imaginary contributions
are negligible to subleading order. 

Using the well-known solution Eq. (\ref{phi02SC}) for the 2SC gap
function it is now straightforward to confirm that the various gluon
sectors occuring in Eq. (\ref{phi1}) indeed contribute at the claimed
orders of magnitude. For the static electric and the non-static magnetic
gluons the integral over $q$ gives for $e^{\prime}=+$ and $m=1$\begin{eqnarray}
g^{2}\phi^{+}\int_{\mu-\delta}^{\mu+\delta}\frac{dq}{\epsilon_{\mathbf{q}}^{+}} & = & 2g^{2}\phi^{+}\ln\left(\frac{\delta+\sqrt{\delta^{2}+(\phi^{+})^{2}}}{\phi^{+}}\right)\nonumber \\
 & \simeq & 2g^{2}\phi^{+}\ln\left(\frac{2\,\delta}{\phi^{+}}\right)\nonumber \\
 & \sim & g^{2}\phi^{+}\ln\left[\frac{g}{\exp(-1/g)}\right]\sim g\phi^{+},\label{BCSlog}\end{eqnarray}
 where for simplicity the gap function is moved out of the integral
and the fact was used that $\delta\sim g\mu\gg\phi^{+}\sim\mu\exp(-1/g)$.
Note that this result does not depend on the specific choice of $\delta$,
as long as $\delta$ is of the order $\delta\sim g^{c}(\mu^{a}\phi^{b})^{\frac{1}{a+b}}$
with some numbers $a\neq0$ and $b,c$. The integral over $p$ in
combination with the factor $1/k\sim\mu$ is of order $O(1)$. It
follows that static electric and non-static magnetic gluons actually
contribute at subleading order to the gap function. The above integral
and the logarithm that follows from it also appear in normal BCS theory
\cite{fetwal}. Therefore we will refer to it as the \textit{BCS-logarithm}.
For $e'=-$, one has $\epsilon_{\mathbf{q}}^{-}\simeq q+\mu$ and
the contribution corresponding to Eq. (\ref{BCSlog}) is of order
$g^{2}\phi\delta/\mu\sim g^{3}\phi$, i.e. beyond subleading order. 

In the case of almost static, Landau-damped magnetic gluons the integration
over $q$ can be estimated as in Eq. (\ref{BCSlog}). Choosing the
term with $e^{\prime}=+$ and $m=0$ and canceling $q\simeq\mu$ in
the integral with the factor $1/k\sim1/\mu$ at the front one obtains
the BCS-logarithm. In order to render the total contribution of leading
order, the $p-$integral must yield an additional factor $1/g$. Using
the notation $\omega_{\pm}\equiv\epsilon_{\mathbf{q}}^{+}\pm\epsilon_{\mathbf{k}}^{+}$
one finds as the dominant contribution the term \begin{eqnarray}
\int_{k-q}^{M}dp\frac{p^{5}}{p^{6}+M^{4}\omega_{\pm}^{2}} & = & \frac{1}{6}\ln\left[\frac{M^{6}+M^{4}\omega_{\pm}^{2}}{(k-q)^{6}+M^{4}\omega_{\pm}^{2}}\right]\nonumber \\
 & \sim & \ln\left(\frac{M^{2}}{\omega_{\pm}^{2}}\right)\sim\frac{1}{g}\,\,.\label{ll}\end{eqnarray}
 The last two steps are justified if one chooses for the cut-off $\delta$
in the $q$ integral $\delta=(M^{2}\phi)^{1/3}$. Consequently, in
combination with the BCS-logarithm (\ref{BCSlog}) this indeed is
a leading order contribution. 

After having identified the leading and subleading order contributions,
one evaluates the $p$ integral in Eq. (\ref{phi1}) and finds to
subleading order (supressing all the superscripts $e=e^{\prime}=+$)
\begin{eqnarray}
\phi(\epsilon_{\mathbf{k}},k) & = & \bar{g}^{2}\int_{0}^{\delta}d(q-\mu)Z(\epsilon_{\mathbf{q}})\frac{\phi(\epsilon_{\mathbf{q}},q)}{\epsilon_{\mathbf{q}}}\nonumber \\
 &  & \times\tanh\left(\frac{\epsilon_{\mathbf{q}}}{2T}\right)\frac{1}{2}\ln\left(\frac{b^{2}\mu^{2}}{|\epsilon_{\mathbf{q}}^{2}-\epsilon_{\mathbf{k}}^{2}|}\right)\,\,.\label{gap1}\end{eqnarray}
 How to solve Eq. (\ref{gap1}) to subleading order is demonstrated
explicitly in \cite{Wang:2001aq}. In the following, only some important
steps are presented. As first proposed by Son \cite{Son:1998uk} one
approximates \begin{eqnarray*}
\frac{1}{2}\ln\left(\frac{b^{2}\mu^{2}}{|\epsilon_{\mathbf{q},s}^{2}-\epsilon_{\mathbf{k},r}^{2}|}\right) & \simeq & \Theta(\epsilon_{\mathbf{q},s}-\epsilon_{\mathbf{k},r})\ln\left(\frac{b\mu}{\epsilon_{\mathbf{q},s}}\right)\\
 &  & +\Theta(\epsilon_{\mathbf{k},r}-\epsilon_{\mathbf{q},s})\ln\left(\frac{b\mu}{\epsilon_{\mathbf{k},r}}\right)\,\,.\end{eqnarray*}
 Rewriting Eq. (\ref{gap1}) in terms of the new variables \cite{Wang:2001aq}\begin{eqnarray}
x & \equiv & \bar{g}\ln\left(\frac{2b\mu}{k-\mu+\epsilon_{k}}\right),\nonumber \\
y & \equiv & \bar{g}\ln\left(\frac{2b\mu}{q-\mu+\epsilon_{q}}\right)\label{vartrans}\end{eqnarray}
 and differentiating twice with respect to $x$ one can transform
the gap equation into the form of Airy's differential equation. Its
solution can be written as \begin{equation}
\phi(x)\equiv\phi_{0}^{\mathrm{2SC}}F(x)\,\,,\label{solution}\end{equation}
 where $\phi_{0}^{\mathrm{2SC}}$ is given by Eq. (\ref{phi02SC})
and $F(x)$ is a combination of different Airy-functions \cite{Wang:2001aq}.
The situation simplifies considerably when one neglects the effect
of the normal quark self-energy, i.e. the factor $Z(\epsilon_{\mathbf{q}})$
in Eq. (\ref{gap1}). Besides $b_{0}^{\prime}\equiv1$ in Eq. (\ref{constants}),
one then finds $F(x)\equiv\sin x$ \cite{Son:1998uk,Pisarski:1999tv},
as the differential equation for $\phi$ simplifies to that of an
harmonic oszillator. For $|k-\mu|\sim\phi$ one obtains $x=\pi/2+O(\bar{g})$,
while for $|k-\mu|\sim g\mu$ it is $x\sim O(\bar{g})$. Hence, for
momenta exponentially close to the Fermi surface, $|k-\mu|\sim\phi$,
the on-shell gap function is $\phi(\epsilon_{k},k)=\phi_{0}^{\mathrm{2SC}}[1+O(\bar{g}^{2})]$,
i.e. constant to subleading order. Otherwise, for example for momenta
$|k-\mu|\sim g\mu$, one has already $\phi(\epsilon_{k},k)=\bar{g}\phi_{0}^{\mathrm{2SC}}$.
Hence, the on-shell gap-function is sharply peaked around the Fermi
surface. 

The gap for on-shell quasi-particles at the Fermi surface in the 2SC
phase as given by Eq. (\ref{phi02SC}) is valid for asymptotic densities
where $g(\mu)\ll1$. At chemical potentials of physical relevance,
however, one has $g(\mu)\sim O(1)$. Then sub-subleading terms which
are corrections of order $O(g)$ to the prefactor of the gap function
in principle have to be accounted for. So far, they have not been
calculated. Keeping this caveat in mind, it is still instructive to
extrapolate the present result down to physical quark-chemical potentials
of several hundred MeV. As explained in more detail in Ref. \cite{Rischke:2003mt}
one finds $\phi_{0}^{\mathrm{2SC}}\simeq10$ MeV. 

As already mentioned before, the solution of the gap equation in all
phases considered here \cite{Schmitt:2002sc,Schmitt:2004et} can be
written in a general form. For all considered phases it is found that
differences arises only in the prefactor in front of the exponential,
i.e. are only of subleading order \cite{Schmitt:2002sc,Schmitt:2004et}:\begin{equation}
\phi_{0}=2\tilde{b}b'_{0}\mu e^{-d}e^{-\varsigma}\exp\left(-\frac{\pi}{2\bar{g}}\right),\label{phi0}\end{equation}
where $d=0$ and $\varsigma=\ln(\langle\lambda_{1}\rangle^{a_{1}}\langle\lambda_{2}\rangle^{a_{2}})^{1/2}$
for all spin-zero phases. Here the coefficients $a_{1},a_{2}$ are
positive constants obeying $a_{1}+a_{2}=1$. In spin-one phases $d\neq0$
and $\varsigma$ are given by \begin{eqnarray*}
\varsigma & = & \frac{1}{2}\frac{\left\langle n_{1}\lambda_{q,1}\ln\lambda_{q,1}+n_{2}\lambda_{q,2}\ln\lambda_{q,2}\right\rangle _{\mathbf{\widehat{q}}}}{\left\langle n_{1}\lambda_{q,1}+n_{2}\lambda_{q,2}\right\rangle _{\mathbf{\widehat{q}}}}\end{eqnarray*}
where $n_{r}$ is the degeneracy factor for the excitation branch.
Generally the eigenvalues $\lambda_{r}$ depend on the direction of
the quark momentum, so the angular average $\langle\rangle$ is taken
for all directions $\mathbf{\widehat{q}}$. For all spin-one phases
considered in this paper, $d$ is about $5$, it means $e^{-d}\sim10^{-2}-10^{-3}$,
which strongly reduce spin-one gaps relative to spin-zero-gaps. Tab.
(\ref{tablephi}) give an overview of the gaps of the different phases
in units of the 2SC gap, $\phi_{0}^{\mathrm{2SC}}$. 

\begin{table}
\caption{\label{tablephi}The value of the gap function at the Fermi surface,
$\phi_{0}$ in units of its value in the 2SC phase, and the critical
temperature, in units of its value expected from BCS theory, and in
units of the critical temperature in the 2SC phase, $T_{c}^{\mathrm{2SC}}$. }

\vspace{0.5cm}

\begin{tabular}{|l|c|c|c|c|}
\hline 
phase~  & ~$d$~  & $\phi_{0}/\phi_{0}^{{\rm 2SC}}$  & $T_{c}/T_{c}^{{\rm BCS}}$  & $T_{c}/T_{c}^{{\rm 2SC}}$ \tabularnewline
\hline 
2SC  & 0  & 1  & 1  & 1 \tabularnewline
\hline 
CFL  & 0  & $2^{-1/3}$  & $2^{1/3}$  & 1 \tabularnewline
\hline 
CSL  & 5  & $2^{-1/6}\, e^{-d}$  & $2^{1/6}$  & $e^{-d}$ \tabularnewline
\hline 
polar  & 5  & $e^{-d}$  & $1$  & $e^{-d}$ \tabularnewline
\hline 
A  & 21/4  & $e^{-d}$  & $1$  & $e^{-d}$ \tabularnewline
\hline
\end{tabular}
\end{table}

Tab. (\ref{tablephi}) also lists the critical temperatures $T_{c}$
for all considered phases \cite{Schmitt:2002sc,Schmitt:2004et,Wang:2001aq,Pisarski:1999tv}.
Generally the ratio of the critical temperatures to the zero temperature
gaps on the Fermi surface is given by \begin{eqnarray}
\frac{T_{c}}{\phi_{0}} & = & \frac{e^{\gamma}}{\pi}e^{-\varsigma}\simeq0.57e^{-\varsigma}\label{Tc}\end{eqnarray}
 where $\gamma\simeq0.577$ is the Euler-Mascheroni constant. An important
feature is that in such single-gapped phases as the 2SC, polar or
A phase, the ratio is the same as in BCS theory (the so-called BCS
ratio $T_{c}/\phi_{0}=0.57$) \cite{Pisarski:1999tv}. For phases
with more than two different excitations, such as the CFL or CSL phase,
one has $e^{-\varsigma}\neq1$, which violats the BCS ratio \cite{Schafer:2000tw,Schmitt:2002sc},
cf. the third column in Tab. (\ref{tablephi}). Another situation
where the BCS ratio is violated is that the gap is anisotropic in
momentum direction. Expressing the respective critical temperatures
$T_{c}$ in terms of $T_{c}^{\mathrm{2SC}}$ the factor $e^{-\varsigma}$
in Eq. (\ref{phi0}) cancels the factor $e^{-\varsigma}$ in Eq. (\ref{Tc})
yielding $T_{c}/T_{c}^{\mathrm{2SC}}=e^{-d}$, which is listed in
the forth column of Tab. (\ref{tablephi}). The transition temperature
of a superconducting phase is normally regarded as the temperature
where the gap disappears in the gap equation. This is justified in
the mean field approximation. If one goes beyond the mean field approximation
by including the fluctuation of the diquark fields, even above the
transition temperature the 'gap' (the so-called pseudogap) is not
vanishing \cite{Kitazawa:2003cs,Kitazawa:2005vr}. The transition
temperature can also be modified by the gluon fluctuation which is
beyond the mean field approximation \cite{Giannakis:2004xt}. 

We know that in the gap equation there are gluon propagators. If one
changes the gauge parameter of the gluon propagator, the question
arises if the gap would be invariant. The answer is yes under the
following conditions \cite{Gerhold:2003js}: (a) the gap equation
is put on the mass-shell, and (b) the poles of the gluon propagator
do not overlap with those of the quark propagator. However in the
mean field approximation the gap depends on the gauge parameter at
the sub-subleading level if one imposes the on-shell condition \cite{Hou:2004bn,Xu:2008xj}. 

In intermediate coupling regime, the superconducting phases can be
investigated within a Dyson-Schwinger approach for the quark propagator
in QCD \cite{Nickel:2006vf,Marhauser:2006hy}. It was found that at
moderate chemical potentials the quasiparticle pairing gaps are several
times larger than the extrapolated weak-coupling results.

\section{Debye and Meissner masses}

In this section we will address another two important energy scales
in superconductivity: Debye and Meissner masses. The Debye mass for
gluons/ photons characterizes the screening of the chromoelectric/electric
charge in the medium, while the Meissner mass leads to expelling the
chromomagnetic/magnetic field out of the type-I color/normal superconductor.
We consider the weak coupling limit where the gap is parametrically
smaller than the Debye or Meissner mass, so the coherence length (inverse
of the gap) is much larger than the penetration length (inverse of
the Meissner mass), which means in the weak coupling limit, all color
superconductors are of type-I. 

The Debye and Meissner masses for gluons or photons can be calculated
through gluon or photon polarization tensors and taking the static
limit ($p_{0}=0$, $\mathbf{p}\rightarrow0$). The polarization tensors
depend on the diquark condensate through the quark propagators in
the quark loop. The self-energies with full momentum dependence and
the spectral densities of longitudinal and transverse gluons at zero
temperature in color-superconducting quark matter was studied in detail
in \cite{Rischke:2001py,Malekzadeh:2006ud,Malekzadeh:2008im}. In
some color superconducting phases the $U(1)$ charge which can propagate
in the medium is the rotated or new photon. It can be found by symmetry
analysis shown in previous section or by diagonalizing polarization
tensors for both the photon and gluons. We can treat the photon in
the same footing as eight gluons by enlarging gluon's color indices
to denote the photon by $a=0$. Diagonalizing the polarization tenors
with respect to new 'color' indices $a=0,1-8$, we can find Deybe
and Meissner masses of the rotated photon and gluon which are eigenvalues
of the resulting mass matrices. The detail calculation is carried
out in Ref. \cite{Schmitt:2003aa}. The results are listed in Tab.
(\ref{tab:debye-mass}-\ref{tab:mixing}). 

\begin{table}
\caption{\label{tab:debye-mass}Zero-temperature Debye masses. All masses are
given in units of $N_{f}\mu^{2}/(6\pi^{2})$, where $N_{f}=2$ in
the 2SC phase, $N_{f}=3$ in the CFL phase, and $N_{f}=1$ in the
polar and CSL phases. We use the abbreviations $\zeta\equiv(21-8\ln2)/54$,
$\alpha\equiv(3+4\ln2)/27$, and $\beta\equiv(6-4\ln2)/9$.}

\vspace{0.5cm}

{\small

\begin{tabular}{|c|c|c|c|c|c|c|c|c|c|c|c|}
\hline 
 & \multicolumn{8}{c|}{$m_{D,aa}^{2}$} & \multicolumn{2}{c|}{$m_{D,a\gamma}^{2}=m_{D,\gamma a}^{2}$} & $m_{D,\gamma\gamma}^{2}$\tabularnewline
\hline 
a & 1 & 2 & 3 & 4 & 5 & 6 & 7 & 8 & 1-7 & 8 & 0\tabularnewline
\hline 
2SC & \multicolumn{3}{c|}{0} & \multicolumn{4}{c|}{$3g^{2}/2$} & $3g^{2}$ & 0 & 0 & $2e^{2}$\tabularnewline
\hline 
CFL & \multicolumn{8}{c|}{$3\zeta g^{2}$} & 0 & $-2\sqrt{3}\zeta eg$ & $4\zeta e^{2}$\tabularnewline
\hline 
polar & \multicolumn{3}{c|}{0} & \multicolumn{4}{c|}{$3g^{2}/2$} & $3g^{2}$ & 0 & 0 & $18q^{2}e^{2}$\tabularnewline
\hline 
CSL & $3\beta g^{2}$ & $3\alpha g^{2}$ & $3\beta g^{2}$ & $3\beta g^{2}$ & $3\alpha g^{2}$ & $3\beta g^{2}$ & $3\alpha g^{2}$ & $3\beta g^{2}$ & 0 & 0 & $18q^{2}e^{2}$\tabularnewline
\hline
\end{tabular}

}
\end{table}

\begin{table}
\caption{\label{tab:meissner-mass}Zero-temperature Meissner masses. All results
are given in the same units as the Debye masses in Table \ref{tab:debye-mass}.
The abbreviations of Table \ref{tab:debye-mass} are used.}

\vspace{0.5cm}

{\small

\begin{tabular}{|c|c|c|c|c|c|c|c|c|c|c|c|}
\hline 
 & \multicolumn{8}{c|}{$m_{M,aa}^{2}$} & \multicolumn{2}{c|}{$m_{M,a\gamma}^{2}=m_{M,\gamma a}^{2}$} & $m_{M,\gamma\gamma}^{2}$\tabularnewline
\hline 
a & 1 & 2 & 3 & 4 & 5 & 6 & 7 & 8 & 1-7 & 8 & 0\tabularnewline
\hline 
2SC & \multicolumn{3}{c|}{0} & \multicolumn{4}{c|}{$g^{2}/2$} & $g^{2}/3$ & 0 & $eg/(3\sqrt{3})$ & $e^{2}/9$\tabularnewline
\hline 
CFL & \multicolumn{8}{c|}{$\zeta g^{2}$} & 0 & $-2\zeta eg/\sqrt{3}$ & $4\zeta e^{2}/3$\tabularnewline
\hline 
polar & \multicolumn{3}{c|}{0} & \multicolumn{4}{c|}{$g^{2}/2$} & $g^{2}/3$ & 0 & $2qeg/\sqrt{3}$ & $4q^{2}e^{2}$\tabularnewline
\hline 
CSL & $\beta g^{2}$ & $\alpha g^{2}$ & $\beta g^{2}$ & $\beta g^{2}$ & $\alpha g^{2}$ & $\beta g^{2}$ & $\alpha g^{2}$ & $\beta g^{2}$ & 0 & 0 & $3q^{2}e^{2}$\tabularnewline
\hline
\end{tabular}

}
\end{table}

We distinguish between the normal-conducting and the superconducting
phase. In the normal-conducting phase, i.e. for temperatures larger
than the critical temperature for the superconducting phase transition,
$T\geq T_{c}$, Meissner mass is vanishing , i.e., as expected, there
is no Meissner effect in the normal-conducting state. However, there
is electric screening for temperatures larger than $T_{c}$. Here,
the Debye mass solely depends on the number of quark flavors and their
electric charge. We find for $T\geq T_{c}$ (with $a=1,\cdots,8$)\begin{eqnarray*}
m_{D,aa}^{2} & = & 3N_{f}\frac{g^{2}\mu^{2}}{6\pi^{2}}\\
m_{D,a\gamma}^{2} & = & 0\\
m_{D,\gamma\gamma}^{2} & = & 18\sum_{n}q_{n}^{2}\frac{e^{2}\mu_{n}^{2}}{6\pi^{2}}.\end{eqnarray*}
where $q_{n}$ is the electric charge for the quark flavor labled
by $n$. Consequently, the $9\times9$ Debye mass matrix is already
diagonal. Electric gluons and electric photons are screened.

The masses in the superconducting phases are more interesting. The
results for all phases are collected in Tab. (\ref{tab:debye-mass})
(Debye masses) and Tab. (\ref{tab:meissner-mass}) (Meissner masses).
The physically relevant, or {}``rotated'', masses are obtained after
a diagonalization of the $9\times9$ mass matrices. We see from these
tables that all off-diagonal gluon masses, $a,b=1-8$, as well as
all mixed masses for $a=1-7$ vanish. Furthermore, in all cases where
the mass matrix is not diagonal, we find $m_{8\gamma}^{2}=m_{88}m_{\gamma\gamma}$.
We collect the rotated masses and mixing angles for all phases in
Tab. (\ref{tab:mixing}).

\begin{table}
\caption{\label{tab:mixing}Zero-temperature rotated Debye and Meissner masses
in units of $N_{f}\mu^{2}/(6\pi^{2})$ and mixing angles for electric
and magnetic gauge bosons. The constants $\zeta$, $\alpha$, and
$\beta$ are defined as in Tables \ref{tab:debye-mass} and \ref{tab:meissner-mass}.}

\vspace{0.5cm}

{\small

\begin{tabular}{|c|c|c|c|c|c|c|}
\hline 
 &  $\tilde{m}_{D,88}^{2}$ &  $\tilde{m}_{D,\gamma\gamma}^{2}$ &  $\cos^{2}\theta_{D}$ &  $\tilde{m}_{M,88}^{2}$ & $\tilde{m}_{M,\gamma\gamma}^{2}$ &  $\cos^{2}\theta_{M}$\tabularnewline
\hline
2SC  &  $3\, g^{2}$ &  $2\, e^{2}$ &  1  &  $\frac{1}{3}g^{2}+\frac{1}{9}e^{2}$ &  0  &  $3g^{2}/(3g^{2}+e^{2})$\tabularnewline
\hline
CFL  &  $(4e^{2}+3g^{2})\zeta$ &  0  &  $3g^{2}/(3g^{2}+4e^{2})$ &  $\left(\frac{4}{3}e^{2}+g^{2}\right)\zeta$ &  0  &  $3g^{2}/(3g^{2}+4e^{2})$\tabularnewline
\hline
polar  &  $3g^{2}$ &  $18q^{2}e^{2}$ &  1  &  $\frac{1}{3}g^{2}+4q^{2}e^{2}$ &  0  &  $g^{2}/(g^{2}+12q^{2}e^{2})$\tabularnewline
\hline
CSL  &  $3\beta g^{2}$ &  $18q^{2}e^{2}$ &  1  &  $\beta g^{2}$ &  $6q^{2}e^{2}$ &  1  \tabularnewline
\hline
\end{tabular}

}
\end{table}

Let us first discuss the spin-zero cases, 2SC and CFL. In the 2SC
phase, due to a cancellation of the normal and anomalous parts, the
Debye and Meissner masses for the gluons 1,2, and 3 vanish. Physically,
this is easy to understand. Since the condensate picks one color direction,
all quarks of the third color, say blue, remain unpaired. The first
three gluons only interact with red and green quarks and thus acquire
neither a Debye nor a Meissner mass. We recover the results of Ref.
\cite{Rischke:2000qz}. For the mixed and photon masses we find the
remarkable result that the mixing angle for the Debye masses is different
from that for the Meissner masses, $\theta_{D}\neq\theta_{M}$. The
Meissner mass matrix is not diagonal. By a rotation with the angle
$\theta_{M}$, given in Tab. (\ref{tab:mixing}), we diagonalize this
matrix and find a vanishing mass for the new photon. Consequently,
there is no electromagnetic Meissner effect in this case. The Debye
mass matrix, however, is diagonal. The off-diagonal elements $m_{D,8\gamma}^{2}$
vanish, since the contribution of the ungapped modes cancels the one
of the gapped modes. Consequently, the mixing angle is zero, $\theta_{D}=0$.
Physically, this means that not only the color-electric eighth gluon
but also the electric photon is screened. Had we considered only the
gapped quarks, we would have found the same mixing angle as for the
Meissner masses and a vanishing Debye mass for the new photon. This
mixing angle is the same as predicted from simple group-theoretical
arguments. The photon Debye mass in the superconducting 2SC phase
differs from that of the normal phase, which, for $q_{1}=2/3$ and
$q_{2}=-1/3$ is $m_{D,\gamma\gamma}^{2}=5\, N_{f}e^{2}\mu^{2}/(6\pi^{2})$.

In the CFL phase, all eight gluon Debye and Meissner masses are equal.
This reflects the symmetry of the condensate where there is no preferred
color direction. The results in Tab. (\ref{tab:debye-mass}) and (\ref{tab:meissner-mass})
show that both Debye and Meissner mass matrices have nonzero off-diagonal
elements, namely $m_{8\gamma}^{2}=m_{\gamma8}^{2}$. Diagonalization
yields a zero eigenvalue in both cases. This means that neither electric
nor magnetic (rotated) photons are screened. Or, in other words, there
is a charge with respect to which the Cooper pairs are neutral. Especially,
there is no electromagnetic Meissner effect in the CFL phase, either.
Note that the CFL phase is the only one considered in this paper in
which electric photons are not screened. Unlike the 2SC phase, both
electric and magnetic gauge fields are rotated with the same mixing
angle $\theta_{D}=\theta_{M}$. This angle is well-known \cite{Alford:1999pb,Manuel:2001mx,Litim:2001mv}. 

Let us now discuss the spin-one phases, i.e., the polar and CSL phases.
For the sake of simplicity, all results in Tab. (\ref{tab:debye-mass}),
(\ref{tab:meissner-mass}), and (\ref{tab:mixing}) refer to a single
quark system, $N_{f}=1$, where the quarks carry the electric charge
$q$. After discussing this most simple case, we will comment on the
situation where $N_{f}>1$ quark flavors separately form Cooper pairs.
The results for the gluon masses show that, up to a factor $N_{f}$,
there is no difference between the polar phase and the 2SC phase regarding
screening of color fields. This was expected since also in the polar
phase the blue quarks remain unpaired. Consequently, the gluons with
adjoint color index $a=1,2,3$ are not screened. Note that the spatial
$z$-direction picked by the spin of the Cooper pairs has no effect
on the screening masses. As in the 2SC phase, electric gluons do not
mix with the photon. There is electromagnetic Debye screening, which,
in this case, yields the same photon Debye mass as in the normal phase.
The Meissner mass matrix is diagonalized by an orthogonal transformation
defined by the mixing angle.

In the CSL phase, we find a special pattern of the gluon Debye and
Meissner masses. In both cases, there is a difference between the
gluons corresponding to the symmetric Gell-Mann matrices with $a=1,3,4,6,8$
and the ones corresponding to the antisymmetric matrices, $a=2,5,7$.
The reason for this is, of course, the residual symmetry group $SO(3)_{c+J}$
that describes joint rotations in color and real space and which is
generated by a combination of the generators of the spin group $SO(3)_{J}$
and the antisymmetric Gell-Mann matrices, $T_{2}$, $T_{5}$, and
$T_{7}$. The remarkable property of the CSL phase is that both Debye
and Meissner mass matrices are diagonal. In the case of the Debye
masses, the mixed entries of the matrix, $m_{D,a\gamma}$, are zero,
indicating that pure symmetry reasons are responsible for this fact
(remember that, in the 2SC phase, the reason for the same fact was
a cancellation of the terms originating from the gapped and ungapped
excitation branches). There is a nonzero photon Debye mass which is
identical to that of the polar and the normal phase which shows that
electric photons are screened in the CSL phase. Moreover, and only
in this phase, also magnetic photons are screened. This means that
there is an electromagnetic Meissner effect. Consequently, there is
no charge, neither electric charge, nor color charge, nor any combination
of these charges, with respect to which the Cooper pairs are neutral. 

Finally, let us discuss the more complicated situation of a many-flavor
system, $N_{f}>1$, which is in a superconducting state with spin-one
Cooper pairs. In both polar and CSL phases, this extension of the
system modifies the results in Tab. (\ref{tab:debye-mass}), (\ref{tab:meissner-mass}),
and (\ref{tab:mixing}). We have to include several different electric
quark charges, $q_{1},\ldots,q_{N_{f}}$, and chemical potentials,
$\mu_{1},\ldots,\mu_{N_{f}}$. In the CSL phase, these modifications
will change the numerical values of all masses, but the qualitative
conclusions, namely that there is no mixing and electric as well as
magnetic screening, remain unchanged. In the case of the polar phase,
however, a many-flavor system might change the conclusions concerning
the Meissner masses. While in the one-flavor case, diagonalization
of the Meissner mass matrix leads to a vanishing photon Meissner mass,
this is no longer true in the general case with arbitrary $N_{f}$.
There is only a zero eigenvalue if the determinant of the matrix vanishes,
i.e., if $m_{M,8\gamma}^{2}=m_{M,88}\, m_{M,\gamma\gamma}$. Generalizing
the results from Tab. (\ref{tab:meissner-mass}), this condition can
be written as \begin{eqnarray}
\sum_{m,n}q_{n}(q_{n}-q_{m})\mu_{n}^{2}\mu_{m}^{2} & = & 0.\label{eq:meissnersurface}\end{eqnarray}
Consequently, in general and for fixed charges $q_{n}$, there is
a hypersurface in the $N_{f}$-dimensional space spanned by the quark
chemical potentials on which there is a vanishing eigenvalue of the
Meissner mass matrix and thus no electromagnetic Meissner effect.
All remaining points in this space correspond to a situation where
the (new) photon Meissner mass is nonzero (although, of course, there
might be a mixing of the eighth gluon and the photon). Eq. (\ref{eq:meissnersurface})
is trivially fulfilled when the electric charges of all quarks are
equal. Then we have no electromagnetic Meissner effect in the polar
phase which is plausible since, regarding electromagnetism, this situation
is similar to the one-flavor case. For specific values of the electric
charges we find very simple conditions for the chemical potentials.
In a two flavor system with $q_{1}=2/3$, $q_{2}=-1/3$, Eq. (\ref{eq:meissnersurface})
reads $\mu_{1}^{2}\mu_{2}^{2}=0$. In a three-flavor system with $q_{1}=-1/3$,
$q_{2}=-1/3$, and $q_{3}=2/3$, we have $(\mu_{1}^{2}+\mu_{2}^{2})\mu_{3}^{2}=0$.
Consequently, these systems \textit{always}, i.e., for all combinations
of the chemical potentials $\mu_{n}$, exhibit the electromagnetic
Meissner effect in the polar phase except when they reduce to the
above discussed simpler cases (same electric charge of all quarks
or a one-flavor system). 

In the weak coupling limit, the gap values are parametrically smaller
than the Meissner masses, therefore the coherence length (inverse
of the gap) is much larger than the penetration length (inverse of
the Meissner mass). So the spin-1 color superconductor is a type-I
superconductor in the weak coupling limit \cite{Schmitt:2003xq}.
In the intermediate densities, this statement might not be valid.
A natural way of studying superconductivity in external magnetic field
is to make use of the Ginzburg-Landau effective theory, see e.g. Ref.
\cite{Iida:2000ha,Iida:2001pg,Iida:2002ev,Giannakis:2001wz,Giannakis:2003am,Blaschke:1999fy,Sedrakian:2000kw}.
An introduction to the Ginzburg-Landau theory can be found in Sec.
\ref{sec:GL-theory}.

\section{General effective theory for color superconductivity}

As we mentioned before, quark matter at small temperature $T$ and
large quark chemical potential $\mu$ is a color superconductor. In
a color superconductor, there are several energy scales: the quark
chemical potential $\mu$, the inverse gluon screening length $m_{g}\sim g\mu$,
and the superconducting gap parameter $\phi$. In weak coupling these
three scales are naturally ordered, $\phi\ll g\mu\ll\mu$. The ordering
of scales implies that the modes near the Fermi surface, which participate
in the formation of Cooper pairs, can be considered to be independent
of the modes deep within the Fermi sea. This suggests that the most
efficient way to compute properties such as the color-superconducting
gap parameter is via an \emph{effective theory for quark modes near
the Fermi surface}. Such an effective theory has been originally proposed
by Hong \cite{Hong:1998tn} and was subsequently refined by others
\cite{Hong:1998tn,Schafer:2003jn,Schafer:2004yx,Nardulli:2002ma}.

In order to construct such an effective theory, we pursue a different
venue and introduce cut-offs in momentum space for quarks, $\Lambda_{q}$,
and gluons, $\Lambda_{g}$. These cut-offs separate relevant from
irrelevant quark modes and soft from hard gluon modes. We then explicitly
integrate out irrelevant quark and hard gluon modes and derive a general
effective action for hot and/or dense quark-gluon matter \cite{Reuter:2004kk,Reuter:2004ze}.

We show that the standard HTL and HDL effective actions are contained
in our general effective action for a certain choice of the quark
and gluon cut-offs $\Lambda_{q},\,\Lambda_{g}$. We also show that
the action of the high-density effective theory derived by Hong and
others is a special case of our general effective action. In this
case, relevant quark modes are located within a layer of width $2\Lambda_{q}$
around the Fermi surface.

The two cut-offs, $\Lambda_{q}$ and $\Lambda_{g}$, introduced in
our approach are in principle different, $\Lambda_{q}\neq\Lambda_{g}$.
We show that in order to produce the correct result for the color-superconducting
gap parameter to subleading order in weak coupling, we have to demand
$\Lambda_{q}\leq g\mu\ll\Lambda_{g}\leq\mu$, so that $\Lambda_{q}/\Lambda_{g}\sim g\ll1$.
Only in this case, the dominant contribution to the QCD gap equation
arises from almost static magnetic gluon exchange, while subleading
contributions are due to electric and non-static magnetic gluon exchange.

The color-superconducting gap parameter is computed from a Dyson-Schwinger
equation for the quark propagator. In general, this equation corresponds
to a self-consistent resummation of all one-particle irreducible (1PI)
diagrams for the quark self-energy. A particularly convenient way
to derive Dyson-Schwinger equations is via the Cornwall-Jackiw-Tomboulis
(CJT) formalism. In this formalism, one constructs the set of all
two-particle irreducible (2PI) vacuum diagrams from the vertices of
a given tree-level action. The functional derivative of this set with
respect to the full propagator then defines the 1PI self-energy entering
the Dyson-Schwinger equation. Since it is technically not feasible
to include all possible diagrams, and thus to solve the Dyson-Schwinger
equation exactly, one has to resort to a many-body approximation scheme,
which takes into account only particular classes of diagrams. The
advantage of the CJT formalism is that such an approximation scheme
is simply defined by a truncation of the set of 2PI diagrams. However,
in principle there is no parameter which controls the accuracy of
this truncation procedure.

The standard QCD gap equation in mean-field approximation studied
in Refs. \cite{Pisarski:1999bf,Hong:1999fh,Brown:2000eh,Wang:2001aq,Schafer:1998na,Pisarski:1999av}
follows from this approach by including just the sunset-type diagram
which is constructed from two quark-gluon vertices of the QCD tree-level
action. We also employ the CJT formalism to derive the gap equation
for the color-superconducting gap parameter. However, we construct
all diagrams of sunset topology from the vertices of the general \emph{effective}
action derived in this work. The resulting gap equation is equivalent
to the gap equation in QCD, and the result for the gap parameter to
subleading order in weak coupling is identical to that in QCD, provided
$\Lambda_{q}\leq g\mu\ll\Lambda_{g}\leq\mu$. The advantage of using
the effective theory is that the appearance of the two scales $\Lambda_{q}$
and $\Lambda_{g}$ considerably facilitates the power counting of
various contributions to the gap equation as compared to full QCD.
We explicitly demonstrate this in the course of the calculation and
suggest that, within this approach, it should be possible to identify
the terms which contribute beyond subleading order to the gap equation.
Of course, for a complete sub-subleading order result one cannot restrict
oneself to the sunset diagram, but would have to investigate other
2PI diagrams as well. This shows that an \emph{a priori} estimate
of the relevance of different contributions on the level of the effective
action does not appear to be feasible for quantities which have to
be computed self-consistently.

\section{Ginzburg-Landau theory for CSC}

\label{sec:GL-theory}In this section we will introduce the Ginzburg-Landau
theory for CSC. The starting point is the thermodynamic potential
$\Omega$, which can be expanded in the second and fourth power of
the diquark condensate in the vicinity of critical temperature. We
have \begin{equation}
\Omega=-\frac{T}{2V}\log\det\mathcal{G}^{-1}=-\frac{T}{2V}\mathrm{Tr}\log\mathcal{G}^{-1},\end{equation}
where $\mathcal{G}^{-1}$ is the inverse of the full propagator and
the trace runs over Nambu-Gorkov, momentum, Dirac, color and flavor
space. We can expand $\Omega$ to the fourth power of $\Sigma$, \begin{eqnarray}
\Omega & = & -\frac{T}{2V}\mathrm{Tr}\log(\mathcal{G}_{0}^{-1}+\Sigma)\nonumber \\
 & = & -\frac{T}{2V}\mathrm{Tr}\log[\mathcal{G}_{0}^{-1}(1+\mathcal{G}_{0}\Sigma)]\nonumber \\
 & = & -\frac{T}{2V}\mathrm{Tr}\log\mathcal{G}_{0}^{-1}-\frac{T}{2V}\mathrm{Tr}\log(1+\mathcal{G}_{0}\Sigma).\label{eq:gl-potential01}\end{eqnarray}
Here the free part $\mathcal{G}_{0}^{-1}$ is given by \begin{eqnarray}
\mathcal{G}_{0}^{-1} & = & \left(\begin{array}{cc}
[G_{0}^{+}]^{-1} & 0\\
0 & [G_{0}^{-}]^{-1}\end{array}\right)=\left(\begin{array}{cc}
\gamma_{\mu}K^{\mu}+\mu\gamma_{0}-m & 0\\
0 & \gamma_{\mu}K^{\mu}-\mu\gamma_{0}-m\end{array}\right)\nonumber \\
 & = & \left(\begin{array}{cc}
(k_{0}+\mu-eE_{k})\gamma_{0}\Lambda_{k}^{e} & 0\\
0 & (k_{0}-\mu+eE_{k})\gamma_{0}\Lambda_{k}^{-e}\end{array}\right)\nonumber \\
 & \approx & \left(\begin{array}{cc}
(k_{0}+\mu-E_{k})\gamma_{0}\Lambda_{k}^{+} & 0\\
0 & (k_{0}-\mu+E_{k})\gamma_{0}\Lambda_{k}^{-}\end{array}\right),\end{eqnarray}
where we only keep the positive energy excitation in the last line.
Here the energy projectors are defined by $\Lambda_{k}^{e}=\frac{1}{2}\left(1+e\frac{\gamma_{0}(\vec{\gamma}\cdot\mathbf{k}+m)}{E_{k}}\right)$
with $E_{k}=\sqrt{k^{2}+m^{2}}$. The free propagator is then \begin{equation}
\mathcal{G}_{0}=\left(\begin{array}{cc}
G_{0}^{+} & 0\\
0 & G_{0}^{-}\end{array}\right)\approx\left(\begin{array}{cc}
(k_{0}+\mu-E_{k})^{-1}\Lambda_{k}^{+}\gamma_{0} & 0\\
0 & (k_{0}-\mu+E_{k})^{-1}\Lambda_{k}^{-}\gamma_{0}\end{array}\right).\end{equation}
The condensate part is \begin{equation}
\Sigma\equiv\left(\begin{array}{cc}
0 & \Phi^{-}\\
\Phi^{+} & 0\end{array}\right),\end{equation}
where $\Phi^{-}\equiv\gamma_{0}[\Phi^{+}]^{\dagger}\gamma_{0}$. To
simplify the last term in Eq. (\ref{eq:gl-potential01}), we use  \begin{equation}
\log(1+\mathcal{G}_{0}\Sigma)=\sum_{n}\frac{(-1)^{n+1}}{n}(\mathcal{G}_{0}\Sigma)^{n},\end{equation}
Eq. (\ref{eq:gl-potential01}) can be expanded as, \begin{equation}
\Omega-\Omega_{n}\approx\frac{T}{V}\left[\frac{1}{4}\mathrm{Tr}(\mathcal{G}_{0}\Sigma)^{2}+\frac{1}{8}\mathrm{Tr}(\mathcal{G}_{0}\Sigma)^{4}\right],\end{equation}
where the odd terms are vanishing due to zero trace in Nambu-Gorkov
space. 

First we evaluate the quartic term of the condensate. Since we consider
the small momentum expansion, we assume that $\Sigma$ is constant
in the quartic term. Then we have \begin{eqnarray}
\frac{1}{8}\frac{T}{V}\mathrm{Tr}(\mathcal{G}_{0}\Sigma)^{4} & = & \frac{1}{4}T\sum_{n}\int\frac{d^{3}k}{(2\pi)^{3}}\frac{\mathrm{Tr}[(\Lambda_{k}^{+}\gamma_{0}\Phi^{-}\Lambda_{k}^{-}\gamma_{0}\Phi^{+})^{2}]}{[k_{0}^{2}-(E_{k}-\mu)^{2}]^{2}}\nonumber \\
 & = & \frac{1}{4}T\sum_{n}\int\frac{dk}{2\pi^{2}}\frac{k^{2}}{[k_{0}^{2}-(E_{k}-\mu)^{2}]^{2}}\int\frac{d\Omega}{4\pi}\mathrm{Tr}[(\Lambda_{k}^{+}\gamma_{0}\Phi^{-}\Lambda_{k}^{-}\gamma_{0}\Phi^{+})^{2}]\nonumber \\
 & = & \int\frac{dk}{2\pi^{2}}\frac{k^{2}}{(E_{k}-\mu)^{3}}\left\{ \tanh\frac{\beta(E_{k}-\mu)}{2}-\frac{\beta(E_{k}-\mu)/2}{\cosh[\beta(E_{k}-\mu)/2]}\right\} \nonumber \\
 &  & \times\left\langle \mathrm{Tr}[(\Lambda_{k}^{+}\gamma_{0}\Phi^{-}\Lambda_{k}^{-}\gamma_{0}\Phi^{+})^{2}]\right\rangle _{\hat{\mathbf{k}}}\label{eq:gl-quatic}\end{eqnarray}
where we have used $k_{0}=i(2n+1)\pi T$ and \[
\sum_{n}\frac{1}{[(2n+1)^{2}\pi^{2}+x^{2}]^{2}}=\frac{1}{4x^{3}}\left[\tanh\frac{x}{2}-\frac{x/2}{\cosh(x/2)}\right].\]
About the momentum integral in Eq. (\ref{eq:gl-quatic}), we see that
the integrand except $k^{2}$ is centered at $\mu$, if we limit the
momentum range around $\mu$ we can replace $k^{2}$ with $\mu^{2}$
and pull it out of the integral. The first term inside the curly bracket
dominates the momentum integral, then we have \begin{eqnarray}
\frac{1}{8}\frac{T}{V}\mathrm{Tr}(\mathcal{G}_{0}\Sigma)^{4} & = & \frac{\mu^{2}}{2\pi^{2}T^{2}}\int dx\frac{\tanh(x/2)}{x^{3}}\left\langle \mathrm{Tr}[(\Lambda_{k}^{+}\gamma_{0}\Phi^{-}\Lambda_{k}^{-}\gamma_{0}\Phi^{+})^{2}]\right\rangle _{\hat{\mathbf{k}}}\nonumber \\
 & \approx & \frac{\mu^{2}}{2\pi^{2}T^{2}}\left\langle \mathrm{Tr}[(\Lambda_{k}^{+}\gamma_{0}\Phi^{-}\Lambda_{k}^{-}\gamma_{0}\Phi^{+})^{2}]\right\rangle _{\hat{\mathbf{k}}}.\label{eq:gl-quatic2}\end{eqnarray}
The dependence on color superconducting phases contains in the average
over momentum direction. 

Now let's evaluate the quadratic term, \begin{eqnarray}
\frac{1}{4}\frac{T}{V}\mathrm{Tr}(\mathcal{G}_{0}\Sigma)^{2} & = & \frac{1}{2}\frac{T}{V}\mathrm{Tr}(G_{0}^{+}\Phi^{-}G_{0}^{-}\Phi^{+})\nonumber \\
 & = & \frac{1}{2}TV\sum_{n}\int\frac{d^{3}k}{(2\pi)^{3}}\int\frac{d^{3}q}{(2\pi)^{3}}\mathrm{Tr}[\Lambda_{k-q/2}^{+}\gamma_{0}\Phi_{q}^{-}\Lambda_{k+q/2}^{-}\gamma_{0}\Phi_{-q}^{+}]\nonumber \\
 &  & \times\frac{1}{(k_{0}+\mu-E_{k-q/2})(k_{0}-\mu+E_{k+q/2})},\label{eq:quadratic-gl}\end{eqnarray}
where we have kept the momentum dependence of condensate. To the leading
order the trace term invloves an angular integral over $\hat{\mathbf{k}}$.
We can make Taylor expansion with respect to $\mathbf{q}$ for the
last line of Eq. (\ref{eq:quadratic-gl}), \begin{eqnarray}
\frac{1}{k_{0}\pm\mu\mp E_{k\mp q/2}} & \approx & \frac{1}{k_{0}\pm\mu\mp E_{k}+\frac{\mathbf{k}\cdot\mathbf{q}}{2E_{k}}\mp\frac{\mathbf{q}^{2}}{8E_{k}}}\nonumber \\
 & \approx & \frac{1}{k_{0}\mp(E_{k}-\mu)}\left[1-\frac{\frac{\mathbf{k}\cdot\mathbf{q}}{2E_{k}}\mp\frac{\mathbf{q}^{2}}{8E_{k}}}{k_{0}\mp(E_{k}-\mu)}\right.\nonumber \\
 &  & \left.+\frac{\left(\frac{\mathbf{k}\cdot\mathbf{q}}{2E_{k}}\right)^{2}}{(k_{0}\mp(E_{k}-\mu))^{2}}\right].\end{eqnarray}
Then the last line of Eq. (\ref{eq:quadratic-gl}) becomes \begin{eqnarray}
 &  & \frac{1}{k_{0}^{2}-(E_{k}-\mu)^{2}}\left[1-\frac{\frac{\mathbf{k}\cdot\mathbf{q}}{2E_{k}}-\frac{\mathbf{q}^{2}}{8E_{k}}}{k_{0}-(E_{k}-\mu)}-\frac{\frac{\mathbf{k}\cdot\mathbf{q}}{2E_{k}}+\frac{\mathbf{q}^{2}}{8E_{k}}}{k_{0}+(E_{k}-\mu)}\right.\nonumber \\
 &  & \left.+\frac{\left(\frac{\mathbf{k}\cdot\mathbf{q}}{2E_{k}}\right)^{2}}{(k_{0}-(E_{k}-\mu))^{2}}+\frac{\left(\frac{\mathbf{k}\cdot\mathbf{q}}{2E_{k}}\right)^{2}}{(k_{0}+(E_{k}-\mu))^{2}}+\frac{\left(\frac{\mathbf{k}\cdot\mathbf{q}}{2E_{k}}\right)^{2}}{k_{0}^{2}-(E_{k}-\mu)^{2}}\right].\label{eq:quadratic-gl1}\end{eqnarray}
The second and third terms inside the square brackets are vanishing
when carrying out integral over $\mathbf{k}$ since the term with
$\frac{\mathbf{k}\cdot\mathbf{q}}{2E_{k}}$ is odd in $\mathbf{k}$
and the term with $\frac{\mathbf{q}^{2}}{8E_{k}}$ is proportional
to $(E_{k}-\mu)$. Then the quadratic term becomes,\begin{eqnarray}
\frac{1}{4}\frac{T}{V}\mathrm{Tr}(\mathcal{G}_{0}\Sigma)^{2} & = & \frac{1}{2}TV\int\frac{d^{3}q}{(2\pi)^{3}}\left\langle \mathrm{Tr}[\Lambda_{k-q/2}^{+}\gamma_{0}\Phi_{q}^{-}\Lambda_{k+q/2}^{-}\gamma_{0}\Phi_{-q}^{+}]\right\rangle _{\mathbf{\hat{k}}}\nonumber \\
 &  & \times\sum_{n}\int\frac{d^{3}k}{(2\pi)^{3}}\frac{1}{k_{0}^{2}-(E_{k}-\mu)^{2}}\left\{ 1+\frac{k^{2}q^{2}}{12E_{k}^{2}}\right.\nonumber \\
 &  & \times\left[\frac{1}{(k_{0}-(E_{k}-\mu))^{2}}+\frac{1}{(k_{0}+(E_{k}-\mu))^{2}}\right.\nonumber \\
 &  & \times\left.\left.+\frac{1}{k_{0}^{2}-(E_{k}-\mu)^{2}}\right]\right\} .\label{eq:quadratic-gl2}\end{eqnarray}
Note that the constant term inside the curly brackets gives the quadratic
term in the condensate. The term proportional to $\frac{k^{2}q^{2}}{12E_{k}^{2}}$
gives the gradient term $\sim|\nabla\Phi|^{2}$. 

By inserting the concrete forms of the diquark condensates into Eq.
(\ref{eq:gl-quatic2}) and (\ref{eq:quadratic-gl2}) one can have
the Ginzburg-Landau free energy from which one can study a variety
of problems \cite{Iida:2000ha,Iida:2001pg,Iida:2002ev,Giannakis:2001wz,Giannakis:2003am,Blaschke:1999fy,Sedrakian:2000kw}.

\section{BCS-BEC crossover in a boson-fermion model}

In weak couplings the Cooper pairing is well described within Bardeen-Cooper-Schrieffer
(BCS) theory. Cooper pairs are typically of a size much larger than
the mean interparticle distance. The picture changes for sufficiently
large interaction strengths. In this case, Cooper pairs become bound
states, and superfluidity is realized by a Bose-Einstein condensation
(BEC) of molecular bosons composed of two fermions. A crossover between
the weak-coupling BCS regime and the strong-coupling BEC regime is
expected. 

In order to describe the crossover from BCS to BEC in color superconductivity,
one can use Nambu-Jona-Lasinio model \cite{Hatsuda:1994pi,Nishida:2005ds,Abuki:2006dv,Kitazawa:2007im,Kitazawa:2007zs,Sun:2007fc,He:2007yj,Brauner:2008td,Blaschke:2008uf},
a purely fermionic model. Here we shall not consider a purely fermionic
model which may describe this crossover as a function of the fermionic
coupling strength. We rather set up a theory with bosonic and fermionic
degrees of freedom. Here, fermions and bosons are coupled through
a Yukawa interaction and required to be in chemical equilibrium, $2\mu=\mu_{b}$,
where $\mu$ and $\mu_{b}$ are the fermion and boson chemical potentials,
respectively. We treat the (renormalized) boson mass $m_{b,r}$ and
the boson-fermion coupling $g$ as free parameters. Then, tuning the
parameter $x=-(m_{b,r}^{2}-\mu_{b}^{2})/(4g^{2})$ drives the system
from the BCS to the BEC regime. The fermionic chemical potential shall
be self-consistently determined from the gap equation and charge conservation.
This picture is inspired by the boson-fermion model of superconductivity
considered in Ref. \cite{Friedberg:1989gj}, which has been used in
the context of cold fermionic atoms. It also has possible applications
for high-temperature superconductivity. For simplicity, we only discuss
the evaluation of the model in a mean-field approximation (MFA) \cite{Deng:2006ed},
for the case beyond MFA, see Ref. \cite{Deng:2008ah}. 

We use a model of fermions $\psi$ and composite bosons $\varphi$
coupled to each other by a Yukawa interaction. With mean field approximation,
the Lagrangian can be written as \begin{eqnarray}
{\cal L} & = & \frac{1}{2}\overline{\Psi}{\cal S}^{-1}\Psi+\mathcal{L}_{b}[\mu_{b}^{2}-m_{b}^{2}]|\phi|^{2}\nonumber \\
 &  & +|(\partial_{t}-i\mu_{b})\varphi|^{2}-|\nabla\varphi|^{2}-m_{b}^{2}|\varphi|^{2}\,,\\
{\cal S}^{-1}(P) & = & \left(\begin{array}{cc}
P_{\mu}\gamma^{\mu}+\mu\gamma_{0}-m & 2ig\gamma_{5}\phi^{*}\\
2ig\gamma_{5}\phi & P_{\mu}\gamma^{\mu}-\mu\gamma_{0}-m\end{array}\right)\,.\end{eqnarray}
Here charge conjugate spinors are defined by $\psi_{C}=C\overline{\psi}^{T}$
and $\overline{\psi}_{C}=\psi^{T}C$ with $C=i\gamma^{2}\gamma^{0}$,
and Nambu-Gorkov spinors are defines by$\Psi=\left(\begin{array}{c}
\psi\\
\psi_{C}\end{array}\right)\,,\:\Psi=(\overline{\psi},\overline{\psi}_{C})$, ${\cal S}^{-1}$ is the inverse fermion propagator which reads in
momentum space, the fermion (boson) mass is denoted by $m$ ($m_{b}$).
We choose the boson chemical potential to be twice the fermion chemical
potential, $\mu_{b}=2\mu.$ Therefore, the system is in chemical equilibrium
with respect to the conversion of two fermions into one boson and
vice versa. This allows us to model the transition from weakly-coupled
Cooper pairs made of two fermions into a molecular difermionic bound
state, described as a boson. The interaction term accounts for a local
interaction between fermions and bosons with coupling constant $g$.
In order to describe BEC of the bosons, we have to separate the zero
mode of the field $\varphi$. Moreover, we shall replace this zero-mode
by its expectation value $\phi\equiv\langle\varphi_{0}\rangle$ and
neglect the interaction between the fermions and the $non-zero$ boson
modes. This corresponds to the mean-field approximation. 

We can compute the partition function \begin{equation}
{\cal Z}=\int[d\Psi][d\overline{\Psi}][d\varphi][d\varphi^{*}]\exp\left[\int_{0}^{1/T}d\tau d^{3}x\,\mathcal{L}\right],\end{equation}
to obtain the thermodynamic potential density $\Omega=-T/V\,\ln{\cal Z}$,
where $T$ is the temperature and $V$ is the volume of the system.
After performing the path integral and the sum over Matsybara frequencies
one may get 

\begin{eqnarray}
\Omega & = & -\sum_{e=\pm}\int\frac{d^{3}k}{(2\pi)^{3}}\left\{ \epsilon_{k}^{e}+2T\ln\left[1+\exp\left(-\frac{\epsilon_{k}^{e}}{T}\right)\right]\right\} \nonumber \\
 &  & +\frac{(m_{b}^{2}-\mu_{b}^{2})\Delta^{2}}{4g^{2}}\nonumber \\
 &  & +\frac{1}{2}\sum_{e=\pm}\int\frac{d^{3}k}{(2\pi)^{3}}\left\{ \omega_{k}^{e}+2T\ln\left[1-\exp\left(-\frac{\omega_{k}^{e}}{T}\right)\right]\right\} \,.\label{eq:tot-the-pot}\end{eqnarray}
with quasi-particle energy for fermions ($e=+1$) and antifermions
($e=-1$) denoted by $\epsilon_{k}^{e}=\sqrt{(\epsilon_{k0}-e\mu)^{2}+\Delta^{2}}$,
with$\epsilon_{k0}=\sqrt{k^{2}+m^{2}}$, and the (anti)boson energy
denoted by $\omega_{k}^{e}=\sqrt{k^{2}+m_{b}^{2}}-e\mu_{b}$.

Then the charge conservation equation and gap equation can be derived
by $n=-\frac{\partial\Omega}{\partial\mu}$, and $0=\frac{\partial\Omega}{\partial\Delta}$,
which can be writen as $n=n_{F}+n_{0}+n_{B}$, with $n_{F},\, n_{0},\, n_{B}$present
the fermionic, condensate bosonic and thermal bosonic contribution
to the total partical density respectively, and for non-zero condensation,
the gap equation reads \begin{equation}
-x=\sum_{e=\pm}\int\frac{d^{3}k}{(2\pi)^{3}}\left(\frac{1}{2\epsilon_{k}^{e}}\tanh\frac{\epsilon_{k}^{e}}{2T}-\frac{1}{2\epsilon_{k0}}\right),\end{equation}
with crossover parameter defined by $x\equiv-\frac{m_{b,r}^{2}-\mu_{b}^{2}}{4g^{2}}$,
and \begin{equation}
m_{b,r}^{2}=4g^{2}\left.\frac{\partial\Omega}{\partial\Delta^{2}}\right|_{\Delta=\mu=T=0}=m_{b}^{2}-4g^{2}\int\frac{d^{3}k}{(2\pi)^{3}}\frac{1}{\epsilon_{k0}}.\end{equation}
This parameter $x$ can be varied from negative values with large
modulus (BCS) to large positive values (BEC). In between, $x=0$ is
the unitary limit. Solving these two coupled equations to determine
the gap $\Delta$ and the chemical potential $\mu$ as functions of
the crossover parameter $x$ and the temperature $T$ at fixed effective
Fermi momentum $p_{F}$, fermion mass $m$, and boson-fermion coupling
$g$ (in numerical calculation, we fix $\frac{p_{F}}{\Lambda}=0.3,\,\frac{m}{\Lambda}=0.2,\, g=4$,
with $\Lambda$ the cutoff in the momentum integrals). The solution
$[\Delta(x,\, T),\,\mu(x\,,T)]$ can then, in turn, be uesed to compute
the densities of fermions and bosons in the $x-T$ plane. 

\begin{figure}
\caption{\label{fig:figzeroT}Crossover at zero temperature from the BCS regime
(small $x$) to the BEC regime (large $x$). Left panel: fermion chemical
potential $\mu_{0}$ (blue dotted) and gap $\Delta_{0}$ (red dashed)
in units of effective Fermi energy $\epsilon_{F}$. Right panel: condensate
fraction (red solid), fermion fraction (blue solid).}

\includegraphics[scale=0.35]{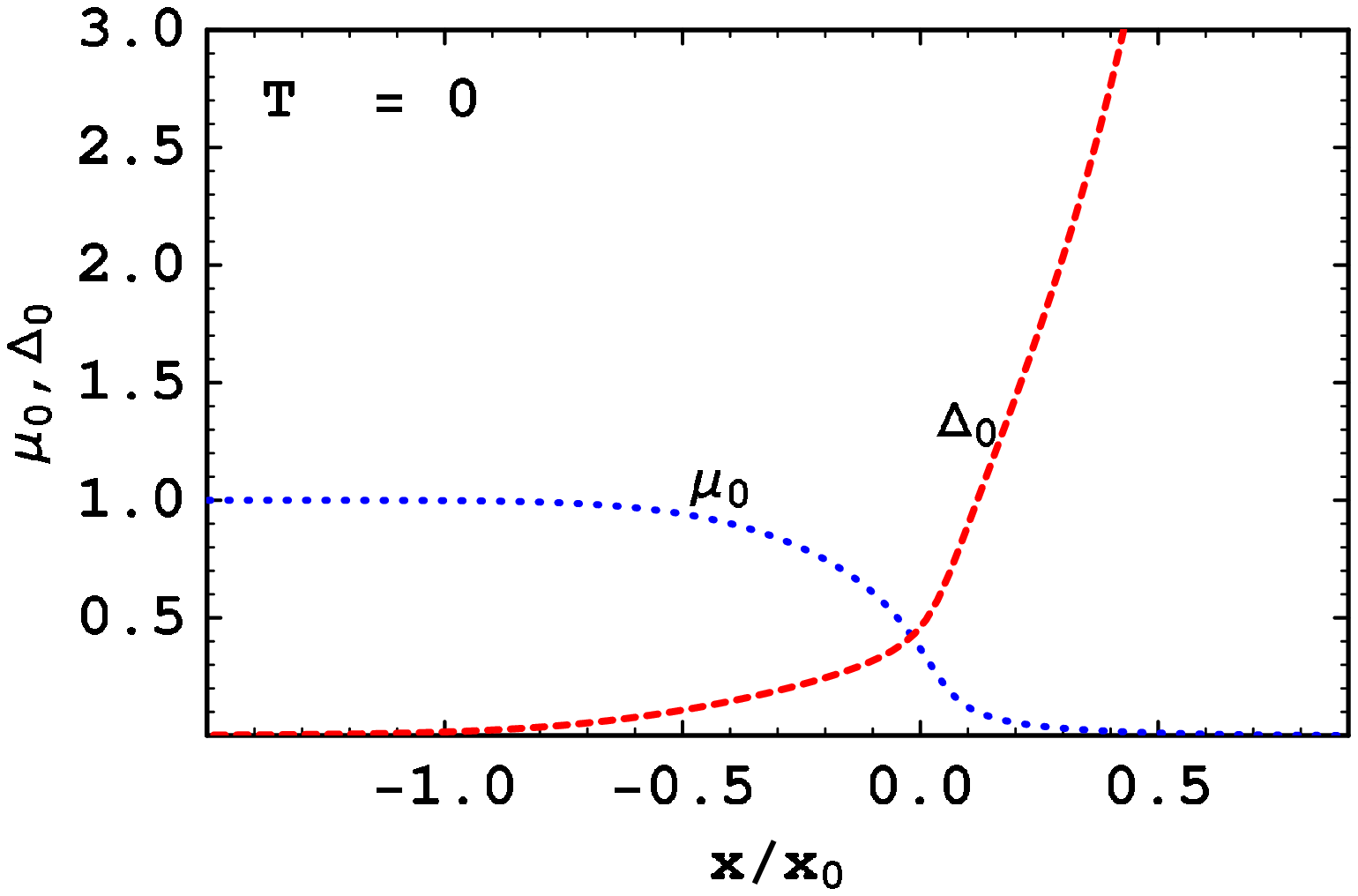}\includegraphics[scale=0.35]{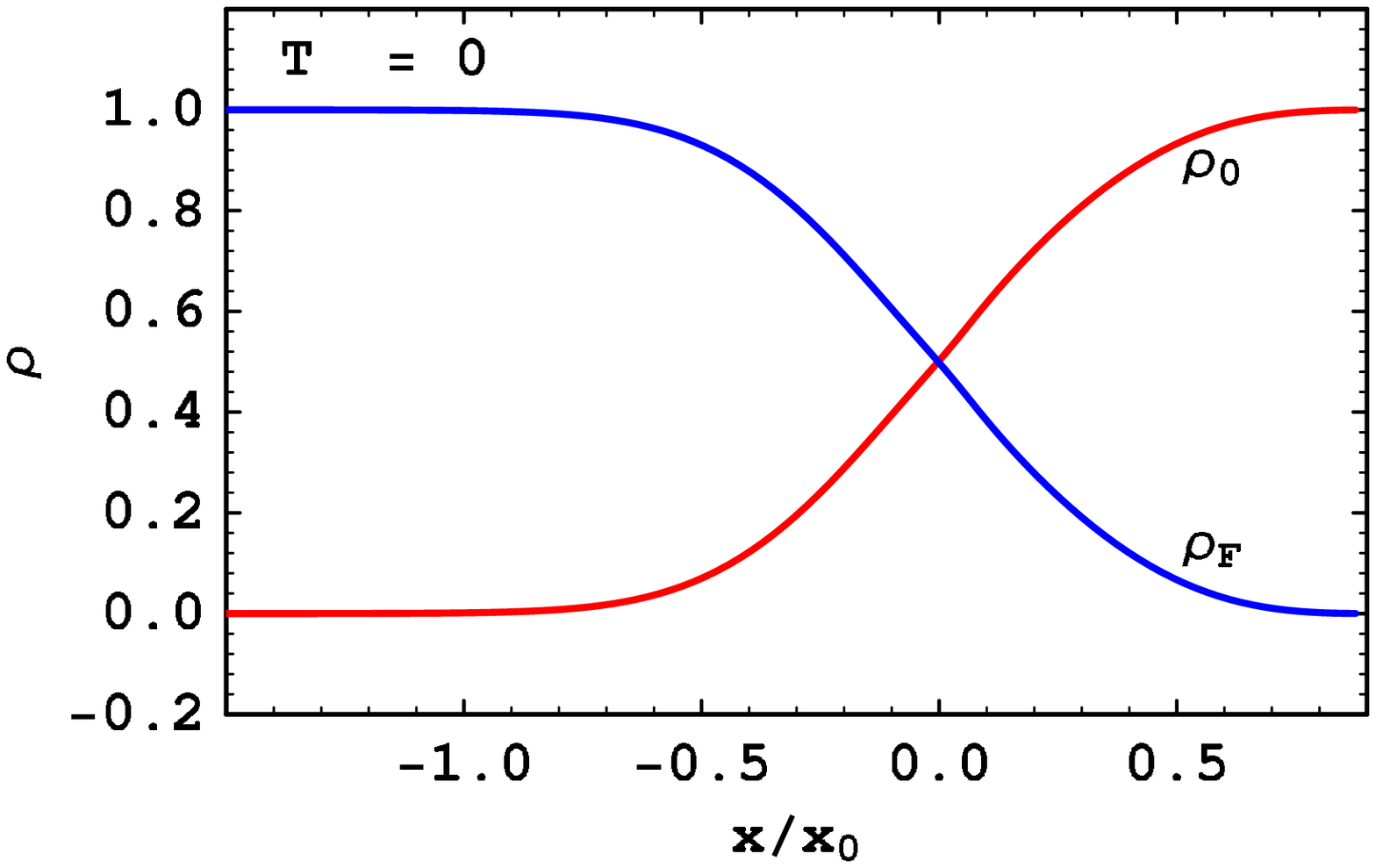}
\end{figure}

The numerical results for the solution of the coupled equations are
shown in Fig. \ref{fig:figzeroT}. The left panel shows the fermion
chemical potential $\mu_{0}$ and the gap $\Delta_{0}$ as functions
of $x$. In the weak-coupling regime (small $x$) we see that the
chemical potential is given by the Fermi energy, $\mu_{0}=\epsilon_{F}$.
For the given parameters, $\epsilon_{F}/\Lambda\simeq0.36$. The chemical
potential decreases with increasing $x$ and approaches zero in the
far BEC region. The gap is exponentially small in the weak-coupling
region, as expected from BCS theory. It becomes of the order of the
chemical potential around the unitary limit, $x=0$, and further increases
monotonically for positive $x$. In the unitary limit, we have $\mu/\epsilon_{F}\simeq0.37$,
while in nonrelativistic fermionic models $\mu/\epsilon_{F}\simeq0.4-0.5$
was obtained \cite{Chang:2004zz,Carlson:2005kg,Nishida:2006br}. 

The corresponding fermion and boson densities are shown in the right
panel of Fig. \ref{fig:figzeroT}. These two curves show the crossover:
at small $x$ all Cooper pairs are resonant states, which is characterized
by a purely fermionic density,$n=n_{F}$; at large $x$, on the other
hand, Cooper pairs are bound states and hence there are no fermions
in the system. The charge density is rather dominated by a bosonic
condensate, $n=n_{0}$. The crossover region is located around $x=0$.
We can characterize this region quantitatively as follows. We write
the boson mass as $m_{b}=2m-E_{{\rm bind}}$. Then, a bound state
appears for positive values of the binding energy $E_{{\rm bind}}$,
i.e., for $2m>m_{b}$. 

\begin{figure}
\caption{\label{fig:figanti}Fermion and antifermion excitation energies $\epsilon_{k}^{+}$
and $\epsilon_{k}^{-}$ for three different values of the crossover
parameter $x/x_{0}$ at $T=0$ as a function of the momentum $k$
(both $\epsilon_{k}^{e}$ and $k$ are given in units of $\epsilon_{F}$).
In the BCS regime (left panel) the energy gap is small and the fermion
excitations are well separated from antifermion excitations. Both
excitations approach each other in the unitary regime (middle panel),
and become indistinguishable in the far BEC regime (right panel).
Note in particular that the minimum of the antiparticle excitation
is not a monotonic function of $x$. }

\includegraphics[scale=0.33]{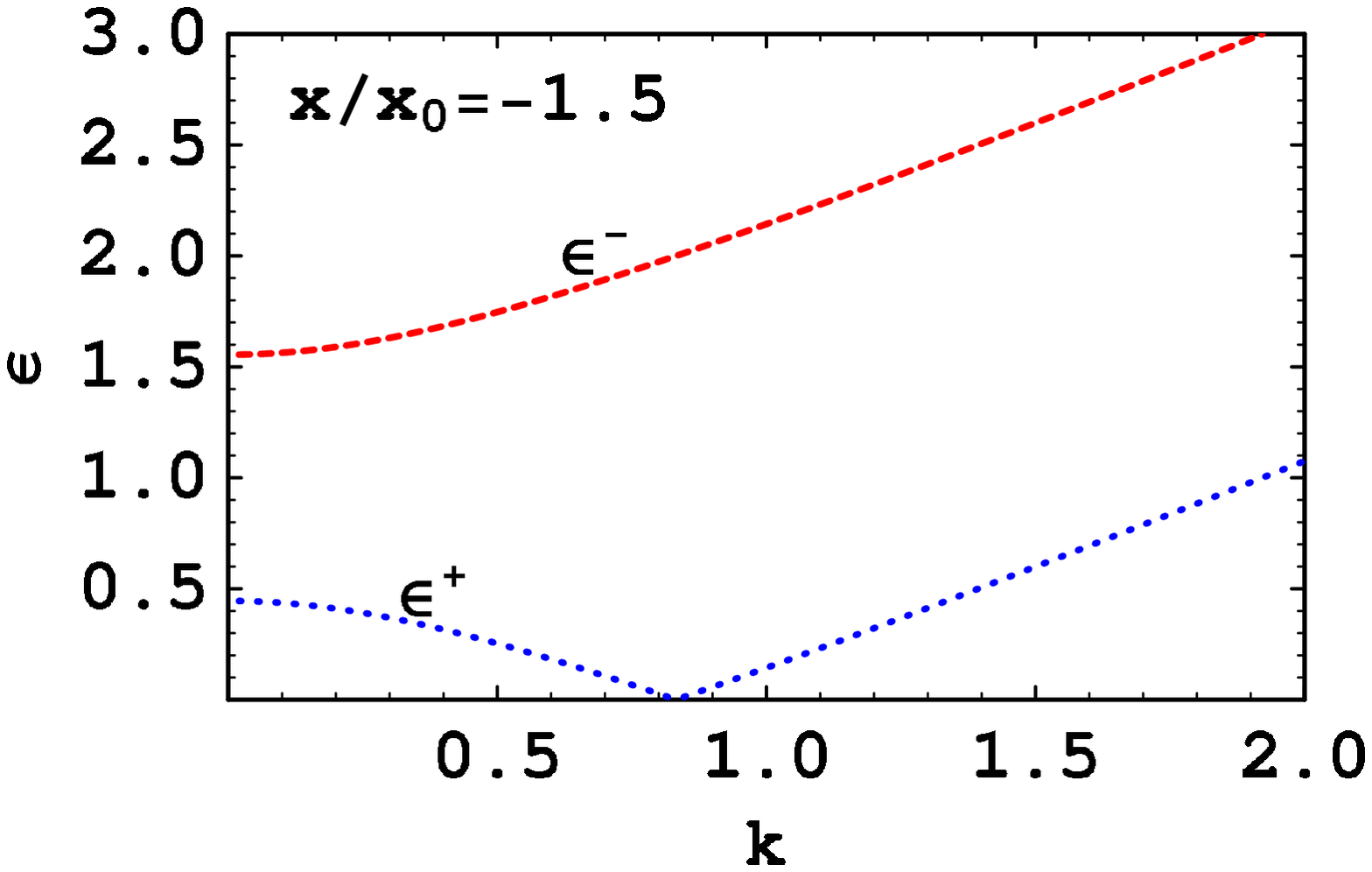}\includegraphics[scale=0.33]{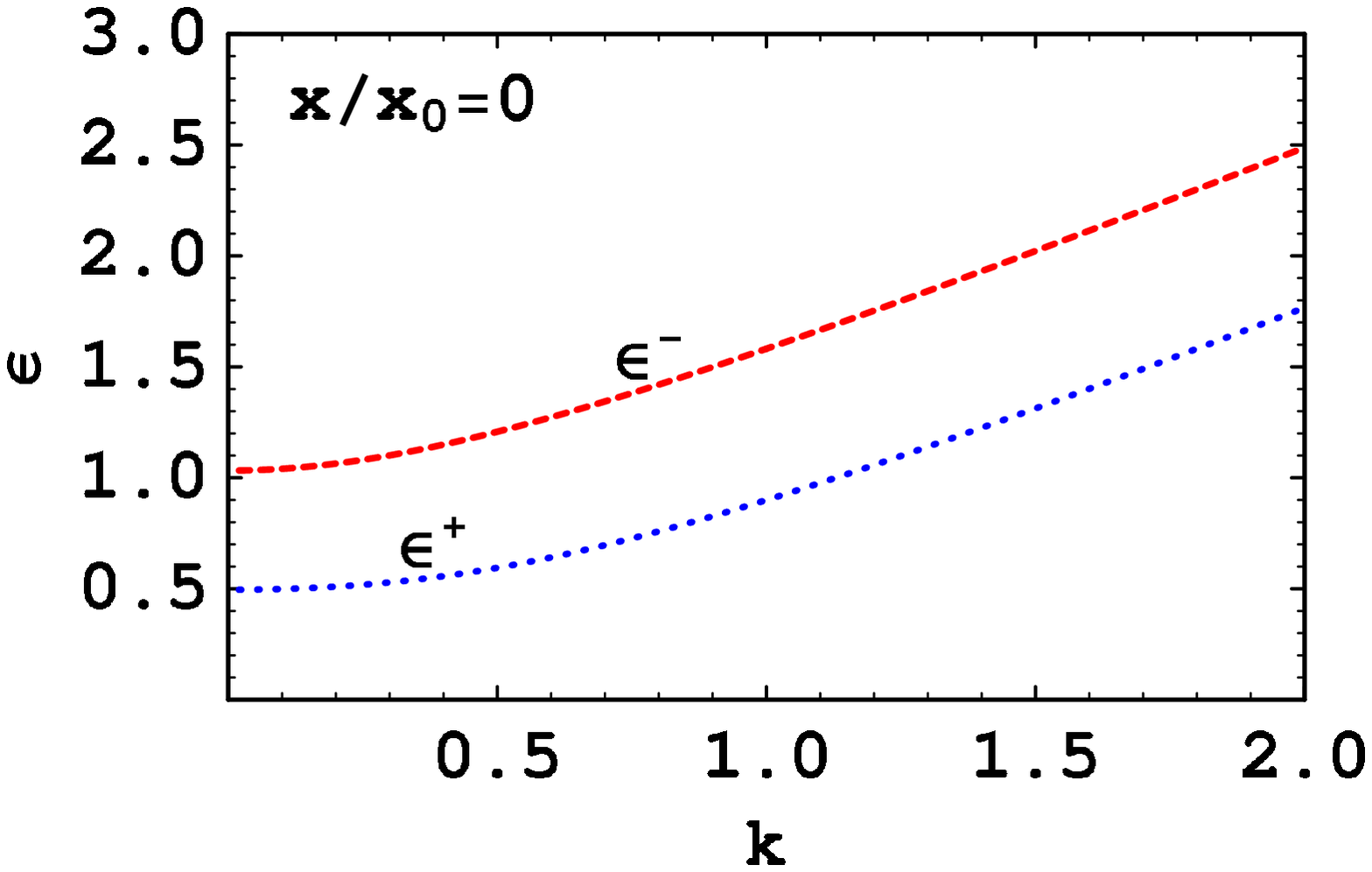}

\includegraphics[scale=0.33]{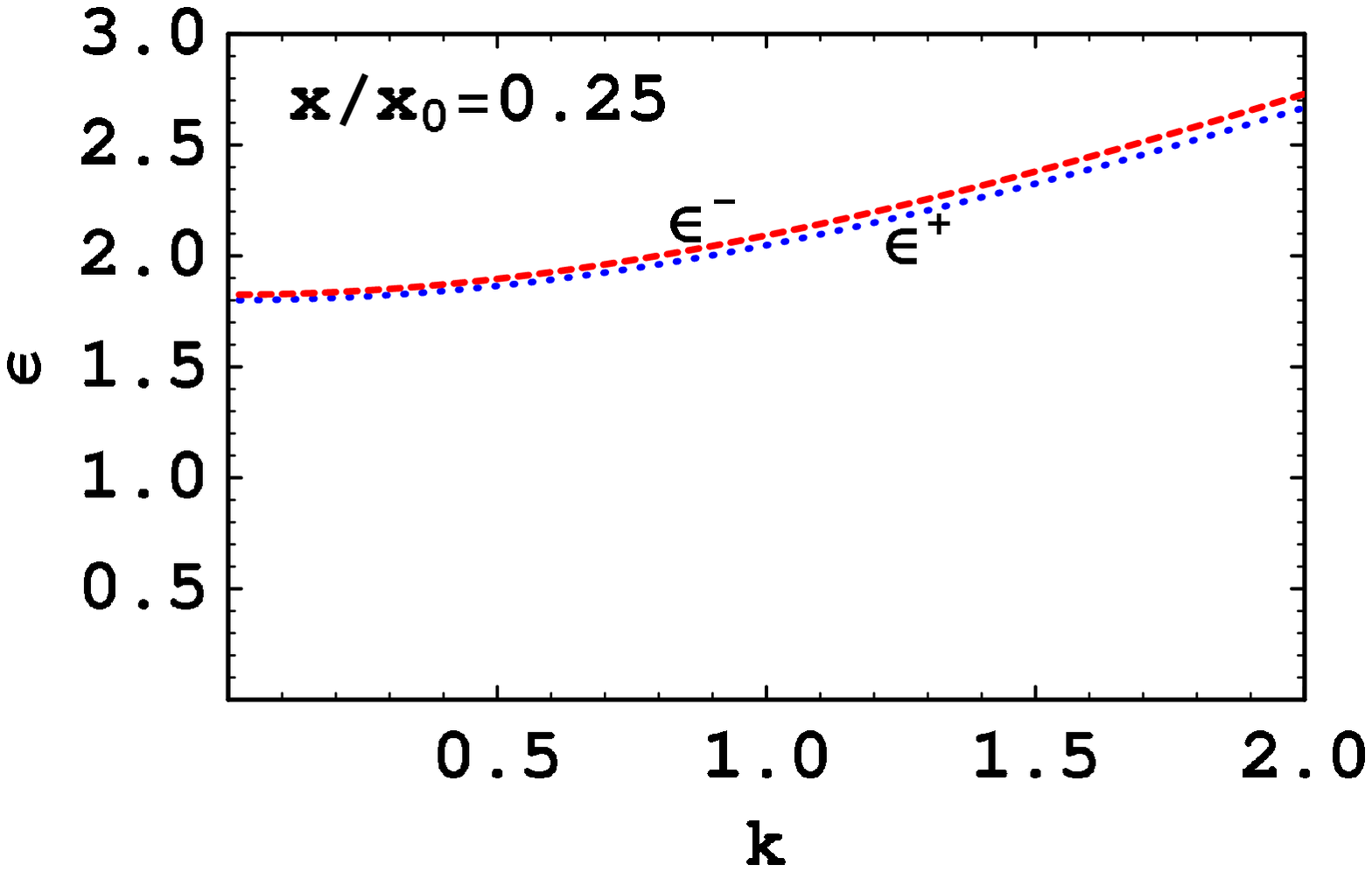}
\end{figure}

Let us discuss the quasi-fermion and quasi-antifermion excitation
energies given by $\epsilon_{k}^{+}$ and $\epsilon_{k}^{-}$. Inserting
the numerical solutions for $\mu$ and $\Delta$ into these energies
results in the curves shown in Fig. \ref{fig:figanti}. These excitation
energies show that, for large values of $x$, quasi-fermions and quasi-antifermions
become degenerate due to the vanishingly small chemical potential.
Because of the large energy gap, we expect neither quasi-fermions
nor quasi-antifermions to be present in the system. 

\begin{figure}
\caption{\label{fig:figallT}Density fractions in the crossover regime at fixed
$x/x_{0}=0.2$ as functions of temperature: condensed bosons (red
solid), fermions and uncondensed bosons (blue dotted and red dashed,
respectively).}

\includegraphics[scale=0.45]{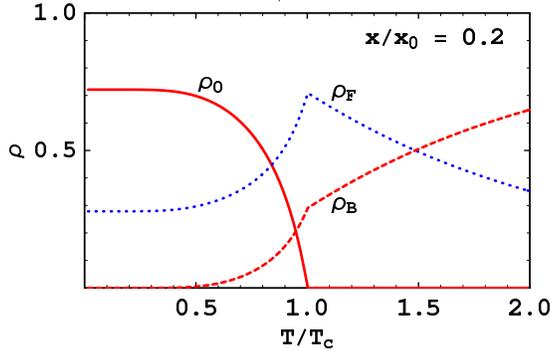}
\end{figure}

The condensate and the fermion and boson density fractions at fixed
crossover parameter $x/x_{0}=0.2$ are shown in Fig. \ref{fig:figallT}.
At the left end, $T=0$, one recovers the results shown in Fig. \ref{fig:figzeroT}
at the particular value $x/x_{0}=0.2$, at the point $T/T_{c}=1$
the second-order phase transition manifests itself in a kink in the
density fractions and a vanishing condensate. Below $T_{c}$ we observe
coexistence of condensed bound states, condensed resonant states,
and, for sufficiently large temperatures, uncondensed bound states.
We obtain thermal bosons even above the phase transition. They can
be interpreted as ``preformed'' pairs, just as the uncondensed pairs
below $T_{c}$. This phenomenon is also called ``pseudogap'' in the
literature \cite{Kitazawa:2005vr}. It suggests that there is a temperature
$T^{*}(x)$ which marks the onset of pair formation. This temperature
is not necessarily identical to $T_{c}$. In the BCS regime, $T^{*}(x)=T_{c}(x)$,
while for $x\gtrsim0$, $T^{*}(x)>T_{c}(x)$. Of course, our model
does not predict any quantitative value for $T^{*}$ because thermal
bosons are present for all temperatures. Therefore, we expect the
model to be valid only for a limited temperature range above $T_{c}$.

With the mean field appoximation, the boson-fermion model can be employed
to describe the BCS-BEC crossover. An important property of the model
is the coexistence of weakly-coupled Cooper pairs with condensed and
uncondensed bosonic bound states. In the crossover regime as well
as in the BEC regime, strongly-bound molecular Cooper pairs exist
below and above the critical temperature $T_{c}$. Above $T_{c}$,
they are all uncondensed (``preformed'' Cooper pairs) while below
$T_{c}$ a certain fraction of them forms a Bose-Einstein condensate.
In contrast, in the BCS regime, pairing and condensation of fermionic
degrees of freedom (in the absence of bosons) both set in at $T_{c}$. 

It is straightforward to extend our boson-fermion model to two fermion
species with cross-species pairing \cite{Shovkovy:2003uu,Huang:2004bg}.
This allows us to introduce a mismatch in fermion numbers and chemical
potentials which imposes a stress on the pairing \cite{Gubankova:2006gj,Rajagopal:2005dg}.
This kind of stressed pairing takes place in a variety of real systems.
For example, quark matter in a compact star is unlikely to exhibit
standard BCS pairing in the color-flavor locked (CFL) phase, i.e.,
pairing of quarks at a common Fermi surface. The cross-flavor (and
cross-color) pairing pattern of the CFL phase rather suffers a mismatch
in chemical potentials in the pairing sectors $bu-rs$ and $bd-gs$
($r,g,b$ meaning red, green, blue, and $u,d,s$ meaning up, down,
strange). This mismatch is induced by the explicit flavor symmetry
breaking through the heaviness of the strange quark and by the conditions
of color and electric neutrality. Our system shall only be an idealized
and simplified model of this complicated scenario. However, as in
the previous sections, we shall allow for arbitrary values of the
crossover parameter and thus model the strong coupling regime of quark
matter. We shall fix the overall charge and the difference in the
two charges. This is comparable to the effect of neutrality conditions
for matter inside a compact star, which also impose constraints on
the various color and flavor densities. Our focus will be to find
stable homogeneous superfluids in the crossover region and, by discarding
the unstable solutions, identify parameter values where the crossover
in fact becomes a phase transition.

We denote the average chemical potential and the mismatch in chemical
potentials by $\overline{\mu}\equiv\frac{\mu_{1}+\mu_{2}}{2},\,\delta\mu\equiv\frac{\mu_{1}-\mu_{2}}{2}.$
Then, the bosonic chemical potential is $\mu_{b}=2\overline{\mu}\,.$
The thermodynamic potential differs from the one-fermion case in the
dispersion relation for the fermions, 

\begin{eqnarray}
\Omega & = & \frac{m_{b}^{2}-\mu_{b}^{2}}{4g^{2}}\Delta^{2}+\frac{1}{2}\sum_{e}\int\frac{d^{3}k}{(2\pi)^{3}}\left[\omega_{k}^{e}+2T\ln\left(1-e^{-\omega_{k}^{e}/T}\right)\right]\nonumber \\
 &  & -\sum_{e}\int\frac{d^{3}k}{(2\pi)^{3}}\left\{ \epsilon_{k}^{e}+T\ln\left[1+e^{-(\epsilon_{k}^{e}+\delta\mu)/T}\right]\right.\nonumber \\
 &  & \left.+T\ln\left[1+e^{-(\epsilon_{k}^{e}-\delta\mu)/T}\right]\right\} ,\end{eqnarray}
With coupled gap equaiton ($0=\frac{\partial\Omega}{\partial\Delta}$)
and fixed sum ($\bar{n}\equiv n_{1}+n_{2}=-\frac{\partial\Omega}{\partial\bar{\mu}}$)
and difference ($\delta n\equiv n_{1}-n_{2}=-\frac{\partial\Omega}{\partial\delta\mu}$)
of the particle number density equations, the variables $\bar{\mu\,,}\,\delta\mu\:$
and$\Delta$ can be solved out.

In order to check the gapless states for their stability, we have
to compute the number susceptibility matrix$\chi_{ij}\equiv\frac{dn_{i}}{d\mu_{j}}\,,\, i=1,2$.
Note that we fix $\overline{n}$ and $\delta n$ (or equivalently
$n_{1}$ and $n_{2}$) in our solution. Hence $\chi$ can be regarded
as measuring the response of the system to a small perturbation away
from this solution. In particular, a stable solution requires the
mismatch in density to increase for an increasing mismatch in chemical
potentials. Therefore, a negative eigenvalue of this 2$\times$2 matrix
indicates the instability of a given solution. 

\begin{figure}
\caption{\label{fig:phase}The phase diagram in the plane of the crossover
parameter $x$ and the density difference $\delta n/\overline{n}$.
Shaded areas have unstable homogeneous solutions with negative number
susceptibility. NS denotes ``normal state''; in this region, no solution
for the gap equation is found. gSF denotes ``gapless superfluid'';
in this region a stable gapless superfluid state is found with two
different Fermi surface topologies, divided by the right dashed-dotted
line. Roman numbers II, III and IV denote three types of Fermi surface
topologies \cite{Deng:2006ed}. }

\includegraphics[scale=0.45]{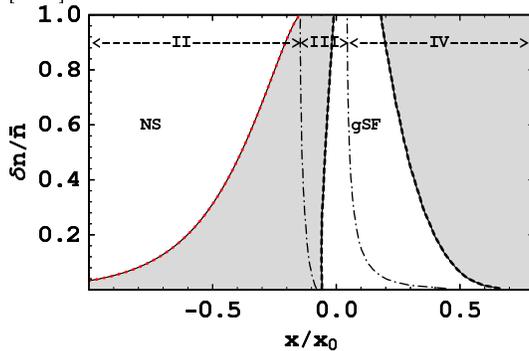}
\end{figure}

We finally present a phase diagram in Fig. \ref{fig:phase} for arbitrary
(positive) values of $\delta n/\overline{n}$. Since we do not consider
spatially inhomogeneous phases, this phase diagram is incomplete.
Its main point is to identify regions where homogeneous gapless superfluids
may exist. We find that for sufficiently large mismatches, $\delta n/\overline{n}\gtrsim0.02$
there is a region where no solution with nonzero $\Delta$ can be
found. We see that the region of stable superfluids shrinks with increasing
$\delta n/\overline{n}$, as expected. Note that the horizontal axis$\delta n/\overline{n}=0$
is not continuously connected to the rest of the phase diagram. For
vanishing mismatch in densities a stable, fully gapped superfluid
exists for all $x$. One should thus not be misled by the instability
for arbitrarily small mismatches.

We conclude with emphasizing the two main qualitative differences
to analogous phase diagrams in nonrelativistic systems: $(i)$ within
the stable region of homogeneous gapless superfluids there is a curve
that separates two different Fermi surface topologies; this is the
right dashed-dotted line in Fig. \ref{fig:phase}. $(ii)$ for large
$x$ the gapless superfluid becomes unstable even in the far BEC region;
this is the shaded area on the right side in Fig. \ref{fig:phase}.

\section{Phenomenological implication: neutrino processes in compact stars}

Generally the color superconducting quark matter cannot be easily
produced in laboratory. The most possible place for the color superconductor
is in the core of some compact stars where the high baryon density
is realized. In recent years, people have been looking for the connection
of color superconductivity to observations. The compact stars are
born from supernova explosions where neutrino processes play crucial
roles. Looking for evidences of color superconductivity from neutrino
processes is a natural choice. One possible way is to study the compact
star cooling through the neutrino emission, which is the major cooling
process for proto-neutron stars and compact stars with relatively
high internal temperature in the neutrino dominant stage. In this
stage star bodies are transparent to neutrinos while opaque to photons.
When the compact stars cool down by the neutrino emission, the photon
emission gradually takes over for the later cooling. The compact star
cooling models depend heavily on models for the neutron star interior
especially the core. The star core could be nucleon matter, hyperon
matter, pion or kaon condensate, or quark matter. There is huge amount
of literature about compact star cooling since it was first studied
by Tsuruta and Cameron in 1966 \cite{tsuruta}. For a recent review
of the neutron star cooling, see, e.g. \cite{Yakovlev:2004iq}. For
a review of neutron star evolution, see e.g. \cite{Prakash:2000jr}.
A recent systematic work in neutron star cooling based on hadronic
models can be found in Ref. \cite{Blaschke:2004vq}. 

\begin{figure}
\caption{\label{cap:neutrino-process}Neutrino processes in quark matter. (a)
Urca; (b) Modified Urca; (c) Bremsstrahlung}

\vspace{0.5cm}

(a)\includegraphics[bb=0bp 0bp 404bp 149bp,scale=0.6]{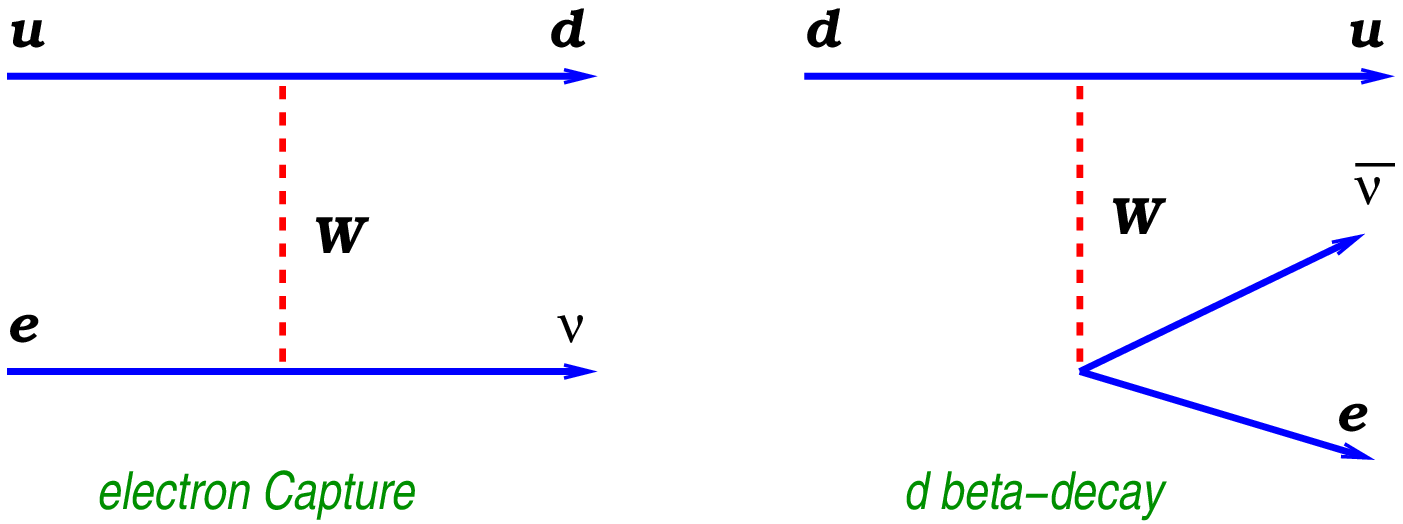}

\vspace{0.5cm}

(b)\includegraphics[scale=0.6]{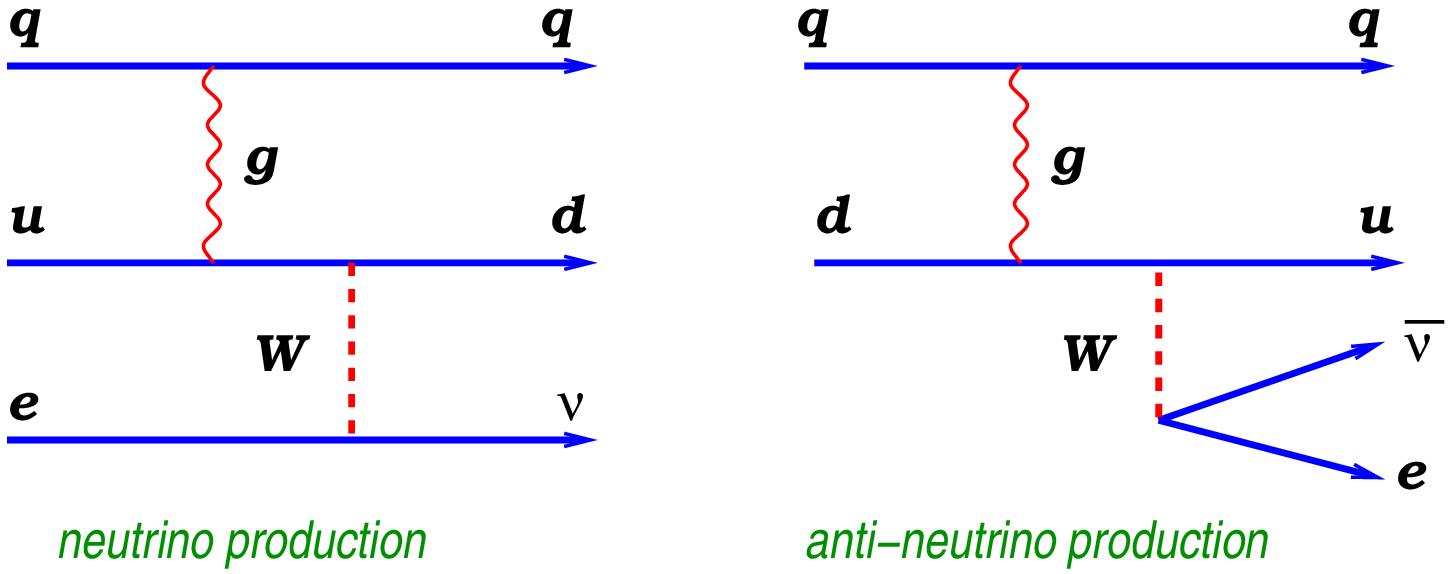}

\vspace{0.5cm}

(c)\includegraphics[scale=0.6]{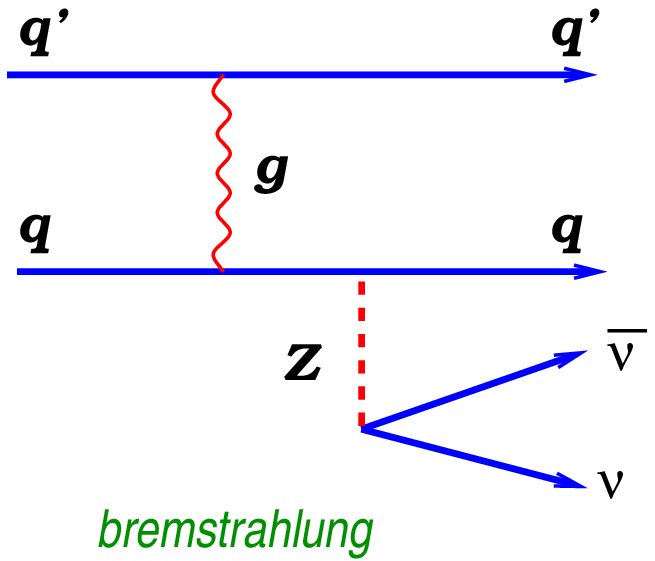}
\end{figure}

Calculating neutrino prcocesses in normal quark matter is part of
effort to describe the cooling behavior of compact stars with quark
matter cores. Roughly speaking there are three kinds of the neutrino
prcocesses in quark matter: the Urca, the modified Urca and the bremsstrahlung
processes, see Fig. (\ref{cap:neutrino-process}). The first systematic
and complete calculation for the Urca process in normal quark matter
was done by Iwamoto addressing the importance of Fermi liquid behavior
for relaxing the strict constraint of energy-momentum conservation
\cite{Iwamoto:1980eb}. The recent work by Schafer and Schwenzer \cite{Schafer:2004jp}
extends Iwamoto's work by taking into account of the non-Fermi liquid
behaviors \cite{Brown:2000eh,Wang:2001aq,Ipp:2003cj,Gerhold:2004tb}.
For the neutrino emission in fully gapped superconducting phases,
the Urca process is exponentially suppressed \cite{Blaschke:1999qx,Page:2000wt,Schmitt:2005wg,Wang:2006tg}.
The possible relevant processes for the neutrino emission would be
from the Goldstone mode \cite{Jaikumar:2002vg} or from the pair breaking
and recombination processes \cite{Jaikumar:2001hq,Kundu:2004mz}.
For gapless phases the Urca processes are not suppressed and like
in the normal quark matter \cite{Alford:2004zr,Jaikumar:2005hy,Huang:2007jw,Huang:2007cr}. 

In this section we will use close-time-path method to derive the Kadnoff-Baym
equation. Based on the Kadnoff-Baym equation, we will obtain the kinetic
equation for the neutrino emissivity for the normal and superconducting
quark matter. There are other applications of close-time-path method,
see, e.g. Ref. \cite{Wang:2001dm,Wang:2002qe} in the transport equation
of the quark gluon plasma. For the derivation of Kadnoff-Baym equation
and kinetic equation for other neutrino processes in compact star,
see, e.g. Ref. \cite{Sedrakian:1999jh,Sedrakian:2000kc}. If the neutron
star has a color superconducting quark matter core which is in the
so-called spin-1 A-phase, we will show that the neutrino emission
is asymmetric in space.

\subsection{Covariant Kadnoff-Baym equation}

The two-point Green function for a fermion can be defined in the following
way\[
iG(1,2)=\left\langle T[\psi_{H}(1)\overline{\psi}_{H}(2)]\right\rangle ,\]
where '1' and '2' denote all indices including spatial coordinates
of the two fields. $T$ is the causal time-ordering operator which
orders an operator at an earlier time to the right of an operator
at a later time. The field operators $\psi_{H}(t)$ and $\overline{\psi}_{H}(t)$
in the Heisenberg picture are related to those in the Schödinger picture
by\begin{eqnarray*}
\psi_{H}(t) & = & \exp[i\mathcal{H}(t-t_{0})]\psi_{S}(t)\exp[-i\mathcal{H}(t-t_{0})]\\
\overline{\psi}_{H}(t) & = & \exp[i\mathcal{H}(t-t_{0})]\overline{\psi}_{S}(t)\exp[-i\mathcal{H}(t-t_{0})],\end{eqnarray*}
where $\mathcal{H}$ is the Hamiltonian of the system. 

\begin{figure}
\caption{\label{fig:close-time-path}Ordering along the close-time path in
Green function.}

\vspace{0.5cm}

\includegraphics[scale=0.8]{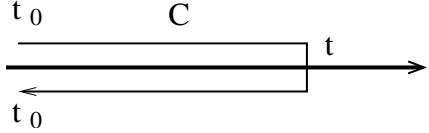}
\end{figure}

In the same way, we can define the two-point Green function on the
close-time path $C$ as shown in Fig. (\ref{fig:close-time-path}),\begin{eqnarray}
iG(1,2) & = & \left\langle T_{C}[\psi_{H}(1)\overline{\psi}_{H}(2)]\right\rangle \nonumber \\
 & = & \left\langle T_{C}\left\{ \exp\left[i\int_{C}d^{4}x\mathcal{L}'_{I}\right]\psi_{I}(1)\overline{\psi}_{I}(2)\right\} \right\rangle ,\label{def-green-ctp}\end{eqnarray}
where $\mathcal{L}'_{I}$ is the interacting part of the Lagrangian
in the interaction picture and $T_{C}$ is the ordering operator along
the close-time path $C$. $G(1,2)$ can be written as\begin{equation}
G(1,2)=\Theta_{C}(t_{1}-t_{2})G^{>}(1,2)+\Theta_{C}(t_{2}-t_{1})G^{<}(1,2),\label{green1}\end{equation}
 where $G^{>,<}$ are correlation functions. We also need the retarded
and advanced Green functions as follows\begin{eqnarray*}
G^{R}(1,2) & = & G(1,2)-G^{<}(1,2)=\Theta_{C}(t_{1}-t_{2})[G^{>}(1,2)-G^{<}(1,2)]\\
G^{A}(1,2) & = & G(1,2)-G^{>}(1,2)=\Theta_{C}(t_{2}-t_{1})[G^{<}(1,2)-G^{>}(1,2)].\end{eqnarray*}

The series expansion of the exponential function in Eq. (\ref{def-green-ctp})
generates the perturbation expansion for the two-point Green function.
In lowest order one obtains the non-interacting Green function $G_{0}$
whose inverse $G_{0}^{-1}$ is\[
G_{0}^{-1}(1,2)=(i\gamma_{\mu}\partial_{1}^{\mu}-m)\delta(1,2).\]
The relation between a Green function and its inverse is \[
\int_{C}d3G^{-1}(1,3)G(3,2)=\int_{C}d3G(1,3)G^{-1}(3,2)=\delta(1,2).\]
Dyson-Schwinger equation is \begin{eqnarray}
G^{-1}(1,2) & = & G_{0}^{-1}(1,2)-\Sigma(1,2),\label{DSE1}\end{eqnarray}
which can be further written as\begin{eqnarray}
G_{0}^{-1}G & = & 1+\Sigma G\nonumber \\
GG_{0}^{-1} & = & 1+G\Sigma,\label{DSE2}\end{eqnarray}
whose explicit form is\begin{eqnarray}
(i\gamma_{\mu}\partial_{1}^{\mu}-m)G(1,2) & = & \delta(1,2)+\int_{C}d3\Sigma(1,3)G(3,2)\label{DSE5}\\
G(1,2)(-i\gamma_{\mu}\overleftarrow{\partial}_{2}^{\mu}-m) & = & \delta(1,2)+\int_{C}d3G(1,3)\Sigma(3,2),\label{DSE4}\end{eqnarray}
After simplification by translating the contour integrals into normal
ones we get Dyson-Schwinger equations\begin{eqnarray}
(i\gamma_{\mu}\partial_{1}^{\mu}-m)G^{<}(1,2) & = & -\int_{t_{0}}^{t_{2}}d3\Sigma^{<}(1,3)\left[G^{>}(3,2)-G^{<}(3,2)\right]\nonumber \\
 &  & -\int_{t_{0}}^{t_{1}}d3\left[\Sigma^{<}(1,3)-\Sigma^{>}(1,3)\right]G^{<}(3,2),\label{DSE6}\end{eqnarray}
 \begin{eqnarray}
G^{<}(1,2)(-i\gamma_{\mu}\overleftarrow{\partial}_{2}^{\mu}-m) & = & -\int_{t_{0}}^{t_{2}}d3G^{<}(1,3)\left[\Sigma^{>}(3,2)-\Sigma^{<}(3,2)\right]\nonumber \\
 &  & -\int_{t_{0}}^{t_{1}}d3\left[G^{<}(1,3)-G^{>}(1,3)\right]\Sigma^{<}(3,2),\label{DSE7}\end{eqnarray}
where we have used the correlation parts of the self-energy $\Sigma^{>}$
and $\Sigma^{<}$ in analogy to Eq. (\ref{green1}). Making gradient
expansion and Fourier transformation, taking the difference of the
above two Dyson-Schwinger equations, and finally taking traces for
the resulting equations, we obtain Kadanoff-Baym equation\begin{equation}
\partial_{X}^{\mu}\mathrm{Tr}[\gamma_{\mu}iG^{<}(X,P)]=-\mathrm{Tr}\left[iG^{>}(-i\Sigma^{<})-(-i\Sigma^{>})iG^{<}\right](X,P).\label{KB-eq}\end{equation}
where $X=(X_{1}+X_{2})/2$ and $P$ is the momentum variable conjugate
to $X$. Note that our convention for the sign of selfenergy is implied
in $G^{-1}=G^{(0)-1}-\Sigma$, which means $-i\Sigma$ in the Feynman
rules.

\subsection{Kinetic equation}

We can derive the kinetic equation from Eq. (\ref{KB-eq}). The fermionic
Green functions (here we mean the electron and the neutrino or their
anti-particles) can be expressed in terms of distribution functions
as follows\begin{eqnarray}
iG^{<}(X,P) & = & -\gamma^{\mu}P_{\mu}\frac{\pi}{p}\left\{ f_{e^{-}/\nu}(t,\mathbf{p})\delta(p_{0}+\mu_{e^{-}/\nu}-p)\right.\nonumber \\
 &  & \left.-[1-f_{e^{+}/\overline{\nu}}(t,-\mathbf{p})]\delta(p_{0}+\mu_{e^{-}/\nu}+p)\right\} .\label{G-less1}\end{eqnarray}
where we have shifted $p_{0}$ by the chemical potential $p_{0}\rightarrow p_{0}+\mu_{e^{-}/\nu}$.
In the same way, we can obtain $G^{>}$ as follows\begin{eqnarray}
iG^{>}(X,P) & = & \gamma^{\mu}P_{\mu}\frac{\pi}{p}\left\{ [1-f_{e^{-}/\nu}(t,\mathbf{p})]\delta(p_{0}+\mu_{e^{-}/\nu}-p)\right.\nonumber \\
 &  & \left.-f_{e^{+}/\overline{\nu}}(t,-\mathbf{p})\delta(p_{0}+\mu_{e^{-}/\nu}+p)\right\} .\label{G-larger1}\end{eqnarray}
We assume that the distribution function $f$ only depends on $X_{0}=t$,
so that $G(X,P)$ can be written as $G(t,P)$. This approximation
is well justified for neutrinos in the compact star with mean free
paths comparable or larger than size of the star.

We are interested in the neutrino cooling stage in the compact star.
When neutrinos are produced they escape the star immediately. The
neutrino density can not be built up in the star. So we can put the
chemical potentials of the neutrino and anti-neutrino zero. Substituting
Eq. (\ref{G-less1}) and (\ref{G-larger1}) into Eq. (\ref{eq:kinetic1}),
and carrying out the integral $\int_{0}^{\infty}dp_{0}$, we obtain
the equation for the time variation rate of the neutrino distribution:\begin{eqnarray}
\frac{\partial}{\partial t}f_{\nu}(t,\mathbf{p}) & = & \frac{1}{4\pi}\int_{0}^{\infty}dp_{0}\mathrm{Tr}\left[iG_{\nu}^{>}(-i\Sigma_{\nu}^{<})\right.\nonumber \\
 &  & \left.-(-i\Sigma_{\nu}^{>})iG_{\nu}^{<}\right](t,P),\label{eq:kinetic1}\end{eqnarray}
which picks up the contribution corresponding to the term with support
at $p_{0}=p$ in the left hand side. In the same way, we can extract
the kinetic equation for the anti-neutrinos by carrying out the integral
$\int_{-\infty}^{0}dp_{0}$\begin{eqnarray}
\frac{\partial}{\partial t}f_{\overline{\nu}}(t,-\mathbf{p}) & = & -\frac{1}{4\pi}\int_{-\infty}^{0}dp_{0}\mathrm{Tr}\left[iG_{\overline{\nu}}^{>}(-i\Sigma_{\overline{\nu}}^{<})\right.\nonumber \\
 &  & \left.-(-i\Sigma_{\overline{\nu}}^{>})iG_{\overline{\nu}}^{<}\right](t,P),\label{eq:kinetic2}\end{eqnarray}
which picks up the contribution corresponding to the term with support
at $p_{0}=-p$. 

In order to evaluate $\Sigma^{>,<}$, we need the Feynman diagram
as shown in Fig. (\ref{fig:urca}). The convention about the order
of fermionic fields in the Feynman diagram is along the direction
of momentum flow. From the diagram Fig. (\ref{fig:urca})(a), the
neutrino self-energy reads:

\begin{figure}
\caption{\label{fig:urca}Self-energy in close-time-path formalism for the
neutrino Urca process.}

\vspace{0.5cm}\psfrag{dquark}{$d$}

\psfrag{uquark}{$u$}

\psfrag{q-mom}{$Q$}

\psfrag{wboson}{$W$}

\psfrag{nu_e}{$\nu _e$}

\psfrag{e-}{$e^-$}

\psfrag{nu}{$\nu$}

\psfrag{nu1}{$\nu '$}

\psfrag{mu}{$\mu$}

\psfrag{mu1}{$\mu '$}

\psfrag{P1}{$P_e$}

\psfrag{P2}{$P_d$}

\psfrag{P3}{$P_u$}

\psfrag{p-mom}{$P_\nu$}

\includegraphics[scale=0.5]{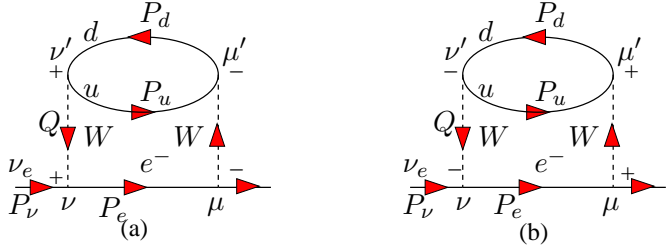}
\end{figure}

\begin{eqnarray*}
-i\Sigma_{\nu}^{<}(P_{\nu}) & = & -\int\frac{d^{4}P_{e}}{(2\pi)^{4}}(-i\Gamma_{e}^{\mu})iG_{e}^{<}(P_{e})i\Gamma_{e}^{\nu}\\
 &  & \times[-i\Pi_{\mu'\nu'}^{>}(P_{e}-P_{\nu})]iD_{\mu}^{\overline{F},\mu'}(P_{e}-P_{\nu})iD_{\nu}^{F,\nu'}(P_{e}-P_{\nu})\\
 & = & \frac{G_{F}^{2}}{2}\int\frac{d^{4}P_{e}}{(2\pi)^{4}}\gamma^{\mu}(1-\gamma_{5})\gamma^{\sigma}P_{e,\sigma}\gamma^{\nu}(1-\gamma_{5})\\
 &  & \times[-i\Pi_{\mu\nu}^{>}(P_{e}-P_{\nu})]\frac{\pi}{p_{e}}f_{e}(t,\mathbf{p}_{e})\delta(p'_{e0}-p_{e}),\end{eqnarray*}
where the overall sign is due to the quark loop. Note that $P_{e}\equiv(p'_{e0},\mathbf{p}_{e})$
with $p'_{e0}\equiv p_{e0}+\mu_{e}$, then integral element is $d^{4}P_{e}\equiv dp'_{e0}d^{3}\mathbf{p}_{e}$.
We have defined the order of '>' and '<' in $\Pi^{>,<}$ along the
momentum direction. For example, in the first Feynman diagram Fig.
(\ref{fig:urca})(a), the W-boson polarization tensor is defined by
from $\mu'$ to $\nu'$ following the momentum flow, i.e. $\mu'$
as the first entry and $\nu'$ as the second one. The first entry
on the negative branch has a time later than the first entry on the
close-time-path contour $C$. Therefore we have $\Pi_{\mu'\nu'}^{>}$.
In the same way, Fig. (\ref{fig:urca})(b) reads\begin{eqnarray*}
-i\Sigma_{\nu}^{>}(P_{\nu}) & = & -\int\frac{d^{4}P_{e}}{(2\pi)^{4}}i\Gamma_{e}^{\mu}iG_{e}^{>}(P_{e})(-i\Gamma_{e}^{\nu})\\
 &  & \times[-i\Pi_{\mu'\nu'}^{<}(P_{e}-P_{\nu})]iD_{\mu}^{F,\mu'}(P_{e}-P_{\nu})iD_{\nu}^{\overline{F},\nu'}(P_{e}-P_{\nu})\\
 & = & -\frac{G_{F}^{2}}{2}\int\frac{d^{4}P_{e}}{(2\pi)^{4}}\gamma^{\mu}(1-\gamma_{5})\gamma^{\sigma}P_{e,\sigma}\gamma^{\nu}(1-\gamma_{5})\\
 &  & \times[-i\Pi_{\mu\nu}^{<}(P_{e}-P_{\nu})]\frac{\pi}{p_{e}}[1-f_{e}(t,\mathbf{p}_{e})]\delta(p'_{e0}-p_{e}).\end{eqnarray*}
Inserting $-i\Sigma_{\nu}^{>}(P_{\nu})$ and $-i\Sigma_{\nu}^{<}(P_{\nu})$
into Eq. (\ref{eq:kinetic1}), we get \begin{eqnarray}
\frac{\partial}{\partial t}f_{\nu}(t,\mathbf{p}_{\nu}) & = & \frac{1}{4\pi}\frac{G_{F}^{2}}{2}\int\frac{d^{3}\mathbf{p}_{e}}{(2\pi)^{4}}\frac{\pi}{p_{\nu}}\frac{\pi}{p_{e}}L^{\mu\nu}(P_{\nu},P_{e})\nonumber \\
 &  & \times[1-f_{\nu}(t,\mathbf{p}_{\nu})]f_{e}(t,\mathbf{p}_{e})[-i\Pi_{\mu\nu}^{>}(P_{e}-P_{\nu})],\label{eq:nu}\end{eqnarray}
where $P_{\nu}=(p_{\nu},\mathbf{p}_{\nu})$ and $P_{e}=(p_{e},\mathbf{p}_{e})$.
Note that the definition for $P_{e}$ has been changed from the original
one. The leptonic tensor $L^{\mu\nu}$ is defined by \[
L^{\mu\nu}(P_{\nu},P_{e})\equiv\mathrm{Tr}\left[\gamma^{\lambda}P_{\nu,\lambda}\gamma^{\mu}(1-\gamma_{5})\gamma^{\sigma}P_{e,\sigma}\gamma^{\nu}(1-\gamma_{5})\right].\]

For the anti-neutrino, we still use $P$ to denote its momentum, only
the second term $\sim(-i\Sigma_{\overline{\nu}}^{>})iG_{\overline{\nu}}^{<}$
in Eq. (\ref{eq:kinetic2}) contributes. Finally we obtain \begin{eqnarray}
\frac{\partial}{\partial t}f_{\overline{\nu}}(t,-\mathbf{p}_{\overline{\nu}}) & = & -\frac{1}{4\pi}\frac{G_{F}^{2}}{2}\int\frac{d^{3}\mathbf{p}_{e}}{(2\pi)^{4}}\frac{\pi}{p_{\overline{\nu}}}\frac{\pi}{p_{e}}L^{\mu\nu}(P_{\overline{\nu}},P_{e})\nonumber \\
 &  & \times[1-f_{\overline{\nu}}(-\mathbf{p}_{\overline{\nu}})][1-f_{e}(\mathbf{p}_{e})][-i\Pi_{\mu\nu}^{<}(P_{e}-P_{\overline{\nu}})],\label{eq:anti-nu}\end{eqnarray}
where $P_{\overline{\nu}}=(-p_{\overline{\nu}},\mathbf{p}_{\overline{\nu}})$.
We can rewrite Eq. (\ref{eq:anti-nu}) by flipping the momentum sign:
$\mathbf{p}_{\overline{\nu}}\rightarrow-\mathbf{p}_{\overline{\nu}}$,
\begin{eqnarray}
\frac{\partial}{\partial t}f_{\overline{\nu}}(t,\mathbf{p}_{\overline{\nu}}) & = & -\frac{1}{4\pi}\frac{G_{F}^{2}}{2}\int\frac{d^{3}\mathbf{p}_{e}}{(2\pi)^{4}}\frac{\pi}{p_{\overline{\nu}}}\frac{\pi}{p_{e}}L^{\mu\nu}(P_{\overline{\nu}},P_{e})\nonumber \\
 &  & \times[1-f_{\overline{\nu}}(\mathbf{p}_{\overline{\nu}})][1-f_{e}(\mathbf{p}_{e})][-i\Pi_{\mu\nu}^{<}(P_{e}-P_{\overline{\nu}})],\label{eq:anti-nu1}\end{eqnarray}
where $P_{\overline{\nu}}$ is changed to $P_{\overline{\nu}}=(-p_{\overline{\nu}},-\mathbf{p}_{\overline{\nu}})$.
We will see that the negative sign in front of Eq. (\ref{eq:anti-nu1})
will cancel the negative sign due to $P_{\overline{\nu}}=(-p_{\overline{\nu}},-\mathbf{p}_{\overline{\nu}})$
in the matrix element. Thus the result for the anti-neutrino is the
identical to the neutrino. 

Using the following relations between $\Pi_{\mu\nu}^{>}$ and $\Pi_{\mu\nu}^{R}$\begin{eqnarray*}
\Pi^{<}(K) & = & -2in_{B}(k_{0})\mathrm{Im}\Pi^{R}(K)\\
\Pi^{>}(K) & = & -2i[1+n_{B}(k_{0})]\mathrm{Im}\Pi^{R}(K),\end{eqnarray*}
where $n_{B}(k_{0})=1/(e^{\beta k_{0}}-1)$ is the Bose-Einstein distribution,
Eq. (\ref{eq:nu}) can be expressed in terms of $\mathrm{Im}\Pi^{R}(P_{e}-P_{\nu})$
as \begin{eqnarray*}
\frac{\partial}{\partial t}f_{\nu}(t,\mathbf{p}_{\nu}) & = & \frac{1}{4\pi}G_{F}^{2}\int\frac{d^{3}\mathbf{p}_{e}}{(2\pi)^{4}}\frac{\pi}{p_{\nu}}\frac{\pi}{p_{e}}L^{\mu\nu}(P_{\nu},P_{e})n_{B}(p_{\nu}-p_{e}+\mu_{e})\\
 &  & \times[1-f_{\nu}(t,\mathbf{p}_{\nu})]f_{e}(t,\mathbf{p}_{e})\mathrm{Im}\Pi_{\mu\nu}^{R}(P_{e}-P_{\nu}),\end{eqnarray*}
where we have used $n_{B}(k_{0})+n_{B}(-k_{0})=-1$. Taking the momentum
integral $\frac{d^{3}\mathbf{p}_{\nu}}{(2\pi)^{3}}$ and noting that
$n_{\nu}=\int\frac{d^{3}\mathbf{p}_{\nu}}{(2\pi)^{3}}f_{\nu}(t,\mathbf{p}_{\nu})$,
we obtain the number emissivity per volume for neutrinos: \begin{eqnarray}
\frac{\partial}{\partial t}n_{\nu}(t) & = & \frac{1}{2}G_{F}^{2}\int\frac{d^{3}\mathbf{p}_{e}}{(2\pi)^{3}}\frac{d^{3}\mathbf{p}_{\nu}}{(2\pi)^{3}}\frac{1}{2p_{\nu}}\frac{1}{2p_{e}}L^{\mu\nu}(P_{\nu},P_{e})n_{B}(p_{\nu}-p_{e}+\mu_{e})\nonumber \\
 &  & \times[1-f_{\nu}(t,\mathbf{p}_{\nu})]f_{e}(t,\mathbf{p}_{e})\mathrm{Im}\Pi_{\mu\nu}^{R}(P_{e}-P_{\nu}).\label{eq:emissivity1}\end{eqnarray}
The emissivity for anti-neutrinos is the same as that of neutrinos.
The energy emissivity can be obtained by including $p_{\nu}$ into
the integrand in Eq. (\ref{eq:emissivity1}).

\subsection{u-d-W vertex in Nambu-Gorkov basis}

\begin{figure}
\caption{\label{fig:u-d-W-vertices}u-d-W vertices. The momenta of W-bosons
flow into the vertices. }

\vspace{0.5cm}\psfrag{W-}{$W^-$}

\psfrag{V-}{$i\Gamma^{NG}_{-}$}

\psfrag{W+}{$W^+$}

\psfrag{V+}{$i\Gamma^{NG}_{+}$}

\includegraphics{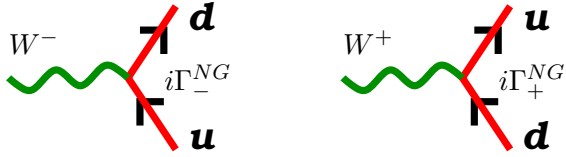}
\end{figure}

Because we are interested in Urca processes, we have to write down
the u-d-W interaction vertex in NG basis. The u-d-W vertex differs
from the quark-gluon one in Nambu-Gorkov forms due to the flavor changing
nature of the W-boson interaction. So we explicitly keep the flavor
indices $f$ for the quark fields:\begin{eqnarray*}
 &  & \psi_{C}(f)=C\overline{\psi}^{T}(f),\;\overline{\psi}_{C}(f)=\psi^{T}(f)C\\
 &  & \psi(f)=C\overline{\psi}_{C}^{T}(f),\;\overline{\psi}(f)=\psi_{C}^{T}(f)C.\end{eqnarray*}
where we assume $f=1,2$ for $u,d$ respectively. We need for the
u-d-W coupling an additional flavor matrix $\tau_{\pm}$ to guarantee
that $W^{\pm}$ couples to $u$ and $d$ instead of the same flavor,
where $\tau_{\pm}$ are defined by \[
\tau_{+}=\left(\begin{array}{cc}
0 & 1\\
0 & 0\end{array}\right),\;\tau_{-}=\left(\begin{array}{cc}
0 & 0\\
1 & 0\end{array}\right).\]
 Here we assume the spinor order of the vertex is $d-W^{-}-u$ and
$u-W^{+}-d$, where the momenta of W-bosons flow into vertices, see
Fig. (\ref{fig:u-d-W-vertices}). Then we have \begin{eqnarray*}
i\Gamma_{\pm}^{\mu} & = & -\frac{ie}{2\sqrt{2}\sin\theta_{W}}\gamma^{\mu}(1-\gamma_{5})\tau_{\pm},\end{eqnarray*}
where $\theta_{W}$ is the Weinberg's angle. Here we have taken CKM
matrix element for $u$ and $d$ quark $V_{ud}\approx1$. The Fermi
weak coupling constant $G_{F}$ is related to $\theta_{W}$ by $\frac{G_{F}}{\sqrt{2}}=\frac{e^{2}}{8M^{2}\sin^{2}\theta_{W}}$. 

Then we can write down the u-d-W interaction term \begin{eqnarray*}
\overline{\psi}(f_{1})\Gamma_{\pm}^{\mu}\psi(f_{2}) & = & \psi_{C}^{T}(f_{1})C\Gamma_{\pm}^{\mu}C\overline{\psi}_{C}^{T}(f_{2})\\
 & = & -\left[\overline{\psi}_{C}(f_{2})C^{T}\Gamma_{\pm}^{\mu T}C^{T}\psi_{C}(f_{1})\right]^{T}\\
 & = & -\overline{\psi}_{C}(f_{2})C^{T}\Gamma_{\pm}^{\mu T}C^{T}\psi_{C}(f_{1})\\
 & \equiv & \overline{\psi}_{C}(f_{2})\overline{\Gamma}_{\pm}^{\mu}\psi_{C}(f_{1}),\end{eqnarray*}
where we have defined $\overline{\Gamma}_{\pm}^{\mu}\equiv-C\Gamma_{\pm}^{\mu T}C$.
Therefore the u-d-W vertex in Nambu-Gorkov basis is given by\begin{eqnarray}
i\Gamma_{\pm;NG}^{\mu} & = & -\frac{ie}{2\sqrt{2}\sin\theta_{W}}\left(\begin{array}{cc}
\gamma^{\mu}(1-\gamma_{5})\tau_{\pm} & 0\\
0 & -\gamma^{\mu}(1+\gamma_{5})\tau_{\mp}\end{array}\right)\nonumber \\
 & \equiv & -\frac{ie}{2\sqrt{2}\sin\theta_{W}}\Gamma_{NG}^{\mu}\tau_{\pm;NG},\label{eq:quark-w1}\end{eqnarray}
where we have defined\begin{eqnarray*}
\Gamma_{NG}^{\mu} & \equiv & \left(\begin{array}{cc}
\gamma^{\mu}(1-\gamma_{5}) & 0\\
0 & -\gamma^{\mu}(1+\gamma_{5})\end{array}\right),\;\tau_{\pm;NG}\equiv\left(\begin{array}{cc}
\tau_{\pm} & 0\\
0 & \tau_{\mp}\end{array}\right).\end{eqnarray*}

\subsection{Polarization tensor for W-boson from quark loop in superconducting
phase}

Now we can calculate the polarization tensor $\Pi_{\mu\nu}(Q)$ for
the W-boson from the quark loop in imaginary time formalism for superconducting
phases. Here $Q=P_{e}-P_{\nu}$. Then we make analytical extension
to get retarded polarization tensor. The quark propagator with condensate
can be written as $\mathcal{G}_{ij}=\mathcal{G}_{ij}^{u}\tau_{11}+\mathcal{G}_{ij}^{d}\tau_{22}$
where $i,j$ are Nambu-Gorkov indices and $\tau_{11}=\mathrm{diag}(1,0)$
and $\tau_{22}=\mathrm{diag}(0,1)$ in the flavor space of $u$ and
$d$ quark. $\mathcal{G}_{ij}^{u}$ and $\mathcal{G}_{ij}^{d}$ are
given by Eq. (\ref{eq:NGsigma}). Let's look at one case where the
left vertex is $\tau_{-;NG}$ while the right must be $\tau_{+;NG}$,
as illustrated corresponding to the fermion loop direction in Fig.
(\ref{fig:urca}). Then the polarization tensor of the W-boson in
imaginary time formalism reads \begin{eqnarray}
\Pi^{\mu\nu}(Q) & = & \frac{1}{2}T\sum_{p_{u0}}\int\frac{d^{3}\mathbf{p}_{u}}{(2\pi)^{3}}\mathrm{Tr}_{NG,c,f,s}[\Gamma_{NG}^{\mu}\tau_{-;NG}\mathcal{G}^{u}(P_{u})\Gamma_{NG}^{\nu}\tau_{+NG}\mathcal{G}^{d}(P_{d})]\nonumber \\
 & = & \frac{1}{2}T\sum_{p_{u0}}\int\frac{d^{3}\mathbf{p}_{u}}{(2\pi)^{3}}\mathrm{Tr}_{c,f,s}[\gamma^{\mu}(1-\gamma_{5})\tau_{-}\mathcal{G}_{11}^{u}(P_{u})\nonumber \\
 &  & \times\gamma^{\nu}(1-\gamma_{5})\tau_{+}\mathcal{G}_{11}^{d}(P_{d})]\nonumber \\
 &  & +\frac{1}{2}T\sum_{p_{u0}}\int\frac{d^{3}\mathbf{p}_{u}}{(2\pi)^{3}}\mathrm{Tr}_{c,f,s}[\gamma^{\mu}(1+\gamma_{5})\tau_{+}\mathcal{G}_{22}^{u}(P_{u})\nonumber \\
 &  & \times\gamma^{\nu}(1+\gamma_{5})\tau_{-}\mathcal{G}_{22}^{d}(P_{d})]\nonumber \\
 &  & -\frac{1}{2}T\sum_{p_{u0}}\int\frac{d^{3}\mathbf{p}_{u}}{(2\pi)^{3}}\mathrm{Tr}_{c,f,s}[\gamma^{\mu}(1-\gamma_{5})\tau_{-}\mathcal{G}_{12}^{u}(P_{u})\nonumber \\
 &  & \times\gamma^{\nu}(1+\gamma_{5})\tau_{-}\mathcal{G}_{21}^{d}(P_{d})]\nonumber \\
 &  & -\frac{1}{2}T\sum_{p_{u0}}\int\frac{d^{3}\mathbf{p}_{u}}{(2\pi)^{3}}\mathrm{Tr}_{c,f,s}[\gamma^{\mu}(1+\gamma_{5})\tau_{+}\mathcal{G}_{21}^{u}(P_{u})\nonumber \\
 &  & \times\gamma^{\nu}(1-\gamma_{5})\tau_{+}\mathcal{G}_{12}^{d}(P_{d})].\label{eq:polar-tensor}\end{eqnarray}
where $P_{d}\equiv P_{u}+Q$, and $p_{u0}=-i(2l+1)\pi T$ with integer
$l$ are Matsubara frequencies. The order in the trace is against
the direction of fermion line in Fig. (\ref{fig:urca}). The last
two terms in Eq. (\ref{eq:polar-tensor}) proportional to the off-diagonal
propagators are zero due to the presence of two identical $\tau_{-}$
or $\tau_{+}$ leading to vanishing flavor traces. This a spectacular
feature for single flavor pairing. Taking the flavor trace and Matsubara
sum, and noting that the first and second terms are identical, we
obtain \begin{eqnarray}
\Pi^{\mu\nu}(Q) & = & T\sum_{p_{u0}}\int\frac{d^{3}\mathbf{p}_{u}}{(2\pi)^{3}}Tr_{c,s}[\gamma^{\mu}(1-\gamma_{5})\mathcal{G}_{11}^{u}(P_{u})\gamma^{\nu}(1-\gamma_{5})\mathcal{G}_{11}^{d}(P_{d})]\nonumber \\
 & = & \int\frac{d^{3}\mathbf{p}_{u}}{(2\pi)^{3}}\mathrm{Tr}_{c,s}[\gamma^{\mu}(1-\gamma_{5})\gamma_{0}\mathcal{P}^{r_{u}}\Lambda_{\mathbf{p}_{u}}^{-e_{u}}\gamma^{\nu}(1-\gamma_{5})\gamma_{0}\mathcal{P}^{r_{d}}\Lambda_{\mathbf{p}_{d}}^{-e_{d}}]\nonumber \\
 &  & \times\left\{ \left[\frac{n_{u}(1-n_{d})}{-q_{0}+\epsilon_{u}+\epsilon_{d}}-\frac{(1-n_{u})n_{d}}{-q_{0}-\epsilon_{u}-\epsilon_{d}}\right]\left[1-n_{F}(\epsilon_{u})-n_{F}(\epsilon_{d})\right]\right.\nonumber \\
 &  & \left.+\left[\frac{(1-n_{u})(1-n_{d})}{-q_{0}-\epsilon_{u}+\epsilon_{d}}-\frac{n_{u}n_{d}}{-q_{0}+\epsilon_{u}-\epsilon_{d}}\right]\left[n_{F}(\epsilon_{u})-n_{F}(\epsilon_{d})\right]\right\} \label{eq:polar-tensor1}\end{eqnarray}
where $\mathcal{P}^{r_{u}}$ is the projector corresponding to the
excitation branch with index $r_{u}$. A sum over the excitation branch
$r_{u}$ is implied (the function inside the curly brackets depends
on the excitation branch $r_{u}$). $\Lambda_{\mathbf{p}}^{e}$ is
the energy projector defined by $\Lambda_{\mathbf{p}}^{e}=(1-e\gamma_{0}\widehat{\mathbf{p}}\cdot\bm{\gamma})/2$.
$n$ and $1-n$ are Bogoliubov coefficients for quasi-hole and quasi-particle
respectively. They are given by\begin{eqnarray}
B_{-}\equiv n & = & \frac{1}{2}-\frac{\xi}{2\epsilon}\nonumber \\
B_{+}\equiv1-n & = & \frac{1}{2}+\frac{\xi}{2\epsilon}\label{eq:bogliubov}\end{eqnarray}
where $\xi=k-e\mu$ and $\epsilon=\sqrt{\xi^{2}+\phi^{2}}$. Note
that the gap parameter $\phi$ depends on the excitation branch. In
Eq. (\ref{eq:polar-tensor1}) we only keep particle excitations, i.e.
with $e_{u}=e_{d}=+$. For the derivation of the Matsubara sum, see,
for example, \cite{Rischke:2000qz,Rischke:2000ra}. 

Then we are in the position of obtaining the imaginary part of the
retarded polarization tensor $\mathrm{Im}\Pi_{R}^{\mu\nu}$ by making
analytic extension $q_{0}\rightarrow q_{0}-i\delta$ in Eq. (\ref{eq:polar-tensor}):\begin{eqnarray}
\mathrm{Im}\Pi_{R}^{\mu\nu}(Q) & = & \pi\int\frac{d^{3}\mathbf{p}_{u}}{(2\pi)^{3}}\mathrm{Tr}_{c,s}[\gamma^{\mu}(1-\gamma_{5})\gamma_{0}\mathcal{P}^{r_{u}}\Lambda_{\mathbf{p}_{u}}^{-}\gamma^{\nu}(1-\gamma_{5})\gamma_{0}\mathcal{P}^{r_{d}}\Lambda_{\mathbf{p}_{d}}^{-}]\nonumber \\
 &  & \times n_{B}^{-1}(-q_{0})B_{a_{u}}B_{a_{d}}\delta(-q_{0}-a_{u}\epsilon_{u}+a_{d}\epsilon_{d})\nonumber \\
 &  & \times n_{F}(a_{u}\epsilon_{u})[1-n_{F}(a_{d}\epsilon_{d})],\label{eq:im-polar}\end{eqnarray}
where $a$ denotes the nature of the excitations: $a=+$ for quasi-particles
and $a=-$ for quasi-holes. The factor $n_{B}^{-1}(-q_{0})$ will
cancel $n_{B}(p-p_{1}+\mu_{e})$ in Eq. (\ref{eq:emissivity1}). A
sum over $a_{u}$ and $a_{d}$ is implied.

\subsection{Evaluation of neutrino emissivity}

Substituting Eq. (\ref{eq:im-polar}) into Eq. (\ref{eq:emissivity1}),
we get the number emissivity for neutrinos:\begin{eqnarray}
\frac{\partial}{\partial t}n_{\nu} & = & 2\pi\int\frac{d^{3}\mathbf{p}_{\nu}}{(2\pi)^{3}2p_{\nu}}\frac{d^{3}\mathbf{p}_{e}}{(2\pi)^{3}2p_{e}}\frac{d^{3}\mathbf{p}_{u}}{(2\pi)^{3}2p_{u}}\frac{1}{2p_{d}}|M|^{2}\nonumber \\
 &  & \times B_{a_{u}}B_{a_{d}}\delta[p_{\nu}-(p_{e}-\mu_{e})-a_{1}\epsilon_{1}^{u}+a_{2}\epsilon_{2}^{d}]\nonumber \\
 &  & \times n_{F}(p_{e}-\mu_{e})n_{F}(a_{u}\epsilon_{u})[1-n_{F}(a_{d}\epsilon_{d})].\label{eq:em-num-super}\end{eqnarray}
We have defined the matrix element\begin{eqnarray}
|M|^{2} & = & G_{F}^{2}p_{u}p_{d}\mathrm{Tr}_{c,s}[\gamma^{\mu}(1-\gamma_{5})\gamma_{0}\mathcal{P}^{r_{u}}\Lambda_{\mathbf{p}_{u}}^{-}\nonumber \\
 &  & \times\gamma^{\nu}(1-\gamma_{5})\gamma_{0}\mathcal{P}^{r_{d}}\Lambda_{\mathbf{p}_{d}}^{-}]L_{\mu\nu}(P_{\nu},P_{e})\nonumber \\
 & \approx & G_{F}^{2}\mathrm{Tr}_{c,s}[\gamma^{\mu}(1-\gamma_{5})\gamma_{0}\mathcal{P}^{r_{u}}P_{u,\sigma}\gamma^{\sigma}\nonumber \\
 &  & \times\gamma^{\nu}(1-\gamma_{5})\gamma_{0}\mathcal{P}^{r_{d}}P_{d,\rho}\gamma^{\rho}]L_{\mu\nu}(P_{\nu},P_{e})\label{eq:matrix1}\end{eqnarray}
where we have used the mass-shell condition for the $u$ and $d$
quark. The energy emissivity for the e-capture process is then \begin{eqnarray}
\frac{\partial E_{\nu}}{\partial t} & = & 2\pi\int\frac{d^{3}\mathbf{p}_{\nu}}{(2\pi)^{3}2p_{\nu}}\frac{d^{3}\mathbf{p}_{e}}{(2\pi)^{3}2p_{e}}\frac{d^{3}\mathbf{p}_{u}}{(2\pi)^{3}2p_{u}}\frac{1}{2p_{d}}|M|^{2}p_{\nu}\nonumber \\
 &  & \times B_{a_{u}}B_{a_{d}}\delta[p_{\nu}-(p_{e}-\mu_{e})-a_{u}\epsilon_{u}+a_{d}\epsilon_{d}]\nonumber \\
 &  & \times n_{F}(p_{e}-\mu_{e})n_{F}(a_{u}\epsilon_{u})[1-n_{F}(a_{d}\epsilon_{d})].\label{eq:em-en-super}\end{eqnarray}
One can also obtain the time rate for the momentum emitted along any
direction by replacing $p_{\nu}$ with $p_{\nu}\cos\theta$, where
$\theta$ is the angle between the neutrino momentum and this direction. 

\begin{figure}
\caption{\label{fig:phase-space}Phase space for Urca process.}

\vspace{0.5cm}\psfrag{d}{$d$}

\psfrag{u}{$u$}

\psfrag{e}{$e$}

\psfrag{mu_u}{$\mu_u$}

\psfrag{mu_d}{$\mu_d$}

\psfrag{mu_e}{$\mu_e$}

\psfrag{pd}{$p^F_{d}$}

\psfrag{pu}{$p^F_{u}$}

\psfrag{pe}{$p^F_{e}$}

\psfrag{pf_e=mu_e}{$p^F_{e}=\mu_{e}$}

\psfrag{pf=(1-xi)mu}{$p^F_{u,d}=(1-\frac{C_F\alpha_s}{2\pi})\mu_{u,d}$}

\psfrag{cos(theta_ue)}{$\mathrm{cos}\theta_{ue}\approx 1-\frac{C_F\alpha_s}{2\pi}$}

\psfrag{cos(theta_ud)}{$\mathrm{cos}\theta_{ud}\approx 1-\frac{C_F\alpha_s}{2\pi} \frac{\mu _e^2}{\mu _u\mu _d}$}

\includegraphics[scale=0.7]{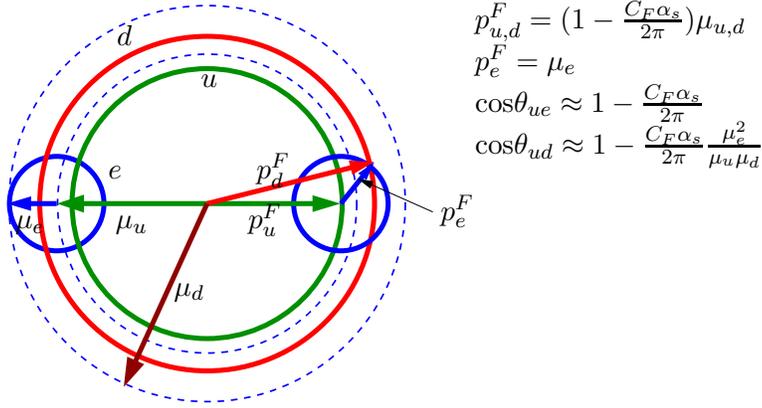}
\end{figure}

Before we evaluate Eq. (\ref{eq:em-en-super}), we analyze the phase
space given by the energy conservation for quasi-particles, i.e. restricted
by the delta function. If we don't take any interactions into account,
for excitations near the Fermi surface, the energy conservation condition
reads\begin{eqnarray*}
0 & = & p_{\nu}-(p_{e}-\mu_{e})-a_{u}\epsilon_{u}+a_{d}\epsilon_{d}\\
 & \approx & p_{\nu}-a_{u}\phi_{u}+a_{d}\phi_{d}\\
 & \approx & p_{\nu},\end{eqnarray*}
where we have used $p_{i}\approx\mu_{i}$ for $i=u,d,e$ and $p_{\nu}\gg\phi_{u,d}$.
We see in the above equation that in order to make the energy conserved,
the neutrino energy must vanish, which means the phase space for the
neutrino emission is very much suppressed. So including the interactions
for quasi-particles is necessary. The Fermi liquid correction to the
non-interacting degenerate Fermi gas is a better description of interacting
degenerate Fermi gas, which opens the phase space for neutrino emission.
The Fermi liquid correction to the quark Fermi momentum is due to
the quark-quark scattering via gluon exchange. The effect is that
it reduces the Fermi momentum by an amount linearly proportional to
the strong coupling constant as below: \begin{eqnarray*}
p_{Fi} & = & (1-\frac{C_{F}\alpha_{S}}{2\pi})\mu_{Fi},\end{eqnarray*}
where $i=u,d$. Actually the Fermi liquid corrections make it possible
to form the triangle by $p_{Fu}$, $p_{Fd}$ and $p_{Fe}$, as shown
in Fig. (\ref{fig:phase-space}). We can then determine the angles
$\theta_{ud}$, $\theta_{ue}$ and $\theta_{de}$ by the triangle
relation. For example, $\cos\theta_{ud}$ is given by\begin{eqnarray*}
\cos\theta_{ud} & = & \frac{p_{Fd}^{2}+p_{Fu}^{2}-p_{Fe}^{2}}{2p_{Fu}p_{Fd}}\\
 & \approx & \frac{(\mu_{d}^{2}+\mu_{u}^{2})(1-2\kappa)-\mu_{e}^{2}}{2\mu_{d}\mu_{u}(1-2\kappa)}\\
 & = & \frac{(2\mu_{u}^{2}+2\mu_{e}\mu_{u})(1-2\kappa)+\mu_{e}^{2}(1-2\kappa)-\mu_{e}^{2}}{2\mu_{d}\mu_{u}(1-2\kappa)}\\
 & \approx & 1-\frac{\mu_{e}^{2}\kappa}{\mu_{d}\mu_{u}},\end{eqnarray*}
where we defined $\kappa\equiv\frac{C_{F}\alpha_{S}}{2\pi}$. One
can see that if one sets $\kappa=0$ by turning off the interaction,
we have $\theta_{ud}=0$, which means the momenta of $u$ and $d$
are collinear. 

\begin{figure}
\caption{\label{fig:collision-integral}One way of simplifying the collision
integral}

\vspace{0.5cm}\psfrag{dpnu}{$d^3\mathbf{p}_\nu$}

\psfrag{dpd}{$d^3\mathbf{p}_d$}

\psfrag{dpu}{$d^3\mathbf{p}_u$}

\psfrag{dpe}{$d^3\mathbf{p}_e$}

\psfrag{momentum}{momentum conservation}

\psfrag{free1}{freedom 1}

\psfrag{free2}{freedom 2}

\psfrag{danu}{$d\phi_\nu d\theta_\nu$}

\psfrag{dad}{$d\phi_d d\theta_d$}

\psfrag{dau}{$d\phi_u d\theta_u$}

\psfrag{dpnudu}{$dp_\nu dp_d dp_u$}

\psfrag{energy}{energy conservation}

\psfrag{dthu}{$d\theta _u$}

\psfrag{dthd}{$d\theta _d$}

\psfrag{colinear}{collinear limit: $\theta _d\approx\theta _u$}

\psfrag{colint}{$dp_\nu dp_d dp_u d\theta _d$}

\psfrag{num}{numerical}

\includegraphics[scale=0.82]{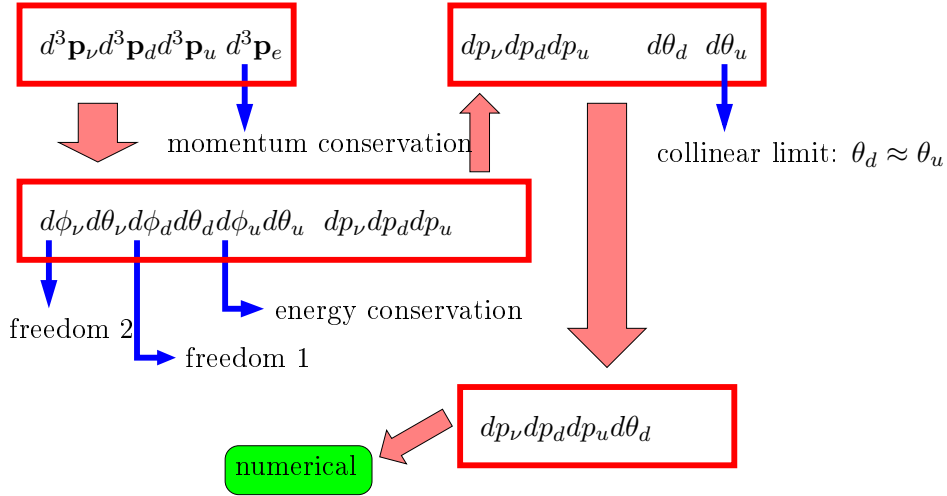}
\end{figure}

Another difficulty we have to tackle is the collision integral. Before
integrating out energy-momentum conservation and making any simplifications,
the integral is of 12 dimension. Carrying out the energy-momentum
conservation, we are still left with an 8-dimensional integral. For
the normal quark matter, the rest integral can be carried out analytically.
But for the color superconducting quark matter, one can only use numerical
method to evaluate it. One way of simplifying the collision integral
is illustrated in Fig. (\ref{fig:collision-integral}). 

\begin{figure}
\caption{\label{fig:g-k}Left panel: The functions $G(\varphi,\varphi)$ of
the neutrino emission contributions due to gapped modes in the CSL,
planar, polar and A phases. Right panel: The functions $K(\varphi)$
for four spin-one color superconductors.}

\includegraphics[scale=0.4]{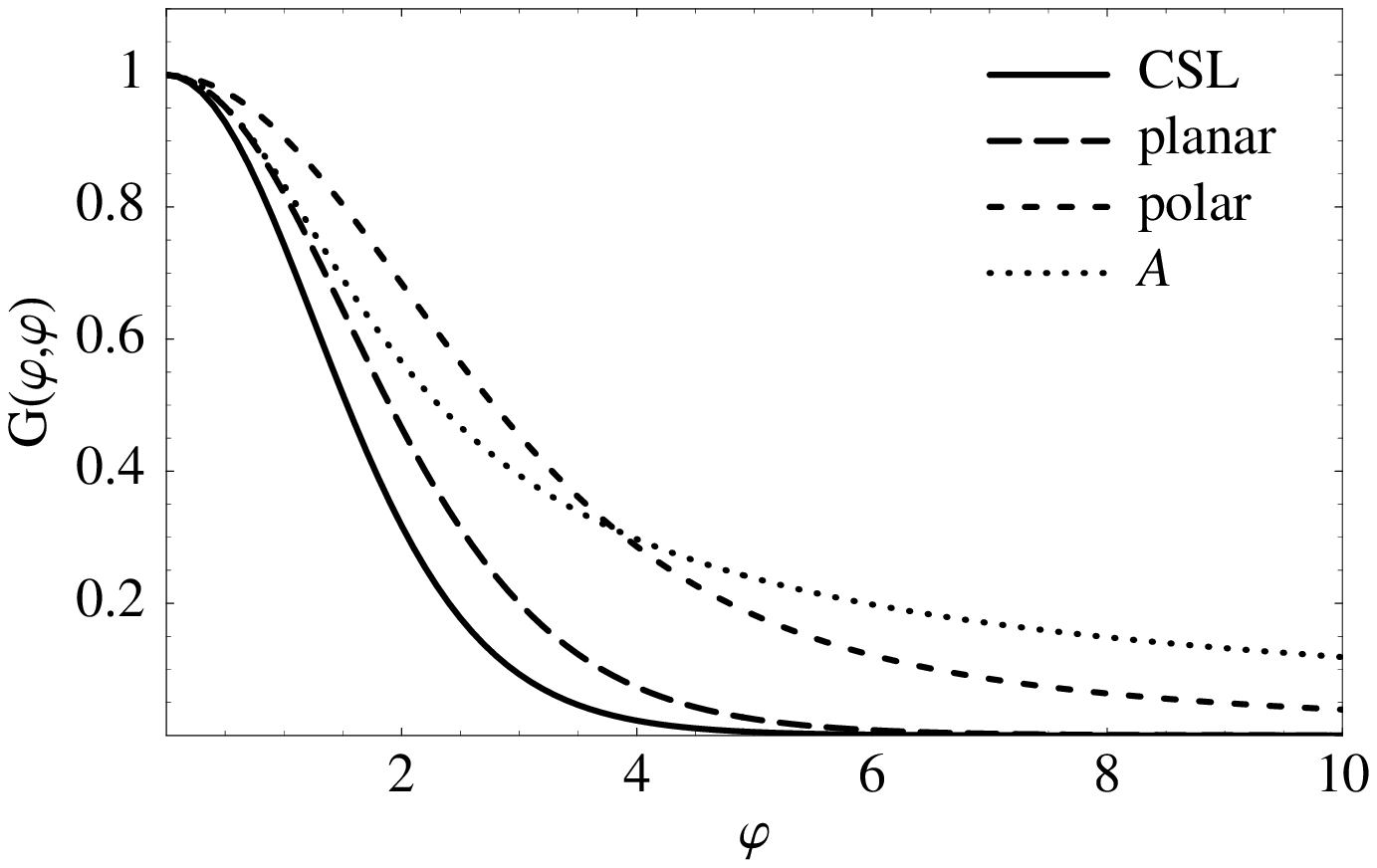}\includegraphics[scale=0.33]{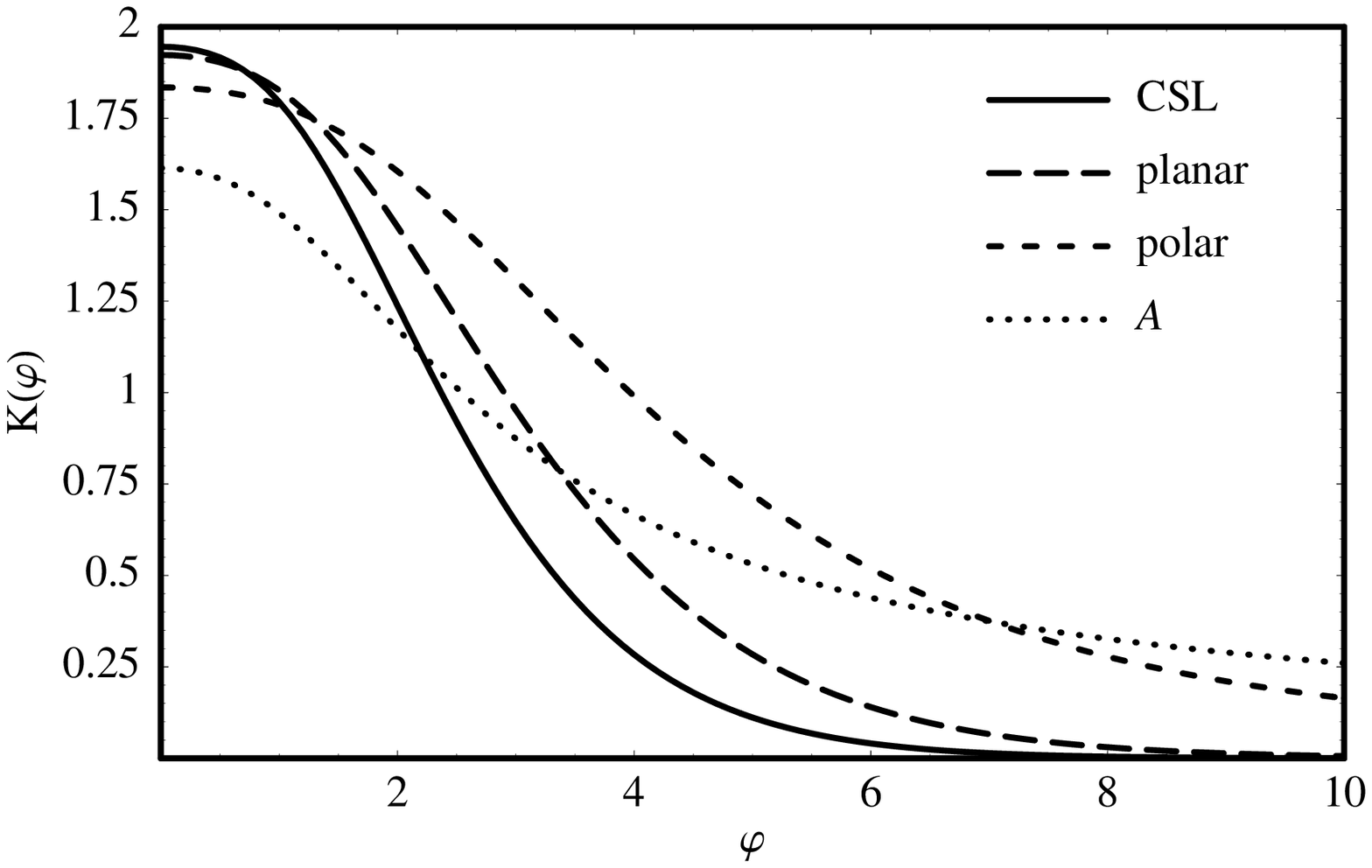}
\end{figure}

The neutrino emission and cooling rates for some of spin-1 phases
are calculated in \cite{Schmitt:2005wg,Wang:2006tg,Wang:2006xfa}.
The observational implication for the CSL phase has been studied in
Ref. \cite{Blaschke:2008vh,Blaschke:2008gd}. The basic ingredients
in the calculation are the quasiparticle dispersion relations, containing
the spin-one gap functions. We have studied in detail the effect of
an isotropic gap function (CSL phase) as well as of anisotropic gap
functions (planar, polar, A phases). The emissivity and the specific
heat can be expressed in terms of the scaled gap parameter, $\lyxmathsym{\textgreek{f}}=\lyxmathsym{\textgreek{f}}/T$
, \begin{eqnarray}
\epsilon _\nu &=& \frac{457}{630}\alpha_s G_F^2 T^6 \mu_e\mu_u\mu_d\,\left[ \frac{1}{3} + \frac{2}{3}\,G(\varphi_u,\varphi_d)\right] \, , \nonumber\\
c_V &=& T\sum_{f=u,d} \mu_f^2 \, \left[\frac 13+\frac 23 K (\varphi_f )\right] \, , 
\end{eqnarray}where  $\epsilon_{\nu}$ and $c_{V}$ are the emissivity and heat
capacity respectively, the numerical results for the functions $G$
and $K$ are presented in Fig. \ref{fig:g-k}.

In all four phases, also analytical expressions have been derived
in the large $\lyxmathsym{\textgreek{f}}$ limit (i.e., in the limit
of small temperatures). In the case of an isotropic gap function (CSL
phase), the well-known exponential suppression of the emissivity and
the specific heat is observed. We find that anisotropic gaps give
rise to different asymptotes in general. For example, the phases in
which the gap has point nodes (i.e., polar and A phases) show a power-law
instead of an exponential suppression at $\lyxmathsym{\textgreek{f}}\rightarrow\infty$.
The actual form of the power-law depends on the behavior of the gap
function in the vicinity of the nodes. While a linear behavior gives
rise to a suppression $\sim1/\lyxmathsym{\textgreek{f}}^{2}$, a quadratic
behavior leads to $\sim1/\lyxmathsym{\textgreek{f}}$.

\subsection{Anisotropic neutrino emissions in spin-1 transverse A phase}

As we mentioned that we can get the time rate of the momentum emitted
along any direction due to the neutrino emission by replacing $p_{\nu}$
with $p_{\nu}\cos\theta$ in Eq. (\ref{eq:em-en-super}). We can use
the resulting formula to calculate the momentum changing rate for
any color superconducting phases. Of particluar interest is the transverse
A phase of spin-1 pairing. This phase shows an asymmetry in neutrino
emission along one spatial direction, which can be regarded as an
example of parity violation in macroscopic stellar objects \cite{Schmitt:2005ee}.
What makes the transverse A phase so special is that it shows a very
spectacular helicity property in coupling with the neutrino emission. 

\begin{figure}
\caption{\label{fig:effective-gap}Effective gap of transverse A phase in neutrino
processes, shaded area shows the magnitude of the gap. }

\includegraphics[scale=0.6]{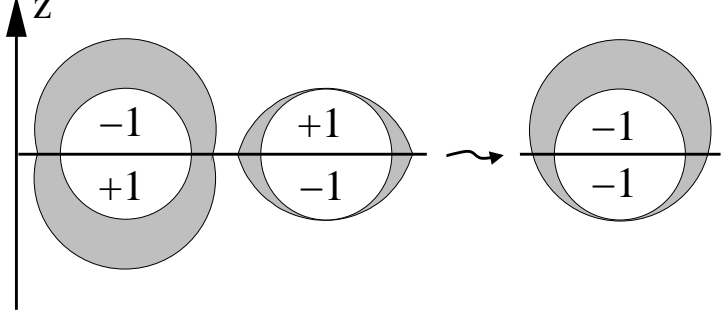}
\end{figure}

Let's look at the order parameter of the A phase: \begin{eqnarray*}
\mathcal{M} & = & J_{3}\left[\bm{\gamma}_{\perp}^{1}(\widehat{\mathbf{k}})+\bm{\gamma}_{\perp}^{2}(\widehat{\mathbf{k}})\right],\end{eqnarray*}
where $\bm{\gamma}_{\perp}=\bm{\gamma}-\widehat{\mathbf{k}}\bm{\gamma}\cdot\widehat{\mathbf{k}}$.
In the A phase, one direction is chosen to break the color and spin
symmetry. This direction is assumed to be along the z-axis. '1' and
'2' are two transverse directions to the z-axis. The excitation branches
are characterized by eigenvalues $\lambda_{r}$ of $\mathcal{M}^{\dagger}\mathcal{M}$.
Here are excitation energies\begin{eqnarray*}
\epsilon_{r} & = & \sqrt{(k-\mu)^{2}+\lambda_{r}\phi^{2}},\end{eqnarray*}
 where $\lambda_{r}$ are given by\begin{eqnarray*}
\lambda_{1,2} & = & (1+|\cos\theta_{k}|^{2})\\
\lambda_{3} & = & 0,\end{eqnarray*}
where $\theta_{k}$ is the angle between the momentum and the z-axis.
So we have spectral representation for $\mathcal{M}^{\dagger}\mathcal{M}$\begin{eqnarray*}
\mathcal{M}^{\dagger}\mathcal{M} & = & \lambda_{1}\mathcal{P}_{1}+\lambda_{2}\mathcal{P}_{2},\end{eqnarray*}
where the projectors are\begin{eqnarray*}
\mathcal{P}_{1,2} & = & \frac{1}{2}J_{3}^{2}\left[1\pm\mathrm{sgn}(\cos\theta_{k})\gamma_{0}\gamma_{5}\bm{\gamma}\cdot\widehat{\mathbf{k}}\right]\\
 & = & \frac{1}{2}J_{3}^{2}\left[1\mp\mathrm{sgn}(\cos\theta_{k})\right]H^{+}(\widehat{\mathbf{k}})\\
 &  & +\frac{1}{2}J_{3}^{2}\left[1\pm\mathrm{sgn}(\cos\theta_{k})\right]H^{-}(\widehat{\mathbf{k}}).\end{eqnarray*}
Here the helicity projectors are given by:\begin{eqnarray*}
H^{\pm}(\widehat{\mathbf{k}}) & = & \frac{1}{2}[1\pm\gamma_{5}\gamma_{0}\bm{\gamma}\cdot\widehat{\mathbf{k}}]\\
 & = & \frac{1}{2}[1\pm\bm{\Sigma}\cdot\widehat{\mathbf{k}}],\end{eqnarray*}
where $\bm{\Sigma}=\mathrm{diag}(\sigma,\sigma)$. We see that for
the first excitation branch $\lambda_{1}$, the mometum modes with
the positive z-component must be negative in helicity, while the modes
with negative z-component is positive in helicity. The second excitation
branch just shows an opposite helicity correlation. Since the Urca
processes involve exchanging W-bosons, only the particles with negative
helicity can participate. This fact can be seen by inserting $H^{h}$
with $h=\pm$ into Eq. (\ref{eq:matrix1}) where the Dirac trace becomes\begin{eqnarray}
\mathrm{Quark}\;\mathrm{Trace} & = & \mathrm{Tr}_{s}[\gamma^{\mu}(1-\gamma_{5})\gamma_{0}H^{h_{u}}(\widehat{\mathbf{p}}_{u})\Lambda_{\mathbf{p}_{u}}^{-}\nonumber \\
 &  & \times\gamma^{\nu}(1-\gamma_{5})\gamma_{0}H^{h_{d}}(\widehat{\mathbf{p}}_{d})\Lambda_{\mathbf{p}_{d}}^{-}].\label{eq:trace-M}\end{eqnarray}
 We use \begin{eqnarray*}
\bm{\Sigma}\cdot\widehat{\mathbf{p}}\Lambda_{\mathbf{p}}^{-} & = & -\gamma_{5}\Lambda_{\mathbf{p}}^{-}\\
H^{h_{u}}(\widehat{\mathbf{p}})\Lambda_{\mathbf{p}}^{-} & = & \frac{1}{2}(1-h_{u}\gamma_{5})\Lambda_{\mathbf{p}}^{-}\end{eqnarray*}
to rewrite the trace (\ref{eq:trace-M}) as \begin{eqnarray*}
\mathrm{Quark}\;\mathrm{Trace} & \sim & \mathrm{Tr}_{s}[\gamma^{\mu}(1-\gamma_{5})\gamma_{0}(1-h_{u}\gamma_{5})(\widehat{\mathbf{p}}_{u})\Lambda_{\mathbf{p}_{u}}^{-}\\
 &  & \times\gamma^{\nu}(1-\gamma_{5})\gamma_{0}(1-h_{d}\gamma_{5})\Lambda_{\mathbf{p}_{d}}^{-}]\\
 & = & 0,\;\;\mathrm{for}\; h_{u,d}=+.\end{eqnarray*}
We see that components with positive helicity $h_{u,d}=+$ vanish.
This selects upper part of the first branch and the lower part of
the second branch, which gives an effective mode showing asymmetry
along the z-direction, see Fig. (\ref{fig:effective-gap}). We know
that the larger the gap is, the more suppression the neutrino emission
gets. So from the right graph of Fig. (\ref{fig:effective-gap}) we
see that the effective gap is larger in the upper half plane (positive
z-component) than in the lower one (negative z-component). This means
that the neutrino emission along the positive z-axis is more suppressed
than in the negative z-axis. This asymmetry can be regarded as an
example for parity violation in macroscopic scales.

\section*{Acknowledgement}

The author thanks D. Blaschke, M. Buballa, J. Deng, P. Reuter, D.
H. Rischke, A. Schmitt, A. Sedrakian and I. A. Shovkovy for helpful
discussions. Special thanks to A. Schmitt for critically reading all
the text with many suggestions and corrections the artcile has been
much improved. The author is supported in part by the '100 talents'
project of Chinese Academy of Sciences (CAS) and by the National Natural
Science Foundation of China (NSFC) under the grants 10675109 and 10735040.

\end{document}